\documentclass[useAMS,usenatbib,babel]{mn2e}

\usepackage[english,english]{babel}
\usepackage{amsmath}
\usepackage{amssymb,amsfonts,textcomp}
\usepackage{array}
\usepackage{supertabular}
\usepackage{hhline}
\usepackage{hyperref}
\usepackage[usenames]{color}
\hypersetup{dvips, colorlinks=true, linkcolor=black, citecolor=black, filecolor=blue, urlcolor=blue}
\usepackage[dvips]{graphicx}

\def\gtrsim{\lower.5ex\hbox{$\; \buildrel > \over \sim \;$}}
\usepackage{graphicx}

\definecolor{grey}{rgb}{0.75,0.75,0.75}
\definecolor{Orange}{rgb}{1.0,0.5,0.15}
\definecolor{brown}{rgb}{0.7,0.25,0.0}
\definecolor{pink}{rgb}{1.0,0.5,0.5}
\definecolor{darkerred}{rgb}{0.8,0,0}
\definecolor{darkerblue}{rgb}{0,0,0.8}
\definecolor{Blue}{rgb}{0,0.08,0.65}
\definecolor{Red}{rgb}{0.65,0.08,0.05}
\definecolor{Green}{rgb}{0.15,0.45,0.25}

\begin{document}

\author[Y. Dubois et al. ]{
\parbox[t]{\textwidth}{
Yohan Dubois$^{1,2}$\thanks{E-mail: dubois@iap.fr}, Marta Volonteri$^{1,2}$, Joseph Silk$^{1,2,3,4}$, Julien Devriendt$^{3,5}$, \\
Adrianne Slyz$^3$ and Romain Teyssier$^6$}
\vspace*{6pt} \\
$^{1}$ Sorbonne Universit\'es, UPMC Univ Paris 06, UMR 7095, Institut d'Astrophysique de Paris, F-75014, Paris, France\\
$^{2}$ CNRS, UMR 7095, Institut d'Astrophysique de Paris, F-75014, Paris, France\\
$^{3}$ Sub-department of Astrophysics, University of Oxford, Keble Road, Oxford OX1 3RH\\
$^{4}$ Department of Physics and Astronomy, The Johns Hopkins University Homewood Campus, Baltimore, MD 21218, USA\\
$^{5}$ Observatoire de Lyon, UMR 5574, 9 avenue Charles Andr\'e, F-69561 Saint Genis Laval, France \\
$^{6}$ Institute f\"ur Theoretische Physik, Universit\"at Z\"urich, Winterthurerstrasse 190, CH-8057 Z\"urich, Switzerland \\
}
\date{Accepted 2015 June 23.  Received 2015 June 15; in original form 2015 March 26.}

\title[SN-regulated BH growth]
{Black hole evolution: I. Supernova-regulated black hole growth}

\maketitle

\begin{abstract}
{The growth of a supermassive black hole (BH) is determined by how much gas the host galaxy is able to feed it, which in turn is controlled by the cosmic environment, through galaxy mergers and accretion of cosmic flows that time how galaxies obtain their gas, but also by internal processes in the galaxy, such as star formation and feedback from stars and the BH itself. 
In this paper, we study the growth of a $10^{12}\, \rm M_\odot$ halo at $z=2$, which is the progenitor of a group of galaxies at $z=0$, and of its central BH by means of a high-resolution zoomed cosmological simulation, the Seth simulation.
We study the evolution of the BH driven by the accretion of cold gas in the galaxy, and explore the  efficiency of  the feedback from supernovae (SNe).
For a relatively inefficient energy input from SNe, the BH grows at the Eddington rate from early times, and reaches self-regulation once it is massive enough. We find that at early cosmic times $z>3.5$, efficient feedback from SNe forbids the formation of a settled disc as well as the accumulation of dense cold gas in the vicinity of the BH and starves the central compact object.
As the galaxy and its halo accumulate mass, they become able to confine the nuclear inflows provided by major mergers and the BH grows at a sustained near-to-Eddington accretion rate. 
We argue that this mechanism should be ubiquitous amongst low-mass galaxies, corresponding to galaxies with a stellar mass below $\lesssim 10^9\, \rm M_\odot$ in our simulations.
}
\end{abstract}


\section{Introduction}

Supermassive black holes (BH) are common compact objects observed in the centre of galaxies.
Their are suspected to grow along with their host galaxy as observations suggest a strong scaling between BH masses and galaxy properties~\citep[e.g.][]{magorrianetal98, tremaineetal02, haring&rix04, kormendy&ho13}.
Gas accretion drives most of the mass growth of black holes at high-redshift, and this accretion mechanism returns a fraction of the rest-mass accreted energy into effective feedback for the host-galaxy, which in turns can explain the observed scaling relation between black holes and galaxies~\citep{silk&rees98, king03, wyithe&loeb03}.
Such a behavior has been successfully implemented in modern cosmological simulations and the impact of active galactic nuclei (AGN) feedback on the galaxy mass content, the circumgalactic gas properties and the halo mass distribution can be dramatic~\citep{dimatteoetal05, crotonetal06, boweretal06, sijackietal07, booth&schaye09, duboisetal10, duboisetal12agnmodel, teyssieretal11}.

AGN feedback is supposed to have a strong impact on the stellar and gaseous content of the most massive galaxies, helping to solve the so-called `over-cooling' problem, i.e. the expected fast cooling times and consequently high star formation rates predicted for galaxies, at odd with observational results.  This was for instance highlighted by comparing theoretical models of galaxy evolution to the observed galaxy mass function \citep{crotonetal06, boweretal06}. Theoretical models that do not include any form of negative feedback from stars or AGN over-predict the mass function at both the low- and high-mass end. Including stellar feedback, namely through supernova (SN) explosions, mitigates the problem at the low-mass end, but the SN energy input is not sufficient to eject gas from galaxies with virial halo velocity above a critical value of $\sim 100\, \rm km\, s^{-1}$~\citep{dekel&silk86}, nor affect the cold stream inflow in high redshift galaxies~\citep{powelletal11}.  The energetics involved in growing BHs are in principle comparable to, if not higher than, the binding energy of the host galaxy \citep[e.g.][]{silk&rees98,Fabian2012}. If a fraction of the rest-mass accreted energy can be tapped, it can profoundly affect the gas distribution in and around the galaxy.
AGN-driven outflows can push the gas in the outskirts of the halo, even impacting the cold streams~\citep{duboisetal13, costaetal14}. It will change the gas properties by increasing its temperature and decreasing its density, thereby, delaying the gas inflows onto galaxies and will impact the star formation rate of the host galaxy.

The masses of BHs at the center of galaxies are observed to scale linearly with their host bulge with a BH to bulge mass ratio of $M_{\rm BH}/M_{\rm b}\sim 10^{-3}$. Assuming that the amount of gas mass impacted by the central AGN wind scales linearly with the bulge mass, the wind velocity $\dot u_{\rm w}\propto L_{\rm AGN}/ M_{\rm gas}\propto M_{\rm BH}/M_{\rm b}$ is constant for BHs accreting at Eddington ($L_{\rm AGN}\propto M_{\rm BH}$).
Therefore, the AGN wind velocity should be independent of the galaxy mass and AGN feedback have a non-negligible impact on high-mass galaxies as well as low-mass galaxies provided that the accretion of gas on the BH is unarrested.
Measurements of the BH mass correlation with the pseudobulge mass show a larger scatter and fall below the correlation between BH mass and classical bulge relation~\citep{hu08, kormendyetal11, sanietal11}. 
A similar deficit in BH mass is also observed for low-mass BHs~\citep{greeneetal10, jiangetal11, graham&scott15}.
SN feedback that is an important driver of gas-mass loss in dwarf galaxies~\citep[e.g.][]{springel&hernquist03, dubois&teyssier08winds, dallavecchia&schaye08, dallavecchia&schaye12, hopkinsetal14} also turns dark matter cusps into cores~\citep{gnedin&zhao02, mashchenkoetal06, pontzen&governato12, teyssieretal13}, and reduces the depth of the gravitational well.
Therefore, the SN feedback activity in low-mass galaxies can probably inhibit the activity of the central BH by depleting the central gas reservoir, and limiting the impact of AGN feedback.
\cite{2013NatSR....E1738B} performed a numerical study of the interplay between SN and AGN feedback in a large-volume cosmological simulation (see also~\citealp{sijackietal07,crainetal15}; and \citealp{newton&kay13} for detailed isolated and merger simulations. 
They found that SN and AGN feedback weaken each other's strength when jointly included in their models, compared to including only one of them at a time.
However, they do not discuss whether SN feedback is able to impact the growth of BHs.

This paper is part of a series of three papers investigating the connection between SN feedback and BH growth (paper I, this paper), SN-driven gas turbulence and BH spin evolution ~\citep[][paper II]{duboisetal14spinturb}, and the impact of cosmic galaxy evolution on the spin of central supermassive BHs~\citep[][paper III]{duboisetal13spinlss}. The main purpose of this paper is to investigate the mass evolution of the central BH of a zoomed cosmological halo at high redshift, called the Seth\footnote{This name refers to the Egyptian god of storms, disorder and violence.} suite of simulations, that have enough resolution to start resolving the substructures of the interstellar medium (ISM) over several Gyrs. We test two implementations of SN feedback with the kinetic modelling of~\cite{dubois&teyssier08winds}, and the prescription from~\cite{teyssieretal13} in order to probe how much the strength of SN activity can alter the growth of the central BH.

In Section~\ref{section:numerics} we introduce the initial conditions and the numerical models for the physics of galaxy formation, the BH mass growth and its associated AGN feedback.
In Section~\ref{section:result} we detail our results on BH mass evolution in the Seth simulations and how SN feedback influences BH growth.
Finally, in Section~\ref{section:conclusion} we summarize our results.

\section{Numerical set-up}
\label{section:numerics}

The simulations are run with the adaptive mesh refinement code {\sc ramses} \citep{teyssier02}.
The evolution of the gas is followed using a second-order unsplit Godunov scheme for the Euler equations. 
The Harten-Lax-Van Leer contact (HLLC,~\citealp{toroetal94}) Riemann solver with a MinMod total variation diminishing scheme to reconstruct the interpolated variables from their cell-centered values is used to compute fluxes at cell interfaces.
Collisionless particles (DM, star and BH particles) are evolved using a particle-mesh solver with a cloud-in-cell interpolation.

\begin{figure*}
  \centering{\resizebox*{!}{8.5cm}{\includegraphics{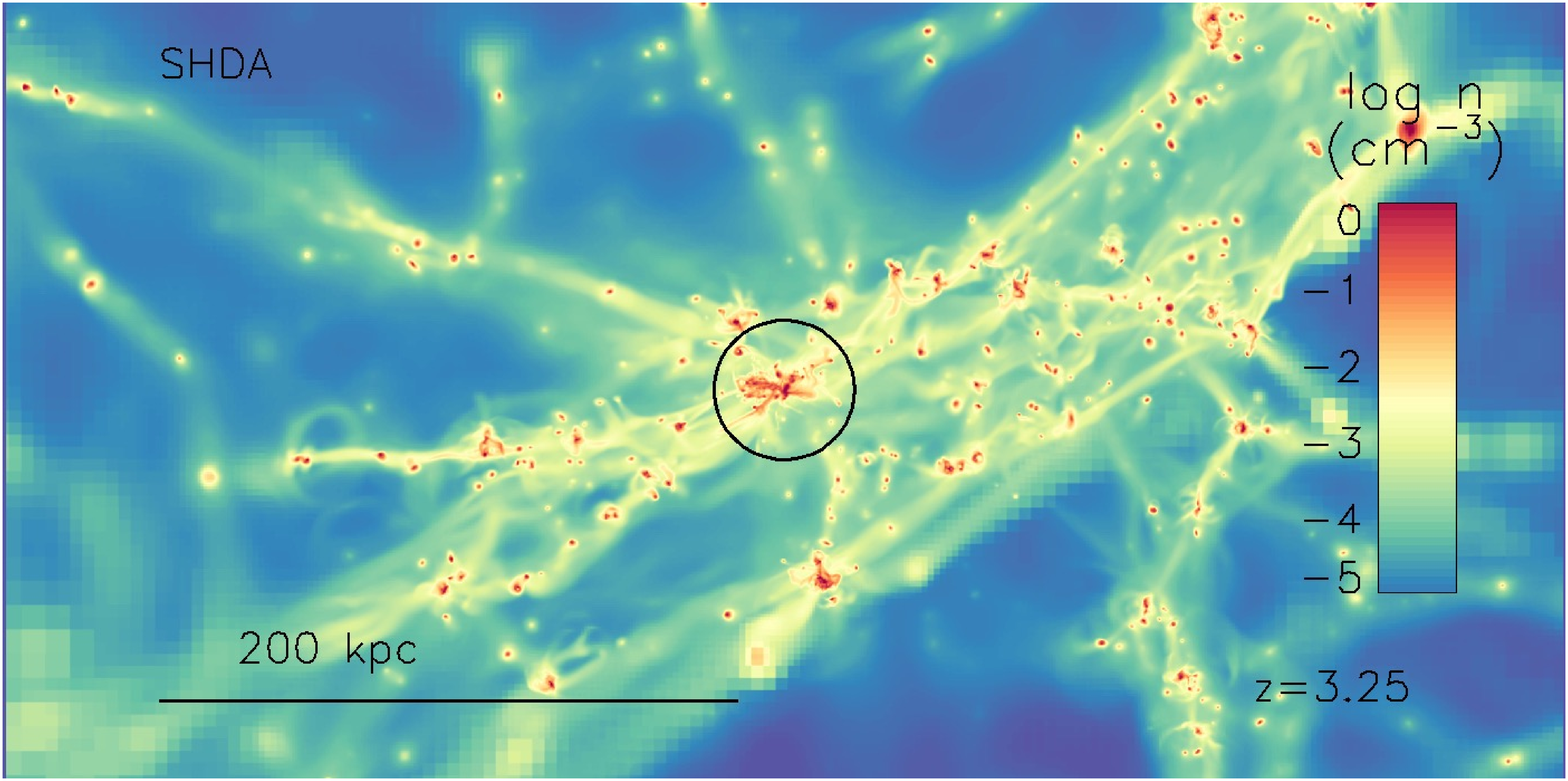}}}
  \centering{\resizebox*{!}{8.5cm}{\includegraphics{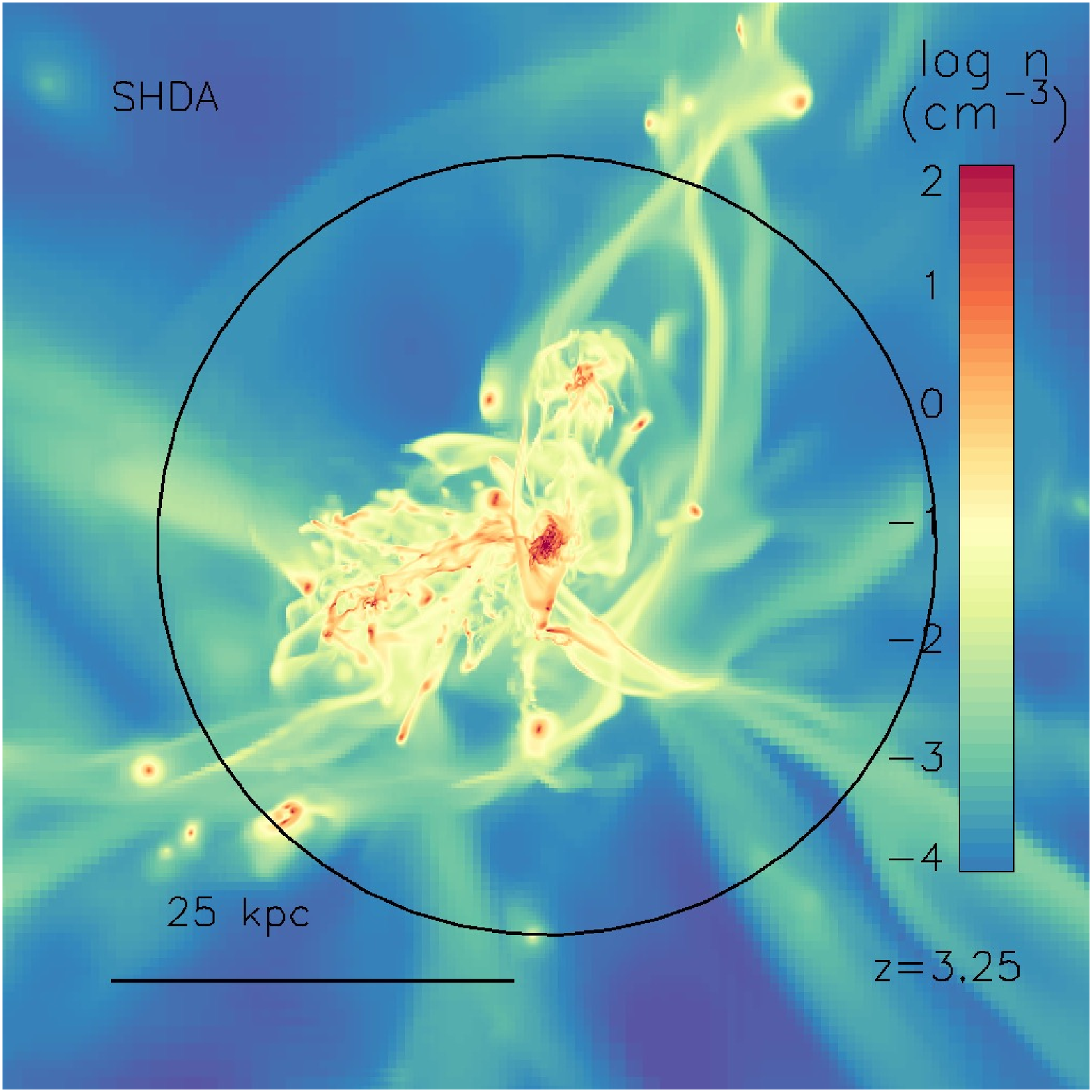}}}
  \centering{\resizebox*{!}{8.5cm}{\includegraphics{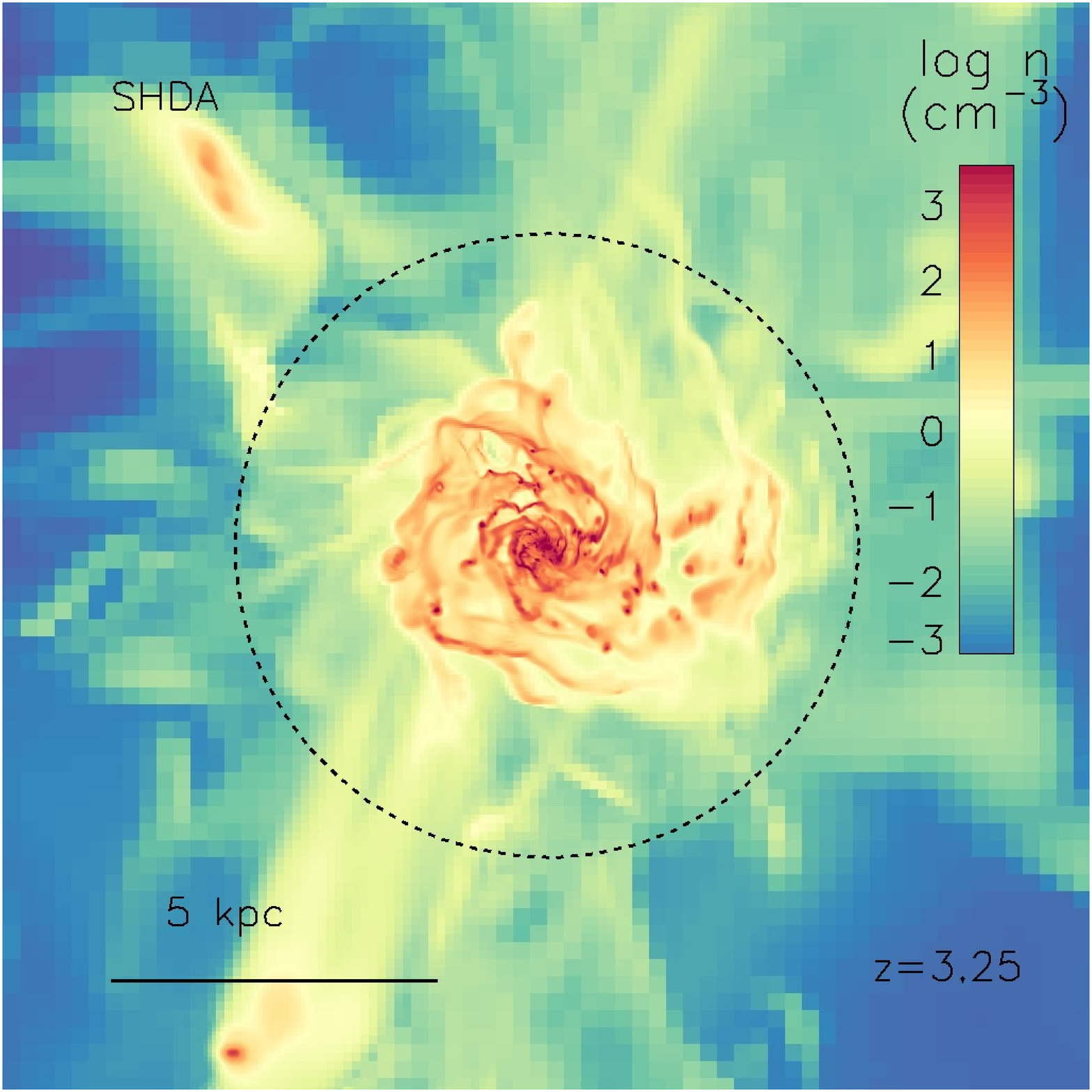}}}
  \caption{Projected gas densities of the Seth simulation (SHDA run) for the zoom region (top panel), the central halo (left panel), and the central galaxy (right panel) at $z=3.25$. The solid circle indicates the virial radius $r_{\rm vir}$ of the dark matter halo, and the dashed circle is for $0.1 r_{\rm vir}$. We can see several cosmic cold filaments that connect to the central halo with a a complex gas distribution once they reach the central galaxy. The galaxy clearly exhibits a rotating disc-like structure with dense star-forming clumps of gas.}
    \label{fig:zoom32}
\end{figure*}

\subsection{The Seth suite of simulations}

We assume a $\Lambda$CDM cosmology with total matter density $\Omega_{m}=0.3$, baryon density $\Omega_b=0.045$, dark energy density $\Omega_{\Lambda}=0.7$, amplitude of the matter power spectrum $\sigma_8=0.8285$, $n_{\rm s}=0.9635$ spectral index and  Hubble constant $H_0=68.14\, \rm km\, s^{-1} \, \rm Mpc^{-1}$ consistent with the Planck data \citep{planckcosmo} for the initial conditions produced with~{\sc music}~\citep{hahn&abel11}.
The box size of our simulation is $L_{\rm box}=50\,  h^{-1}\, \rm Mpc$, with a coarse grid of $256^3$ dark matter (DM) particles corresponding to a DM mass resolution of $M_{\rm res,coarse}=8\times 10^8 \, \rm M_\odot$.
A high-resolution region is defined around a halo of $M_{\rm vir}=10^{12}\, \rm M_\odot$ at $z=2$ that contains only high-resolution DM particles within 2 $r_{\rm vir}$ ($r_{\rm vir}=100$~kpc) with mass $M_{\rm res,high}=2\times 10^5 \, \rm M_\odot$. The halo is a progenitor of a group of galaxies whose mass is $M_{\rm vir}=5 \times 10^{12}\, \rm M_\odot$ at $z=0$.

The mesh is refined up to $\Delta x=8.7\, \rm pc$ (maximum level equals $l_{\rm max}=21$ at redshift $z=3$) or $\Delta x=34.8 \, \rm pc$ ($l_{\rm max}=19$ at redshift $z=3$) using a quasi-Lagrangian strategy: when more than 8 DM particles lie in a cell, or if the baryon density is larger than 8 times the initial DM resolution.
The minimum cell size is kept roughly constant in proper size with redshift, i.e. an additional level of refinement is added every $a_{\rm exp}=n\times0.1$ (where $n=1,2,4,8$ and $a_{\rm exp}$ is the expansion factor of the universe) up to level 21 (or 19) at $a_{\rm exp}=0.33$ ($z=2$). 

\subsection{Physics of galaxy formation}
\label{section:subgridgal}

Gas is allowed to cool by H and He cooling with a contribution from metals using a~\cite{sutherland&dopita93} model for temperatures above $T_0=10^3$~K, which is the minimum temperature of the gas allowed through radiative losses.
Heating from a uniform UV background takes place below redshift $z_{\rm reion}=10$ following~\cite{haardt&madau96}.
Metallicity is modelled as a passive variable for the gas advected with the flow (whose composition is assumed to be solar) and is altered by the injection of gas ejecta during SN explosions and stellar mass losses.
We assume a zero initial metallicity.
The gas follows an equation of state for ideal monoatomic gas with adiabatic index $\gamma=5/3$.

The star formation process is modelled with a Schmidt law:
$\dot \rho_*= \epsilon_* {\rho / t_{\rm ff}}\, ,$ where $\dot \rho_*$ is the star formation rate density, $\epsilon_*$ the constant star formation efficiency, and $t_{\rm ff}$ the local free-fall time of the gas.
We choose a low star formation efficiency $\epsilon_*=0.02$ consistent with observations of giant molecular clouds~\citep{krumholz&tan07} and surface density relations of galaxies~\citep{kennicutt98}.
Star formation is allowed in regions where the gas density exceeds $n_0=250\, \rm H\, cm^{-3}$ for $\Delta x=8.7\, \rm pc$ ($n_0=15\, \rm H\, cm^{-3}$ for $\Delta x=34.8\, \rm pc$) and where the temperature is not larger than $T_0$ (corrected for the contribution of the polytropic equation of state, see below).
The stellar mass resolution is $m_{\rm s, res}=n_0\Delta x^3\simeq5\times 10^3 \, \rm M_\odot$ for the $\Delta x=8.7\, \rm pc$ resolution runs ($m_{\rm s, res}\simeq2\times 10^4 \, \rm M_\odot$ for the $\Delta x=34.8 \, \rm pc$).
The gas pressure is artificially enhanced above $\rho > \rho_0$ assuming a polytropic equation of state $T=T_0(\rho/\rho_0)^{\kappa-1}$ with polytropic index $\kappa=2$ to avoid excessive gas fragmentation.
The value of $n_0$ for both the density threshold of star formation and the polytropic equation of state is chosen so that the Jeans length is always resolved with 4 cells.

Feedback from stars is taken into account assuming a Salpeter initial mass function with $\eta_{\rm SN}=0.1$ of the mass fraction of stars ending up into a type II supernova and releasing $e_{\rm SN}=10^{50} \, \rm erg\, M_\odot^{-1}$.
The numerical implementation is done in two flavors.
\emph{(i) kinetic}: We use the implementation from~\cite{dubois&teyssier08winds} based on a Sedov-blast wave solutions: mass, momentum and energy are deposited in the surrounding gas cells so that it mimics the Sedov solution with a mass loading of $f_w=m_{\rm ej}/m_{\rm s,res}=1$, where $m_{\rm ej}$ is the gas mass of the ejecta taken from the central cell and redistributed to all the cells within the bubble of size $r_{\rm SN}=1.5 \Delta x$ (see~\citealp{dubois&teyssier08winds} for more details).
\emph{(i) delayed cooling}: An alternative modeling is that of~\cite{teyssieretal13} which locally delays the gas cooling after an explosion (in the spirit of~\citealp{stinsonetal06}), where energy is released both in the classical energy component and in a tracer component, which is denoted `non-thermal' and is passively advected with the flow.
The passive non-thermal energy decays on a typical time-scale of $t_{\rm diss}=0.8\,\rm Myr$ for $\Delta x =8.7 \, \rm pc$ ($t_{\rm diss}=2 \, \rm Myr$ for $\Delta x =34.8 \, \rm pc$), and the gas cooling is not allowed until the velocity dispersion associated to the non-thermal energy component is larger than $\sigma_{\rm NT}=50 \, \rm km\, s^{-1}$. 
These values of $t_{\rm diss}$ and $\sigma_{\rm NT}$ are different from the initial choice adopted in~\cite{teyssieretal13} who set them up to the typical molecular cloud lifetimes~\citep{williams&mckee97}.
Here, we prefer to adopt a resolution-dependent approach, so that the explosion is able to propagate over a few cell resolution elements.
The choice of $\sigma_{\rm NT}$ and $t_{\rm diss}$ are discussed in Appendix~\ref{app:delcool}.

\subsection{Model for BH growth and AGN feedback}

\begin{table} 
  \caption{List of the Seth simulation runs. First column indicates the simulation name. Second column indicates the minimum cell size. Third column indicates which SN feedback model is used. Fourth column indicates wether AGN feedback is included. Fifth column indicates the final redshift of the simulation. See the text for details.}
  \label{tbl:Sethsuite}
\begin{center}
  \begin{tabular}{l c c c c c}
     \hline
     Name & $\Delta x$ (pc) & SN & AGN & $z_{\rm end}$\\
     \hline
    SL & 34.8 & no & no & 2\\
    SLD & 34.8 & delayed cooling & no & 2\\
    SLDA & 34.8 & delayed cooling & yes & 2\\
    SHDA & 8.7 & delayed cooling & yes & $2.4$\\
    SLK & 34.8 & kinetic & no & 2\\
    SLKA & 34.8 & kinetic & yes & 2\\
    SHKA & 8.7 & kinetic & yes & $3.5$\\
     \hline
  \end{tabular}
\end{center}
\end{table}

We use the AGN feedback model described in~\cite{duboisetal12agnmodel}.
BHs are created at loci where both the gas density and stellar density is larger than the density threshold for star formation $\rho_0$ with an initial seed mass of $M_{\rm seed}=min(0.75 \rho_{\rm gas}\Delta x^3,10^5\, \rm M_\odot)$ (i.e. between $[1.5\times 10^4,10^5]\, \rm M_\odot$ for $\Delta x=34.8 \, \rm pc$ and $[3.7\times 10^3,10^5]\, \rm M_\odot$ for $\Delta x=8.7 \, \rm pc$).
Thus, the initial seed mass can be smaller than $10^5\, \rm M_\odot$ (BH mass through direct collapse,~\citealp{begelmanetal06, spaans&silk06}) when gas has not accumulated enough to contain this amount of gas mass but has formed enough stars to reach the stellar density criterion.
The accretion rate onto BHs follows the Bondi-Hoyle-Lyttleton~\citep{bondi52} rate
$\dot M_{\rm BH}=4\pi \alpha G^2 M_{\rm BH}^2 \bar \rho / (\bar c_s^2+\bar u^2) ^{3/2},$
where $G$ is the gravitational constant, $M_{\rm BH}$ is the BH mass, $\bar \rho$ is the average gas density, $\bar c_s$ is the average sound speed, $\bar u$ is the average gas velocity relative to the BH velocity, and $\alpha$ is a dimensionless boost factor with $\alpha=(\rho/\rho_0)^2$ when $\rho>\rho_0$ and $\alpha=1$ otherwise~\citep{booth&schaye09} in order to account for our inability to capture the colder and higher density regions of the ISM.
Averaged quantities are summed over 4 cells in radius and kernel-weighted with a Bondi radius-dependancy (see~\citealp{duboisetal12agnmodel} for the exact form of the averaging procedure).
The effective accretion rate onto BHs is capped at the Eddington accretion rate:
$\dot M_{\rm Edd}=4\pi G M_{\rm BH}m_{\rm p} / (\epsilon_{\rm r} \sigma_{\rm T} c),$
where $\sigma_{\rm T}$ is the Thompson cross-section, $c$ is the speed of light, $m_{\rm p}$ is the proton mass, and $\epsilon_{\rm r}$ is the radiative efficiency, assumed to be proportional to the spin of the BH with $\epsilon_{\rm r}=1- E_{\rm isco}=1-\sqrt{1-2/(3r_{\rm isco})}$. Spin magnitude and direction are followed self-consistently, on the fly, in this simulation. We refer the reader to paper II \citep[see also][]{duboisetal13spinlss} for details and we just summarize here the main assumptions.  The spin magnitude of BHs is modified by accretion of gas through the expression derived by~\cite{bardeen70}. If the angular momentum of the accreted gas is misaligned with the direction of the BH spin, Lense-Thirring effect causes the innermost parts of the disc to rotate within the equatorial plane of the BH and a warped disc is created that eventually completely aligns or anti-aligns with the BH spin. The case for anti-alignment of the BH with the disc requires that~\citep{kingetal05} $\cos \theta < - {0.5J_{\rm d}/ J_{\rm BH}}$, where $\bmath{J}_{\rm d}$ and $\bmath{J}_{\rm BH}$ are the angular momenta of disc and BH respectively, and $\theta$ is the angle between $\bmath{J}_{\rm BH}$ and $\bmath{J}_{\rm d}$. To model the unresolved accretion disc we adopt the~\cite{shakura&sunyaev73}  thin accretion disc solution, and define  $J_{\rm d}$ to be the angular momentum calculated at smaller between the warp radius, i.e., the radius within which gas can effectively transfer its angular momentum to the BH within a viscous timescale, or the self-gravity radius, i.e. the radius beyond which the disc becomes prone to fragmentation. Finally, we update BH spins after BH-BH coalescences  using the analytical fit of~\cite{rezzollaetal08} derived from relativistic numerical simulations of BH binaries.

In order to avoid spurious oscillations of the BH in the gravitational potential well due to external perturbations and finite resolution effects, we introduce a drag force that mimics the dynamical friction exerted by the gas onto a massive particle.
This dynamical friction is proportional to $F_{\rm DF}=f_{\rm gas} 4 \pi \alpha \rho (G M_{\rm BH}/\bar c_s)^2$, where $f_{\rm gas}$ is a fudge factor whose value is between 0 and 2 and is a function of the Mach number ${\mathcal M}=\bar u/\bar c_s<1$~\citep{ostriker99, chaponetal13}, and where we introduce the boost factor $\alpha$ for the same reasons than stated above.
A complementary approach, that we do not model here since gas dynamical friction is sufficient, is to explicitly model the unresolved dynamical friction exerted by DM particles onto BHs~\citep{tremmeletal15}.

The AGN feedback is a combination of two different modes, the so-called \emph{radio} mode operating when $\chi=\dot M_{\rm BH}/\dot M_{\rm Edd}< 0.01$ and the \emph{quasar} mode active otherwise.
The quasar mode corresponds to an isotropic injection of thermal energy into the gas within a sphere of radius $\Delta x$, at an energy deposition rate: $\dot E_{\rm AGN}=\epsilon_{\rm f} \epsilon_{\rm r} \dot M_{\rm BH}c^2$,
where $\epsilon_{\rm f}=0.15$ for the quasar mode is a free parameter chosen to reproduce the $M_{\rm BH}$-$M_{\rm b}$, $M_{\rm BH}$-$\sigma_{\rm b}$, and BH density in our local Universe (see \citealp{duboisetal12agnmodel}).
At low accretion rates on the other hand, the radio mode deposits the AGN feedback energy into a bipolar outflow with a jet velocity of $10^4\,\rm km\, s^{-1}$ into a cylinder with a cross-section of radius $\Delta x$ and height $2 \, \Delta x$ following~\cite{ommaetal04} (more details about the jet implementation are given in~\citealp{duboisetal10}).
The efficiency of the radio mode is larger with $\epsilon_{\rm f}=1$.
Though this quasar/radio mode matches the observed radiatively efficient/inefficient activities of AGN feedback~\citep{churazovetal05, russelletal13}, it is possible that some accretion rate-driven jets becomes indistinguishable from large opening angle AGN-driven nuclear outflows after they have propagated into the multiphase gas of the galaxy~\citep{wagneretal12}.

The list of the Seth simulation runs is listed in Table~\ref{tbl:Sethsuite}. All simulations with $\Delta x=34.8\, \rm pc$ resolution are run up to $z=2$, while the simulations with $\Delta x=8.7\, \rm pc$ resolution are stopped at more intermediate redshift after the BH of the central simulation galaxy has reached self-regulation. We perform a simulation with no feedback from SNe nor from AGN (and without BH formation) denoted SL ($\Delta x=34.8\, \rm pc$). We have two simulations with SN feedback, one version with kinetic implementation SLK and another one with delayed cooling SLD, which do not include BHs ($\Delta x=34.8\, \rm pc$). Finally, we run four simulations including AGN feedback, two with kinetic SN feedback SLKA and SHKA (respectively with $\Delta x=34.8\, \rm pc$ and $\Delta x=8.7\, \rm pc$), and two with the delayed cooling SN feedback SLDA and SHDA.

Fig.~\ref{fig:zoom32} shows three projections of the gas density for the SHDA run at $z=3.25$, one at large scale where cosmic filaments are clearly visibles (within a large-scale cosmic wall), one at the halo scale where these cosmic filaments penetrate the central halo in the form of dense and cold gas mixing with the diffuse ambient medium, and another projection at galactic scale where we see the detailed structure of the gas within the galaxy in the form of dense star-forming clumps embedded into gas spiral arms. In those projections, we have indicated the virial radius $r_{\rm vir}$ of the dark matter halo (as the radius at which the average density in the halo is $200$ times the average cosmic density). With zoom simulations, we can resolve at the same time the complex pattern of the large-scale inflows and outflows of gas into the halo and onto the galaxy, and the multiphase structure of the interstellar gas.

\section{Results}
\label{section:result}

\begin{figure}
  \centering{\resizebox*{!}{6.cm}{\includegraphics{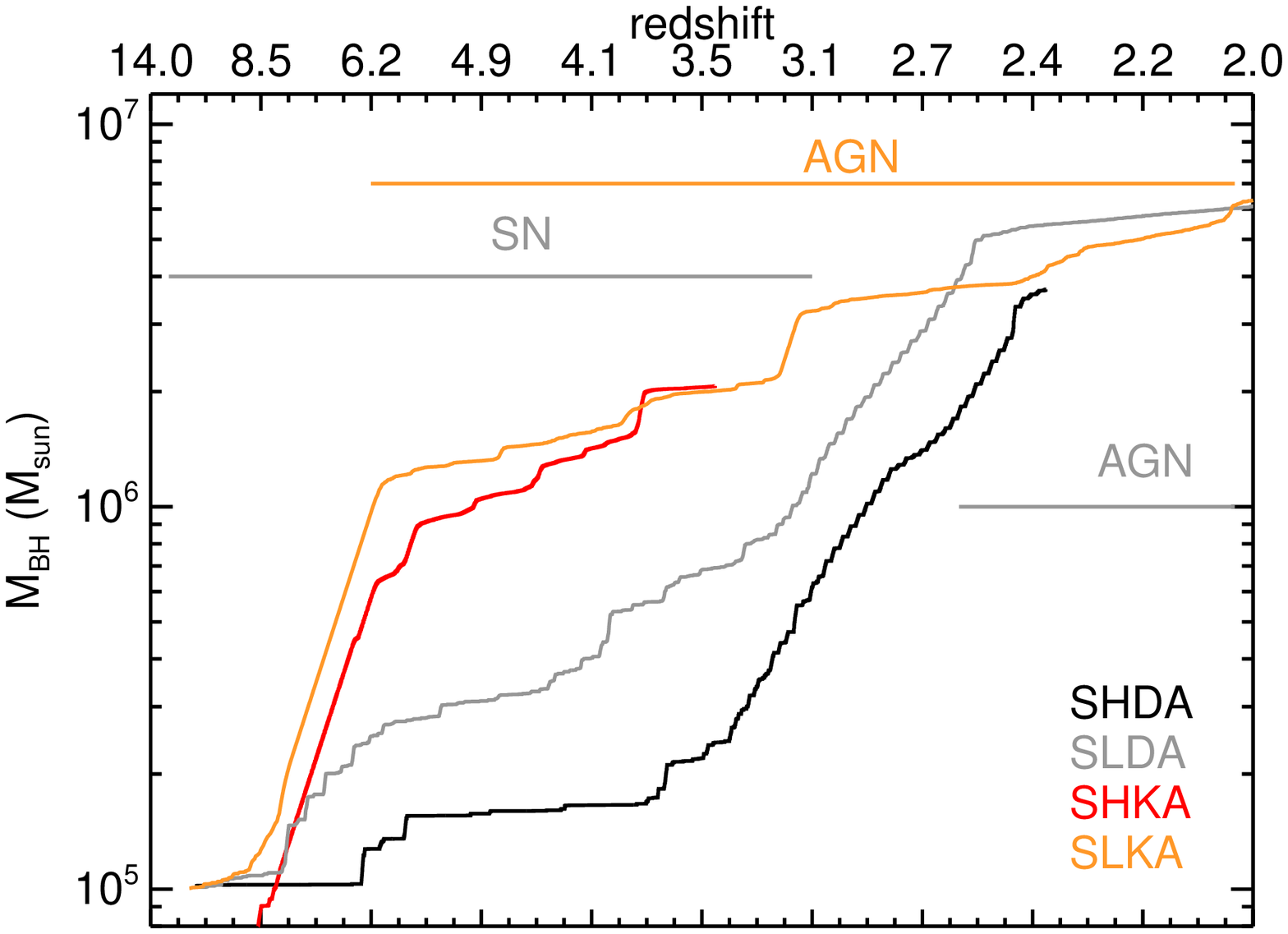}}}\vspace{-1.25cm}
  \centering{\resizebox*{!}{6.cm}{\includegraphics{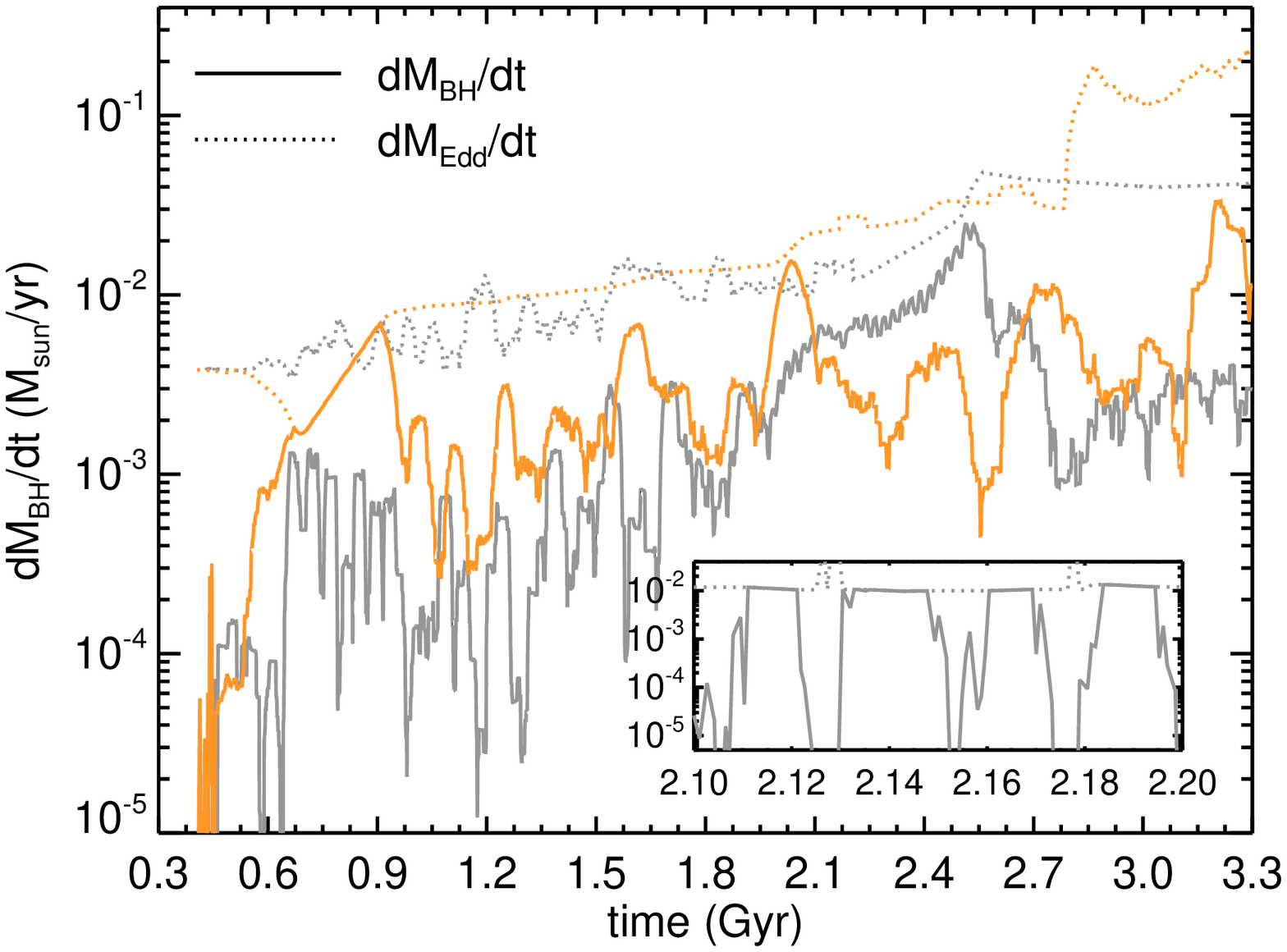}}}
  \caption{Top: BH mass as a function of time (redshift) for SHDA (black), SLDA (grey), SHKA (red) and SLKA (orange). Bottom: BH mass accretion rate (solid lines) and Eddington mass accretion rate (dotted lines) for SLDA and SLKA simulations only and values are smoothed for sake of readability with a $5 \, \rm Myr$ time window. The small inset shows the exact BH mass accretion of the SLDA simulation between $2.1<t<2.2\, \rm Gyr$. The horizontal lines indicate the SN and AGN regulation phases of the BH growth with a color that corresponds to the simulation it refers to. The simulations with kinetic SN feedback (SLKA and SHKA) show a rapid growth of the BH mass at early times and quickly regulate the BH mass growth. The simulations with delayed cooling SN feedback (SLDA and SHDA), instead inhibit the early BH growth and delay its Eddington-limited phase to later times (below redshift $z<3.5$).}
    \label{fig:mbhevol}
\end{figure}

\subsection{Quenching early BH growth with efficient SN feedback}

\begin{figure}
  \centering{\resizebox*{!}{6.cm}{\includegraphics{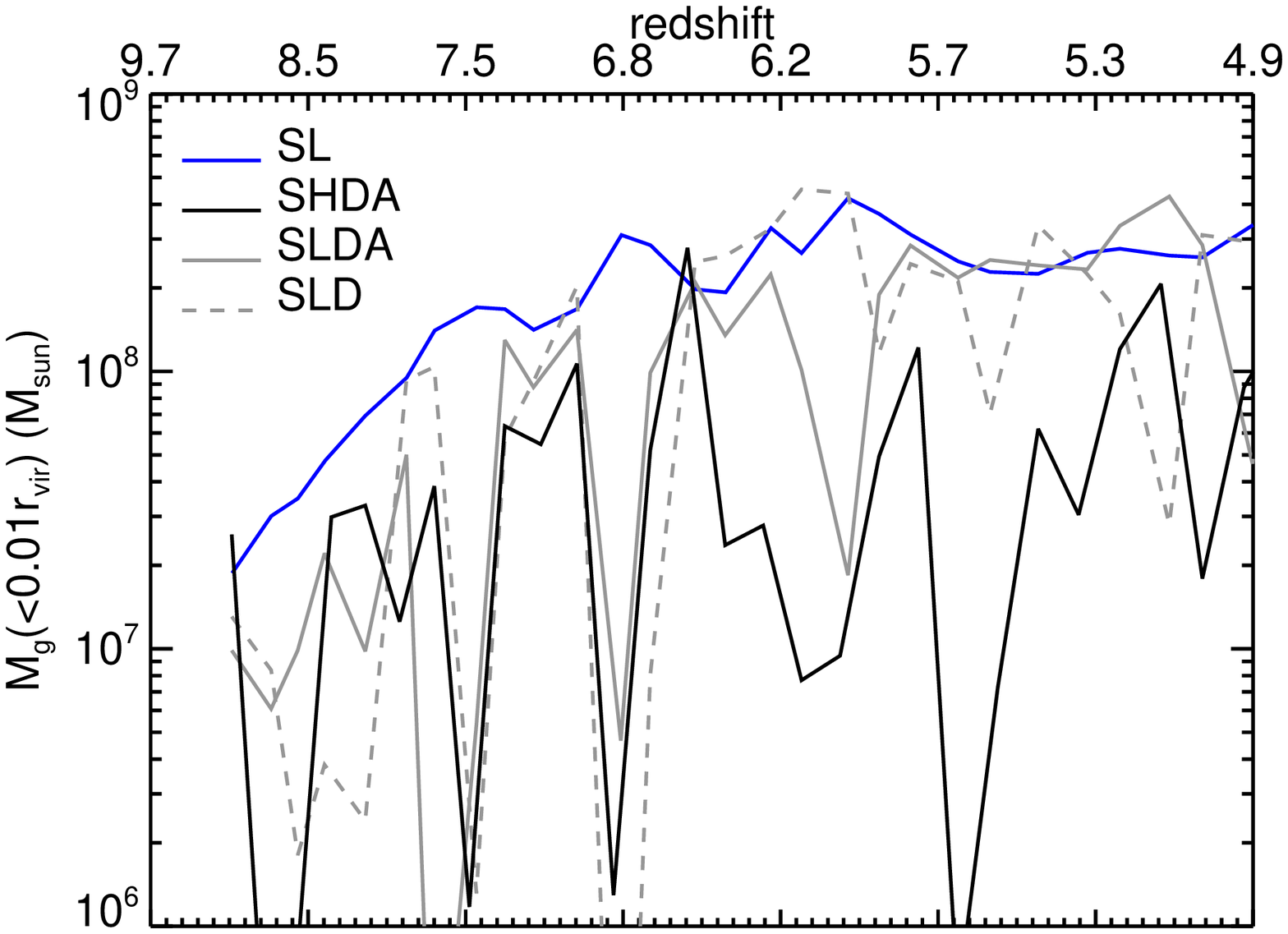}}}\vspace{-1.25cm}
  \centering{\resizebox*{!}{6.cm}{\includegraphics{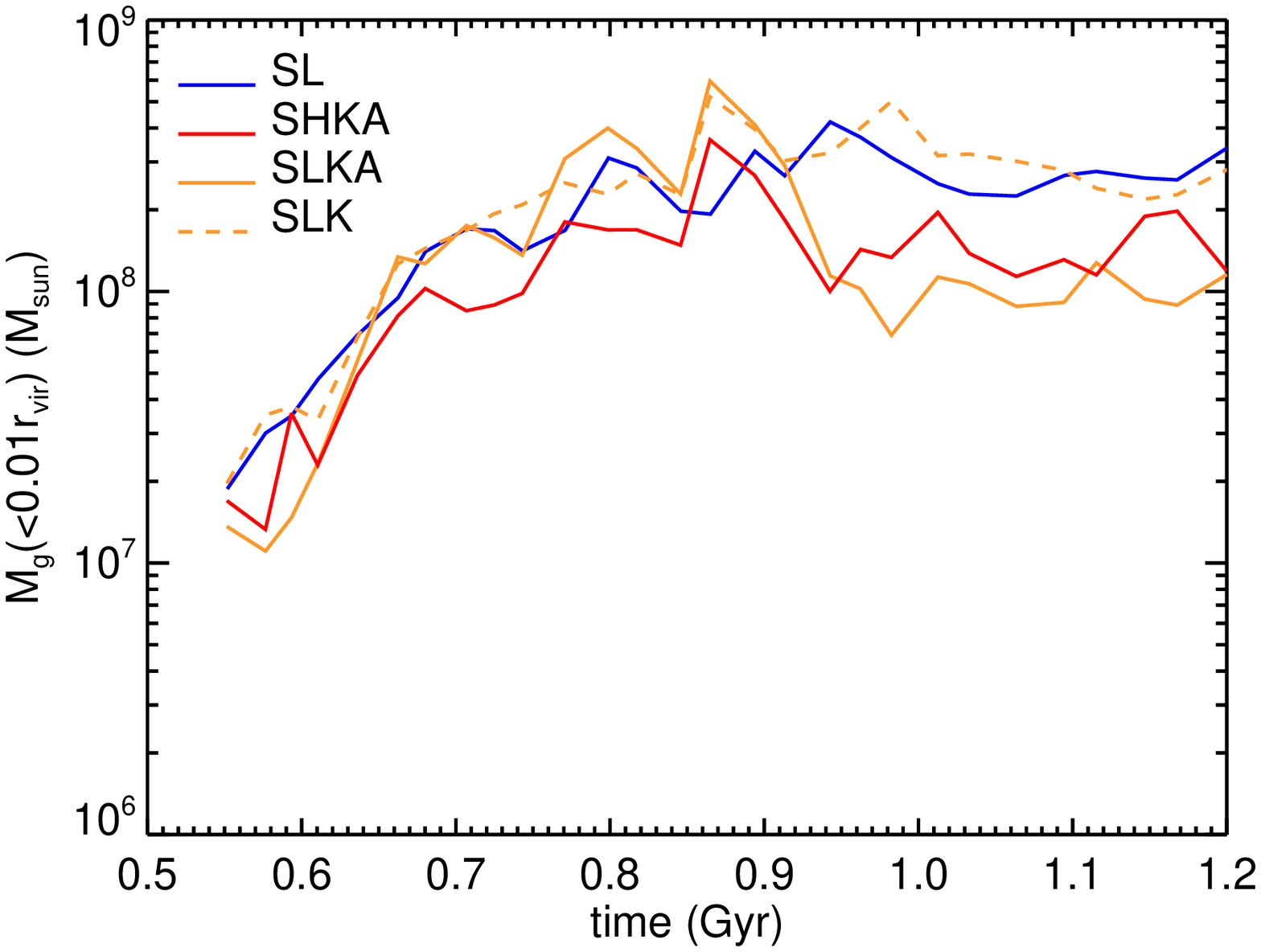}}}
  \caption{Mass of gas within 1 per cent of the halo virial radius as a function of redshift for SHDA (black), SLDA (grey solid), SLD (grey dashed), SHKA (red), SLKA (orange solid), SLK (orange dashed), and SL (blue). Only simulations with delayed cooling SN feedback exhibit rapid variations in the gas mass content at early times due to the efficient evacuation of the cold gas from the galaxy center.}
    \label{fig:menc}
\end{figure}

The top panel of Fig.~\ref{fig:mbhevol} shows the BH mass evolution for the four simulations including AGN feedback, and the bottom panel of Fig.~\ref{fig:mbhevol} shows the BH mass accretion rate for SLDA and SLKA only.
Both simulations with kinetic SN feedback, SLKA and SHKA, show a rapid initial growth of the BH mass at the Eddington rate between $6<z<9$, with a plateau below $z<6$ when the BH has reached self-regulation. 
Viceversa, both simulations with delayed cooling SN feedback, SLDA and SHDA, have an extended period of time, down to $z\simeq 3.5$, where the BH grows quiescently, far from the Eddington limit.
It is only below this redshift, that the BH growth takes off at approximately half the Eddington rate from $2.5<z<3.5$ and then settles in a self-regulated state where the BH mass is comparable in runs SLDA, SHDA and SLKA. 
Note that the evolution of BH mass is relatively independent of resolution for the same set of SN sub-grid modeling: Eddington-limited phases appear at the same epoch and their time extensions are similar, and BH masses at self-regulation are the same to within less than a factor of 2.

At $z\simeq3.5$, a significant galaxy merger with stellar mass ratio of $\sim$1:3 drives a large inflow of gas towards the BH. 
The three simulations that have reached this redshift (SLKA, SHDA and SLDA) show a rapid increase of the BH mass.
In particular, the simulations with delayed cooling SN feedback have an extended near-Eddington growth for the central BH ($2.5<z<3.5$), due to the absence of BH self-regulation prior to the merger. 
Thus, the BH grows until it reaches a mass sufficiently large to pump enough energy into the gas to unbind it from the galaxy gravitational potential well.

\begin{figure*}
  \centering{\resizebox*{!}{4.1cm}{\includegraphics{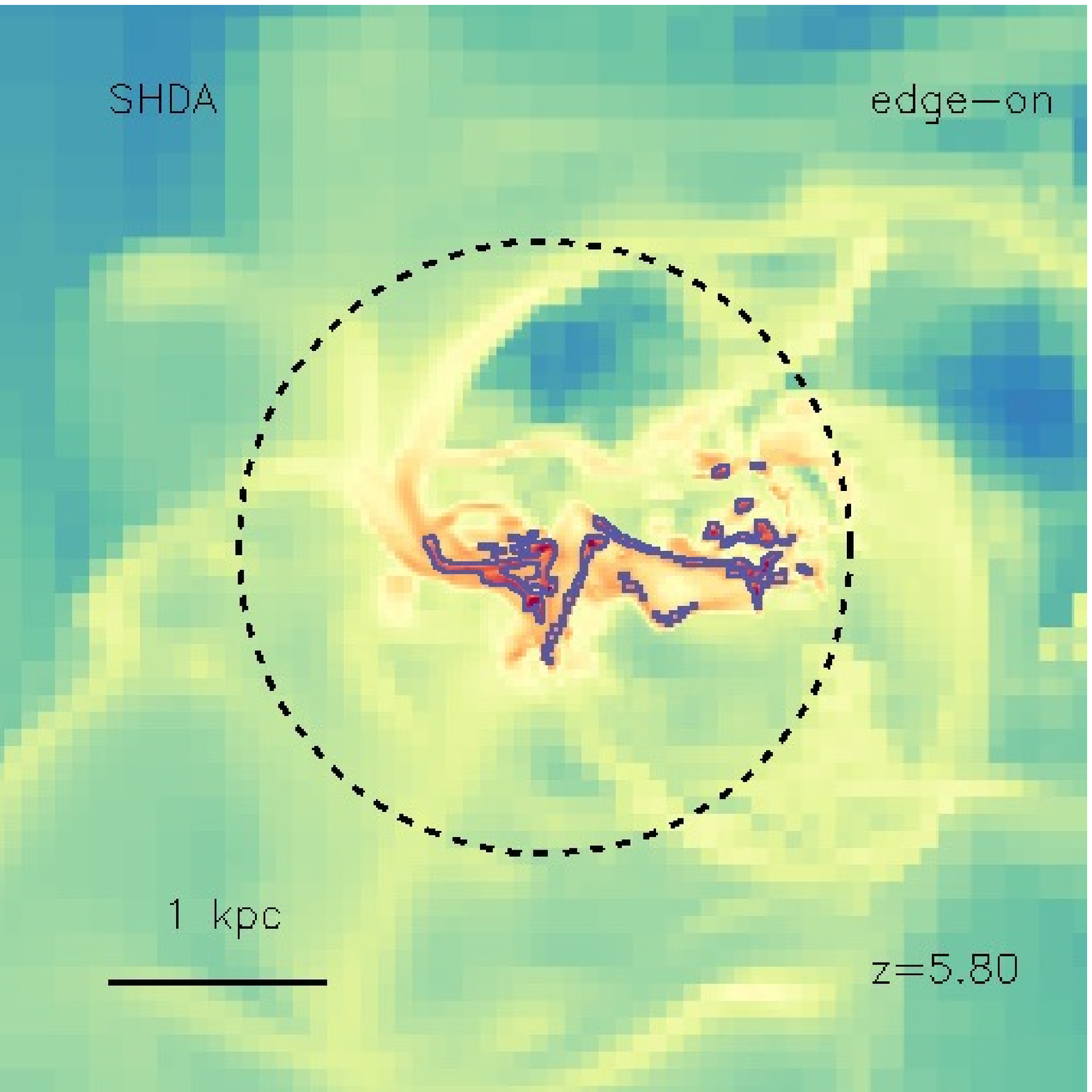}}}
  \centering{\resizebox*{!}{4.1cm}{\includegraphics{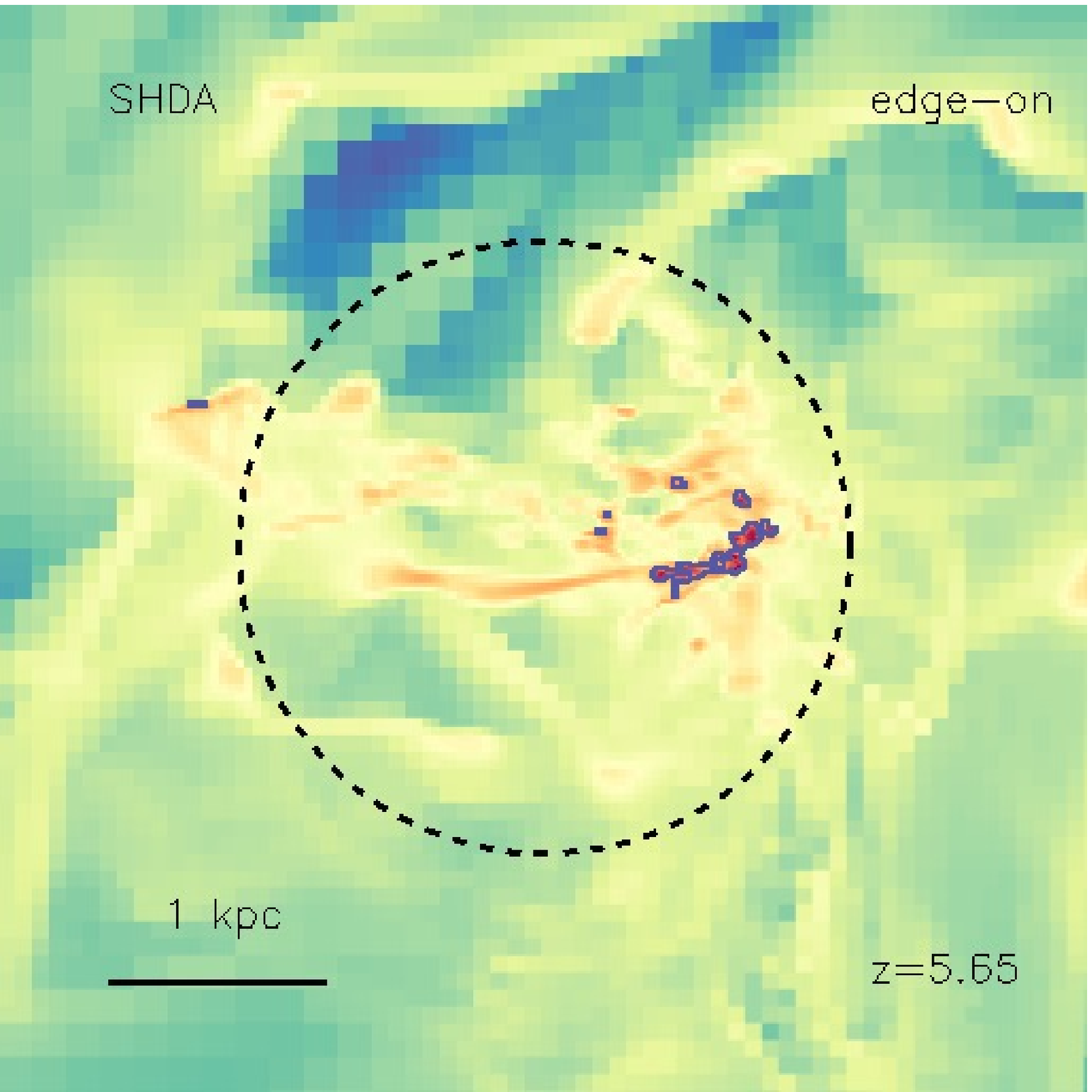}}}\hspace{0.2cm}
  \centering{\resizebox*{!}{4.1cm}{\includegraphics{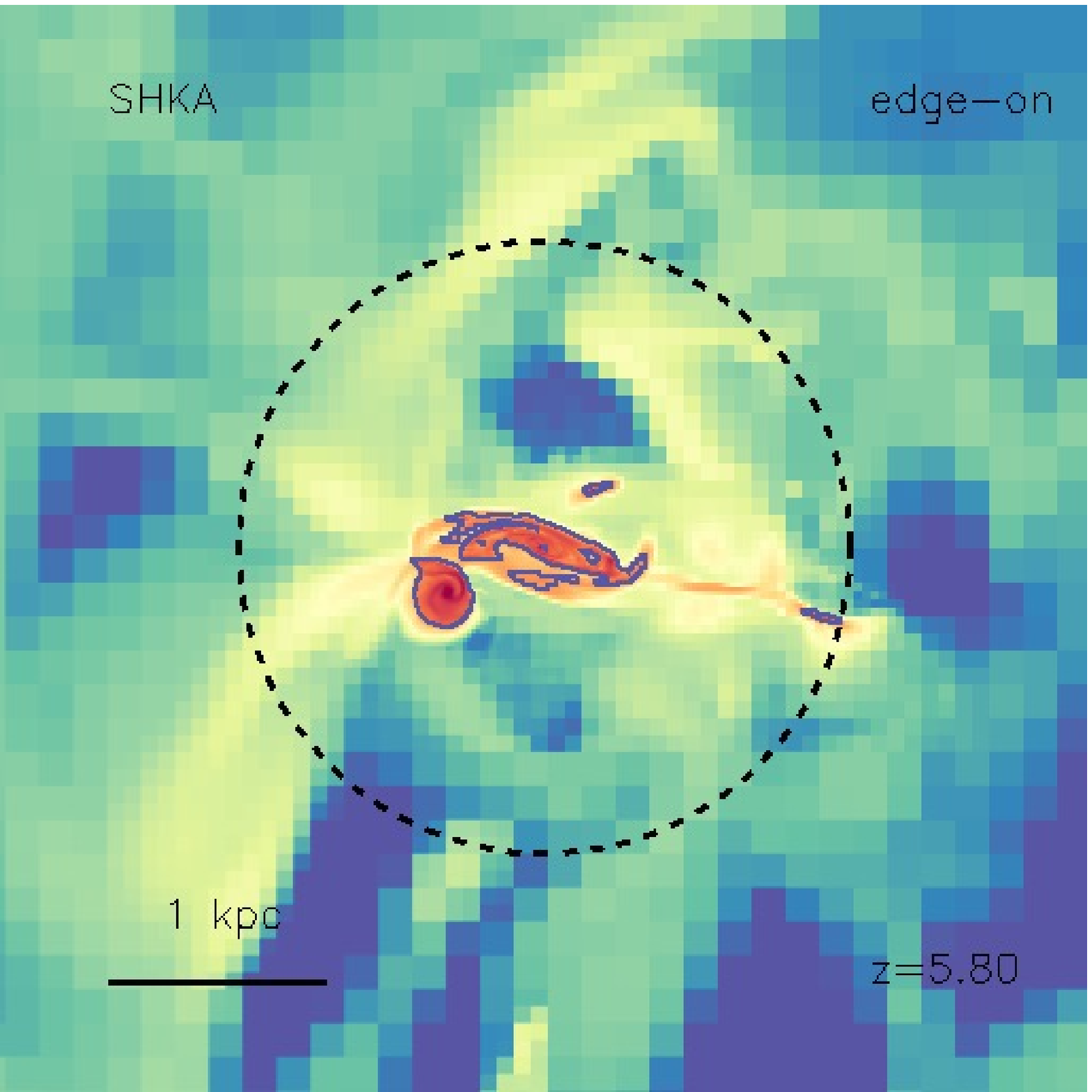}}}
  \centering{\resizebox*{!}{4.1cm}{\includegraphics{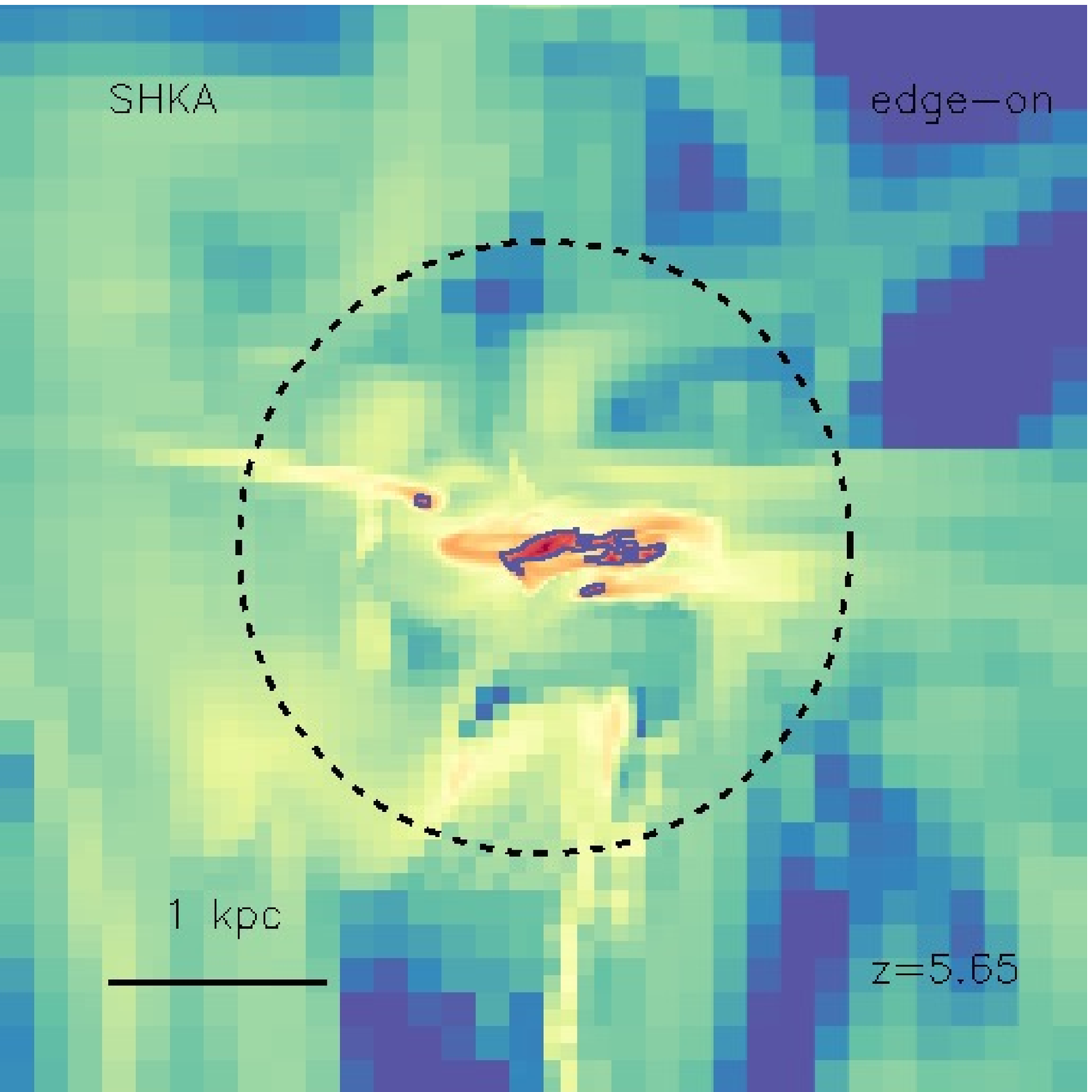}}}
  \centering{\resizebox*{!}{4.1cm}{\includegraphics{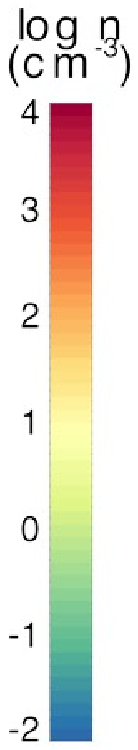}}}
  \centering{\resizebox*{!}{4.1cm}{\includegraphics{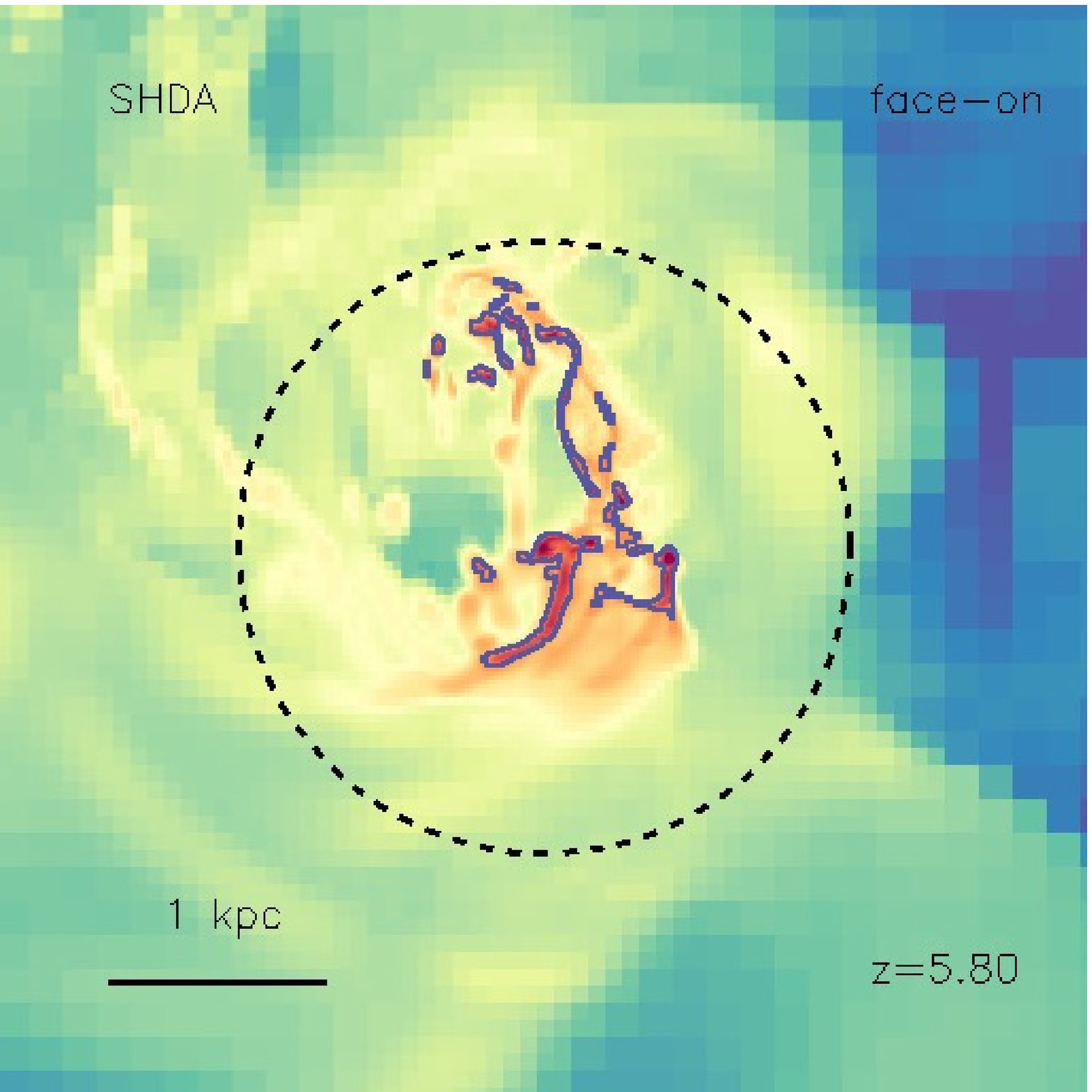}}}
  \centering{\resizebox*{!}{4.1cm}{\includegraphics{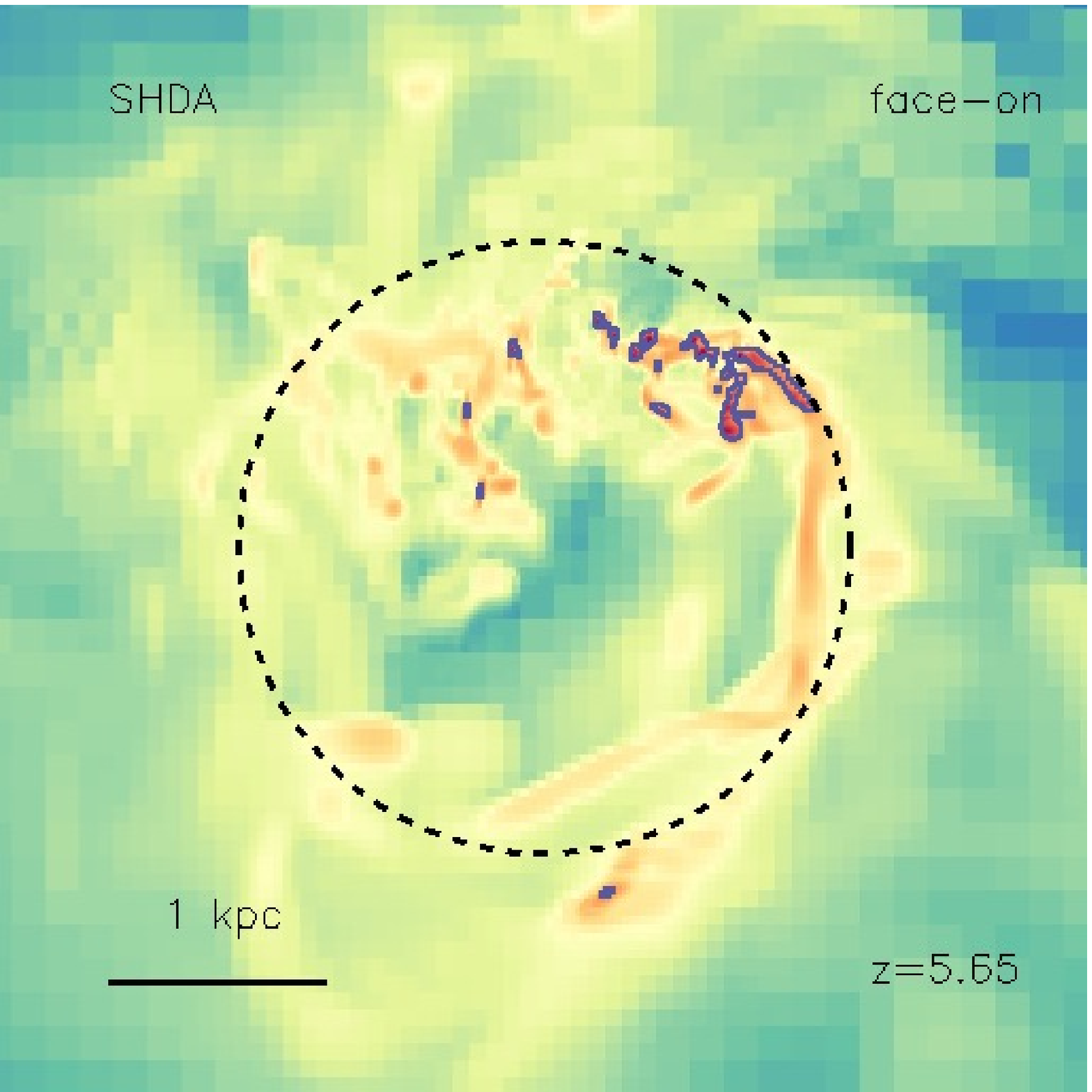}}}\hspace{0.2cm}
  \centering{\resizebox*{!}{4.1cm}{\includegraphics{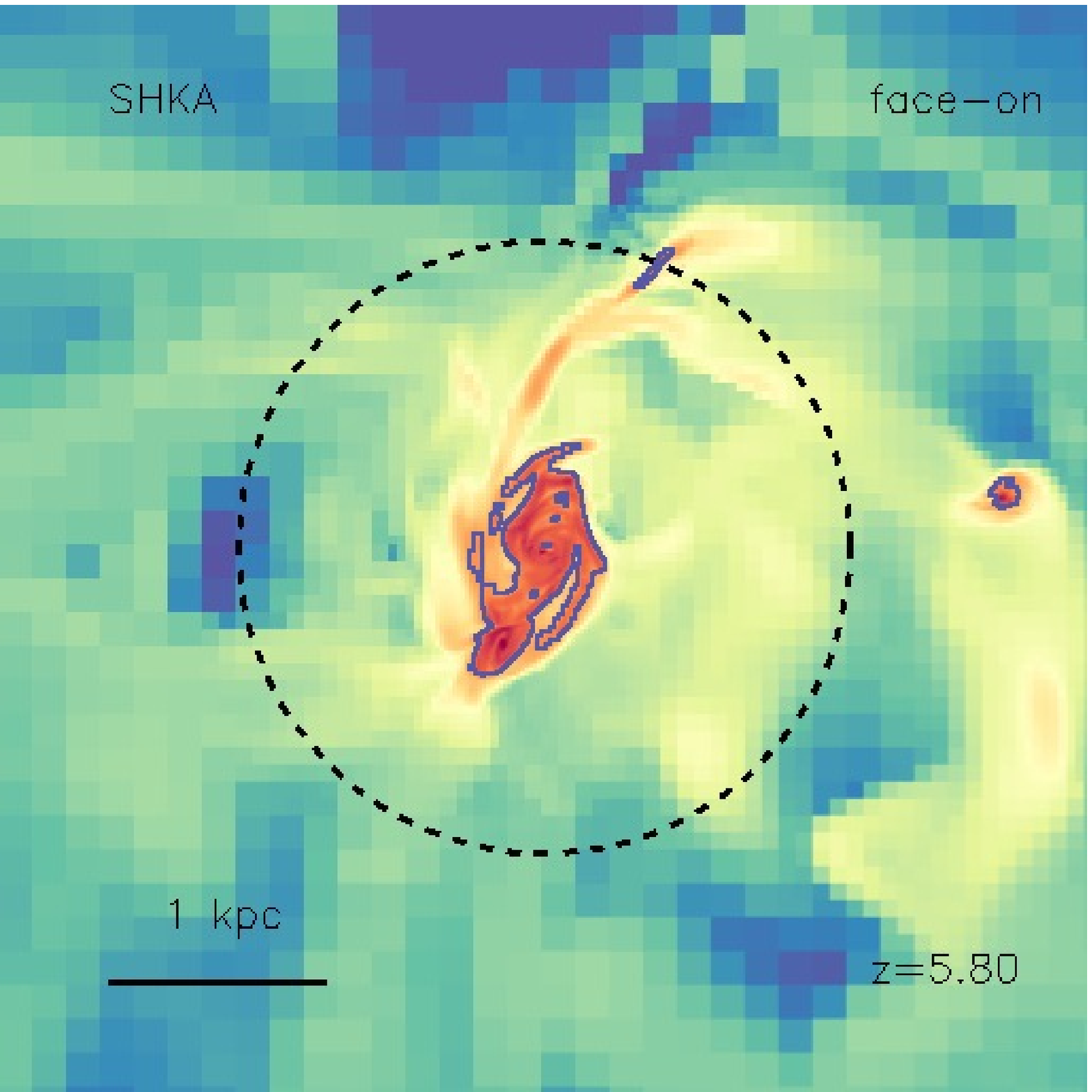}}}
  \centering{\resizebox*{!}{4.1cm}{\includegraphics{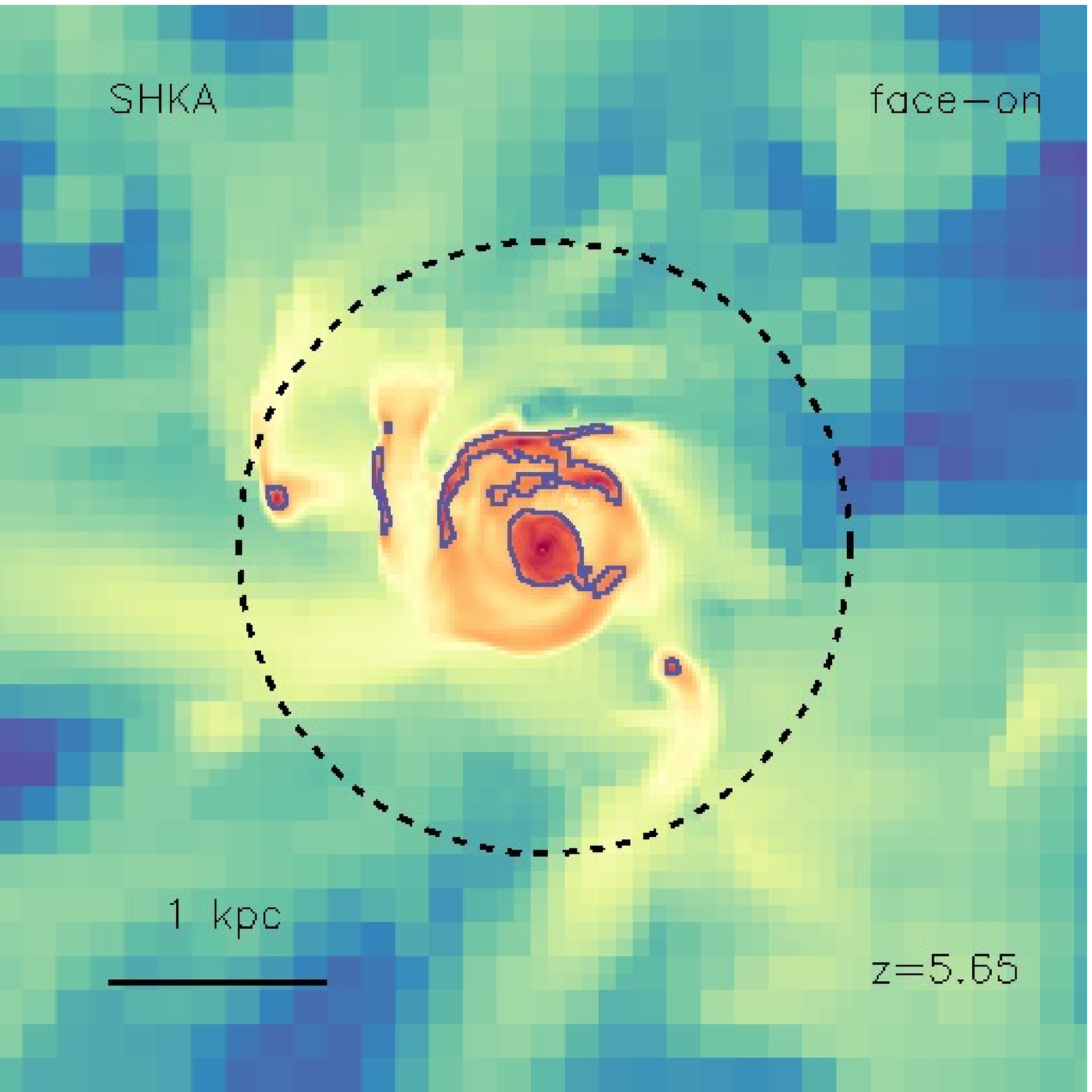}}}
  \centering{\resizebox*{!}{4.1cm}{\includegraphics{./figs/ctable_gdens2.eps}}}
  \caption{Gas density projections of the central galaxy at $z=5.8$ (first and third columns) and $z=5.65$ (second and fourth columns) for the SHDA (four left-most panels) and SHKA (four right-most panels) simulations, seen edge-on (first row) or face-on (second row). The blue contours surround the regions of star formation with gas density above $250 \, \rm H\, cm^{-3}$. The dashed circles indicate the 10 per cent virial radius of the halo. In the simulation with the delayed cooling SN feedback, dense star-forming clumps in the centre of the galaxy are regularly destroyed and reprocessed, while in the simulation with kinetic SN feedback the star-forming gas settles into a long-lived compact rotating disc.}
    \label{fig:gdens2}
\end{figure*}

As explained in detail in paper II, delayed cooling SN feedback causes cold gas clumps in the galaxy to be regularly destroyed by SN explosions, every $\sim$10 Myr. Periods of intense star formation are followed by phases where stellar feedback drives large-scale outflows and hot bubbles, and removes cold gas from the galaxy at the same time.  During these phases, the BH accretion rate is well below the Eddington limit\footnote{Note how the Eddington accretion rate exhibits negative variations with time, while the BH mass can only increase. This apparent paradox is explained by variations in the spin parameter of the BH which in turn change the Eddington accretion rate by modifying the radiative efficiency $\epsilon_{\rm r}$ (see paper II and paper III for a complete overview of the BH spin evolution).}, while it is at Eddington when the gas cools down again and re-accumulates around the BH. This behaviour is highlighted in the inset in Fig.~\ref{fig:mbhevol}. The net effect is to overall reduce the Eddington-limited growth phase by a factor of 2, as reflected by the time-smoothed curve (for sake of clarity) shown in the bottom panel of Fig.~\ref{fig:mbhevol}: the average BH mass accretion rate is a factor 2 below the Eddington accretion rate between $2.5<z<3.5$ for the SLDA simulation (as well as for the SHDA, not represented here). 

In summary, the delayed cooling implementation of SN feedback quenches the initial growth of the BH, while the kinetic SN feedback allows for a rapid Eddington-limited growth at early times.
The fundamental reason is that the two different approaches to SN feedback affect the density and temperature of gas in the galaxy differently. Kinetic SN feedback imparts only momentum to the gas (with lower magnitude than the expected analytical solution, see~\citealp{kimmetal15}). The gas quickly thermalizes if the conditions to create a shock are met, and such conditions are effectively met each time there is a SN explosion. This can be understood as follows. In the early phase, SN explosions are similar to Sedov solutions, therefore in a few time-steps, the solution will approach that of the adiabatic expansion phase, where 20 per cent of the energy is in kinetic form and 80 per cent in internal energy~\citep[e.g.][]{cox72}.  In dense star-forming regions, the temperature reached after a SN explosion is not high enough to avoid over-cooling of the internal component since the Sedov scale remains unresolved at our spatial resolution, and 80 per cent of the total energy of the explosion is lost, limiting the total amount of momentum that would have been transferred by the hot pressurised bubble by the end of the energy-conserving phase. Thus, if gas is not artificially prevented from cooling, dense gas can survive SN explosions, contrary to expectations. Delayed cooling SN feedback obviates  this shortcoming by explicitly forcing the gas to not cool in the dense regions of star-formation, mimicking the effect of  internal energy deposition (however, it does not guarantee that the correct amount of momentum is imparted to the gas, only that energy is efficiently transferred to the gas). This is the reason why simulations with delayed cooling SN feedback show that dense star-forming gas is destroyed after the ignition of the first stars into SNe (see paper II). 

Fig.~\ref{fig:menc} clearly shows these different behaviours.
The gas mass within 1 per cent of the virial radius ($\simeq 100\, \rm pc$ at $z=6$) has a relatively smooth time-evolution with the kinetic mode of SN feedback (SHKA, SLKA and SLK runs), comparable to the simulation without SNe (SL run). Simulations with delayed cooling SN feedback (SHDA, SLDA and SLD runs) instead show strong variations in the amount of gas mass relatively close to the BH. These variations are an intrinsic feature related to the modelling of SNe as they are detected independently of the presence of the AGN feedback and the choice of resolution.

Fig.~\ref{fig:gdens2} highlights two specific times representative of the two phases characteristic of the delayed cooling SN feedback: intense star formation driven by cold gas followed by SN activity which in turns removes the cold dense gas. The four left-most panels of Fig.~\ref{fig:gdens2} show the projected gas density at a maximum ($z=5.80$) and near a minimum ($z=5.65$) of the gas mass within 1 per cent of the halo virial radius for the SHDA simulation (cf. Fig.~\ref{fig:menc}, top panel).
The dense gas in the galaxy appear edge-on with some cold clumps near the very center of the galaxy at $z=5.8$.
However, at $z=5.65$, the dense gas has disappeared from the center of the image due to SN explosions that have disrupted this cold gas component.
For comparison, the four right-most panels of Fig.~\ref{fig:gdens2} show the same gas density projections for the SHKA simulation, where a thin disc of gas also appears nearly edge-on, and this disc is present in both time steps.
Therefore, contrary to the case with delayed cooling SN feedback (SHDA), the simulation with kinetic SN feedback (SHKA) does not show any sign of the destruction of the cold star-forming gas in the galaxy.
Finally, the presence at all times of this cold gas reservoir near the BH for SHKA (and SLKA) is the reason for the rapid early growth of the BH, while in SHDA (and SLDA) disruption of dense gas is efficient enough to starve the BH.

\begin{figure}
  \centering{\resizebox*{!}{5.5cm}{\includegraphics{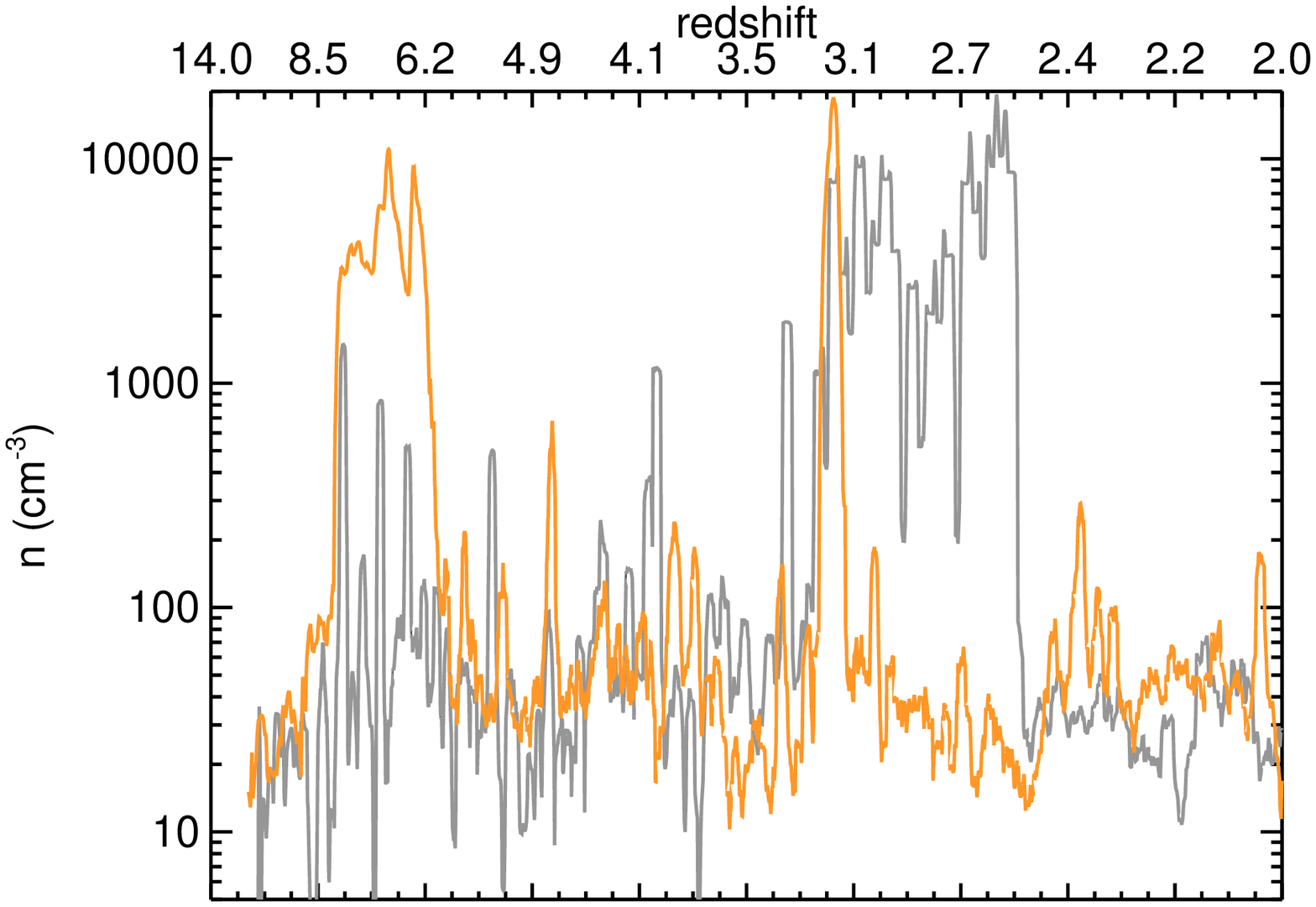}}}\vspace{-1.15cm}
  \centering{\resizebox*{!}{5.5cm}{\includegraphics{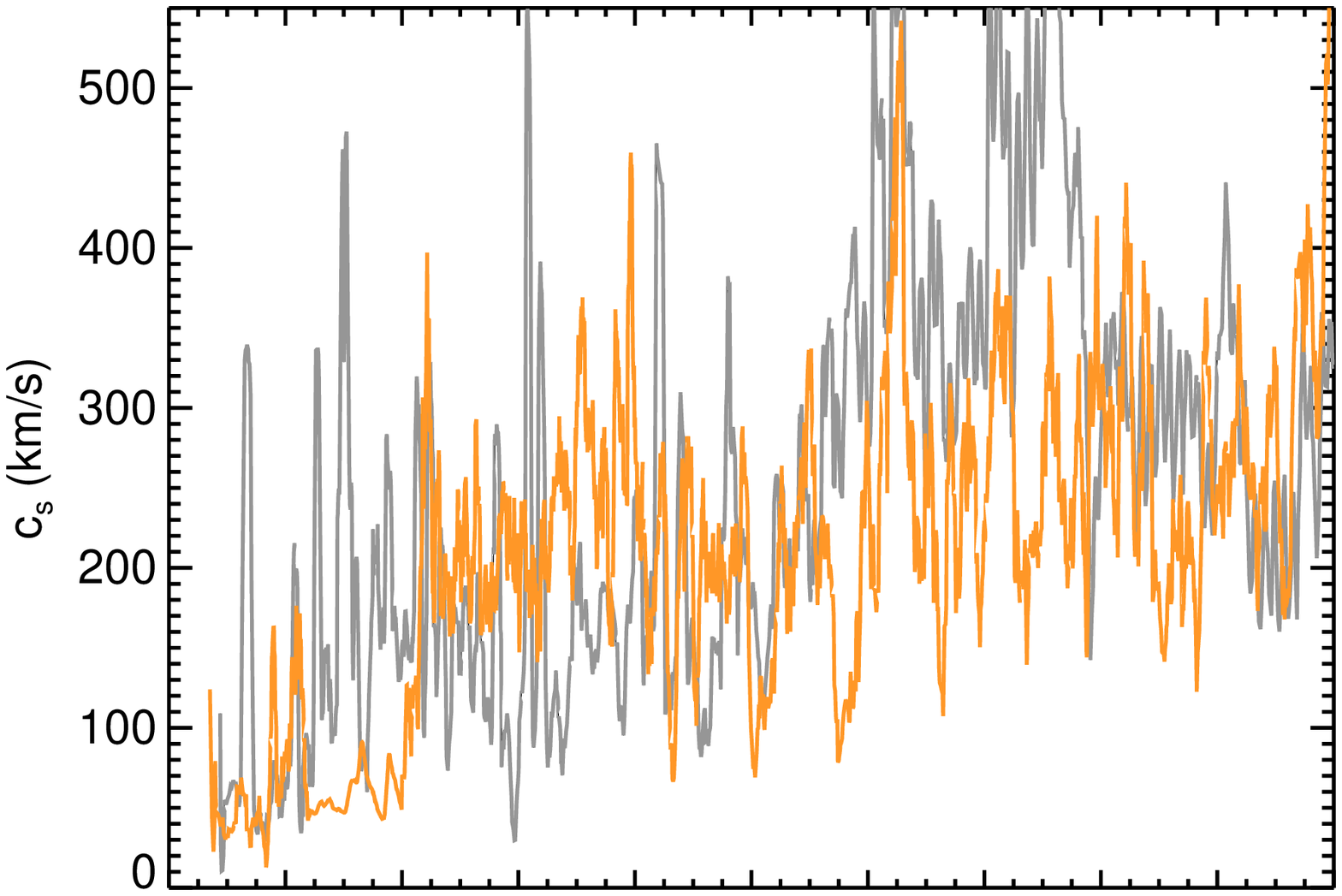}}}\vspace{-1.15cm}
  \centering{\resizebox*{!}{5.5cm}{\includegraphics{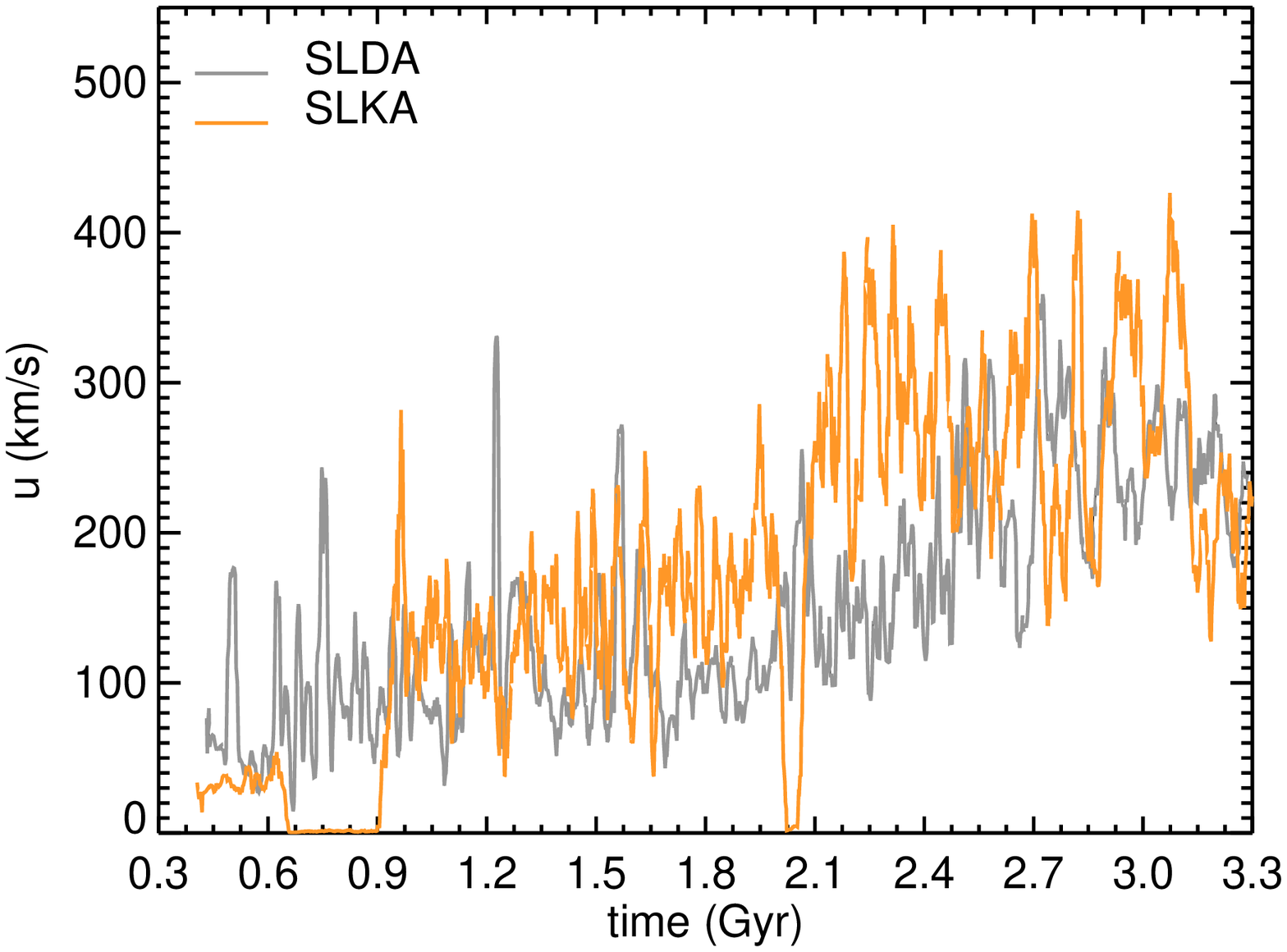}}}
  \caption{Gas number density (top), gas sound speed (middle), and gas relative velocity (bottom) near the BH as a function of time (redshift) for SLDA (grey) and SLKA (orange). }
    \label{fig:bhrho}
\end{figure}

Fig.~\ref{fig:bhrho} shows the gas number density in the vicinity of the BH as a function of time, as well as the gas sound speed and its relative velocity with respect to the BH.
During the early growth of the BH in the SLKA run ($6<z<9$), the gas density near the BH is above $n>10^3 \, \rm cm^{-3}$, the gas sound speed is $\sim 50 \, \rm  km\, s^{-1}$, and the gas relative velocity is $u<10\, \rm km\, s^{-1}$. 
Therefore, the gas is sufficiently dense, cold, and with very little turbulence (relative gas motion), to allow the Bondi accretion rate to be Eddington-limited, and the BH to grow to reach self-regulation in $\sim200 \, \rm Myr$.
During the out-of-Eddington phases, gas density drops to lower values ($n\simeq20-100\, \rm cm^{-3}$), the gas sound speed increases ($c_{\rm s}\simeq200\, \rm km\, s^{-1}$), as does the gas relative velocity ($u\simeq 100-300 \, \rm km\, s^{-1}$), all being consequences of the energy input from the AGN while the BH is at self-regulation. 
It is particularly striking that these quantities change abruptly (gas density increases and velocities decrease) with the new inflow of gas triggered by the merger at $z\simeq 3.5$ ($t\simeq2 \, \rm Gyr$).

For the SLDA run, gas density, sound speed and gas relative velocity values up to $z=3.5$ are comparable to the values obtained in the SLKA run during the BH self-regulation phase (i.e., after $z\sim6$; $t\simeq1 \, \rm Gyr$).
The main difference is that, in the SLDA run, the absence of Eddington-limited phase is due to the energy input from SNe that destroy the dense star-forming gas, keeping the density low, the temperature high and the interstellar medium turbulent.
The gas number density in SLDA between $5<z<9$ shows rapid variations with peaks at $n\simeq 500 \, \rm cm^{-3}$ followed by drops to $n\simeq 50-100 \, \rm cm^{-3}$.
It illustrates well the ``breathing" of the galaxy when delayed cooling SNe feedback is employed (see also Fig.~\ref{fig:menc}): dense gas accumulates in the galaxy (and near the BH) for $\Delta t \sim 10 \, \rm Myr$ until the first SNe explode and drive the cold gas in an expanding wind that falls back onto the galaxy in a dynamical time.
At $z\simeq3.5$, a merger triggers fresh gas near the BH that accumulates and leads to the near-Eddington phase: gas sound speed and gas relative velocity are still high (respectively $\simeq$300 and 100$\, \rm km\, s^{-1}$), but the gas density remains steadily at $n\simeq 500\, \rm cm^{-3}$. 
Thus, even though velocities are on average large due to the starburst activity induced by the merger, the gas density is sufficiently high to keep the BH growing near the Eddington rate.
Note that these quantities have strong small-scale time variations (on a time-scale of 10 Myr) that cannot be represented in Fig.~\ref{fig:bhrho}.
After the merger-driven triggering of the BH growth in the SLDA simulation, the BH ends up self-regulating at $z\simeq2.5$, and the gas density drops, while the gas sound speed and relative velocity increase to values comparable to that of the SLKA simulation at the same epoch and for the same reason: its AGN activity is strong enough to remove the cold gas from its direct surroundings.

\subsection{Why SNe are able to quench BH growth in low-mass galaxies}

\begin{figure}
  \centering{\resizebox*{!}{6.cm}{\includegraphics{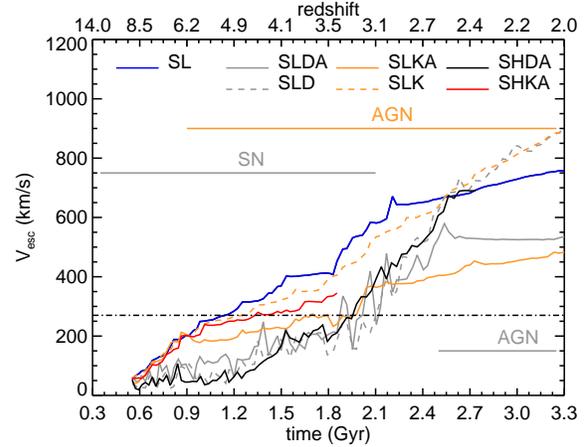}}}
  \caption{Escape velocity from the bulge (at $r_{\rm b}=100\, \rm pc$ from the galactic centre) as a function of time (redshift) for SL (blue), SLDA (grey solid), SLD (grey dashed), SLKA (orange solid) and SLK (orange dashed). The horizontal dot-dashed line corresponds to the velocity of SN-driven winds in the bulge with $f_{\rm blast}=1$ (corresponding to the delayed cooling SN model). Grey horizontal lines mark the SN- and AGN-dominated evolution in the delayed cooling case. The orange horizontal line marks AGN-dominated evolution in the kinetic feedback case.   In SLDA and SLD, the end of the SN regulated growth of the BH corresponds to the time when the escape velocity of the bulge exceeds the velocity of SN-driven wind. In the SLKA and SLK runs, the BH grows as soon as it is created, and there is no SN-dominated phase, which indicates that the SN driven-winds for the kinetic model have a very low energy-conversion efficiency ($f_{\rm blast}\ll 1$) and, thus, an ejection velocity below the horizontal dot-dashed line.}
    \label{fig:vesc}
\end{figure}

\begin{figure*}
  \centering{\resizebox*{!}{4.cm}{\includegraphics{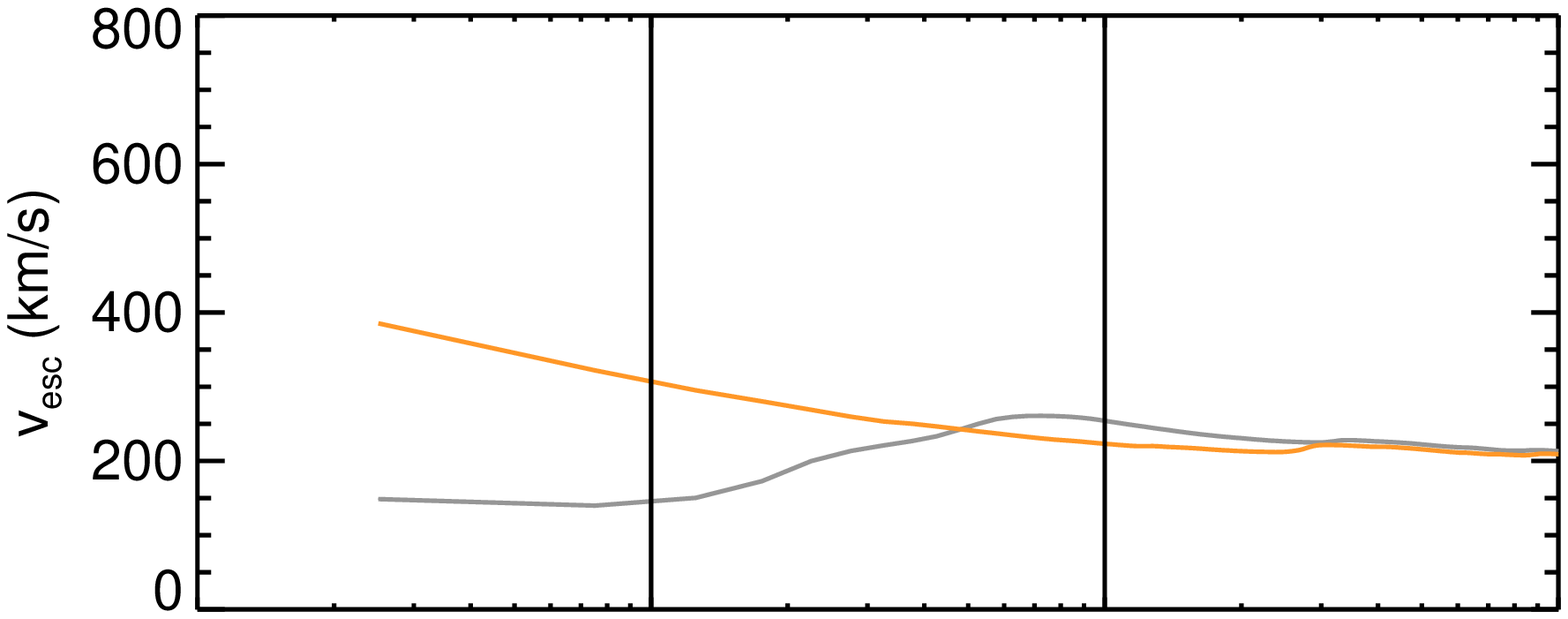}}}
  \centering{\resizebox*{!}{4.cm}{\includegraphics{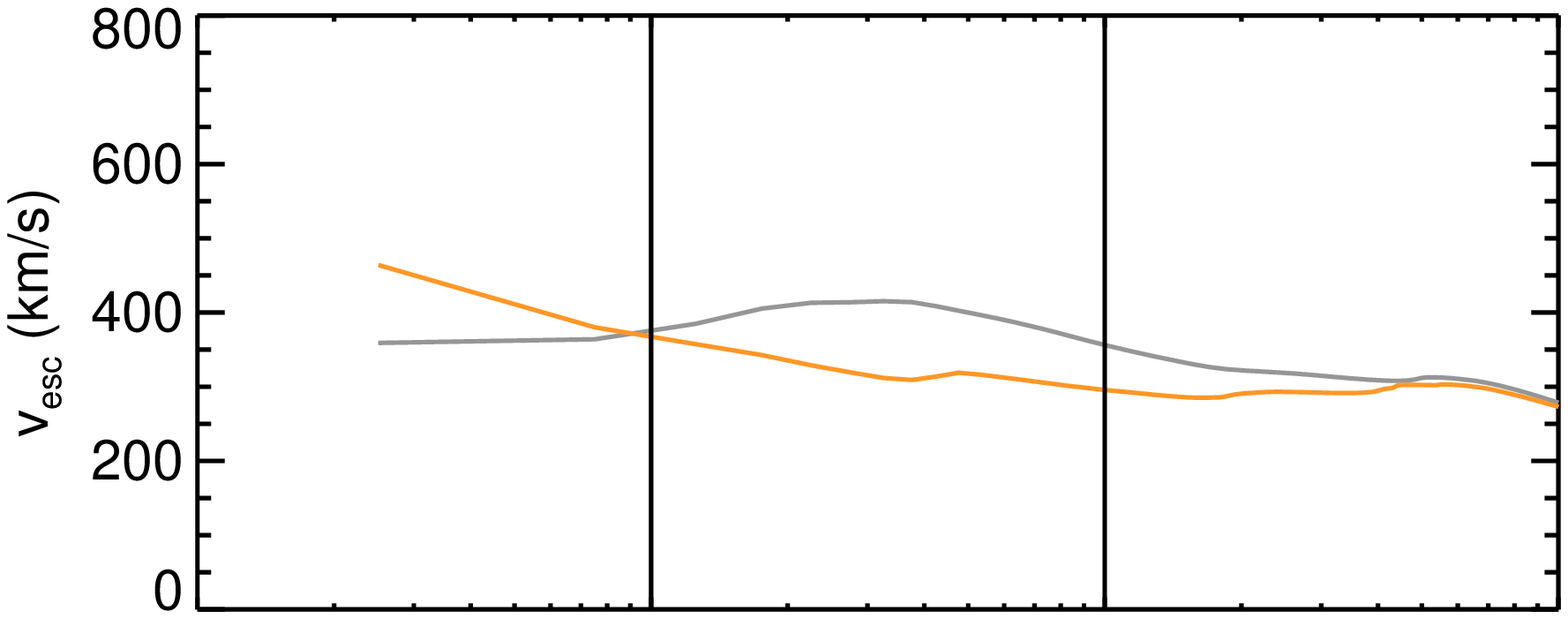}}}\vspace{-1.25cm}\\
  \centering{\resizebox*{!}{6.cm}{\includegraphics{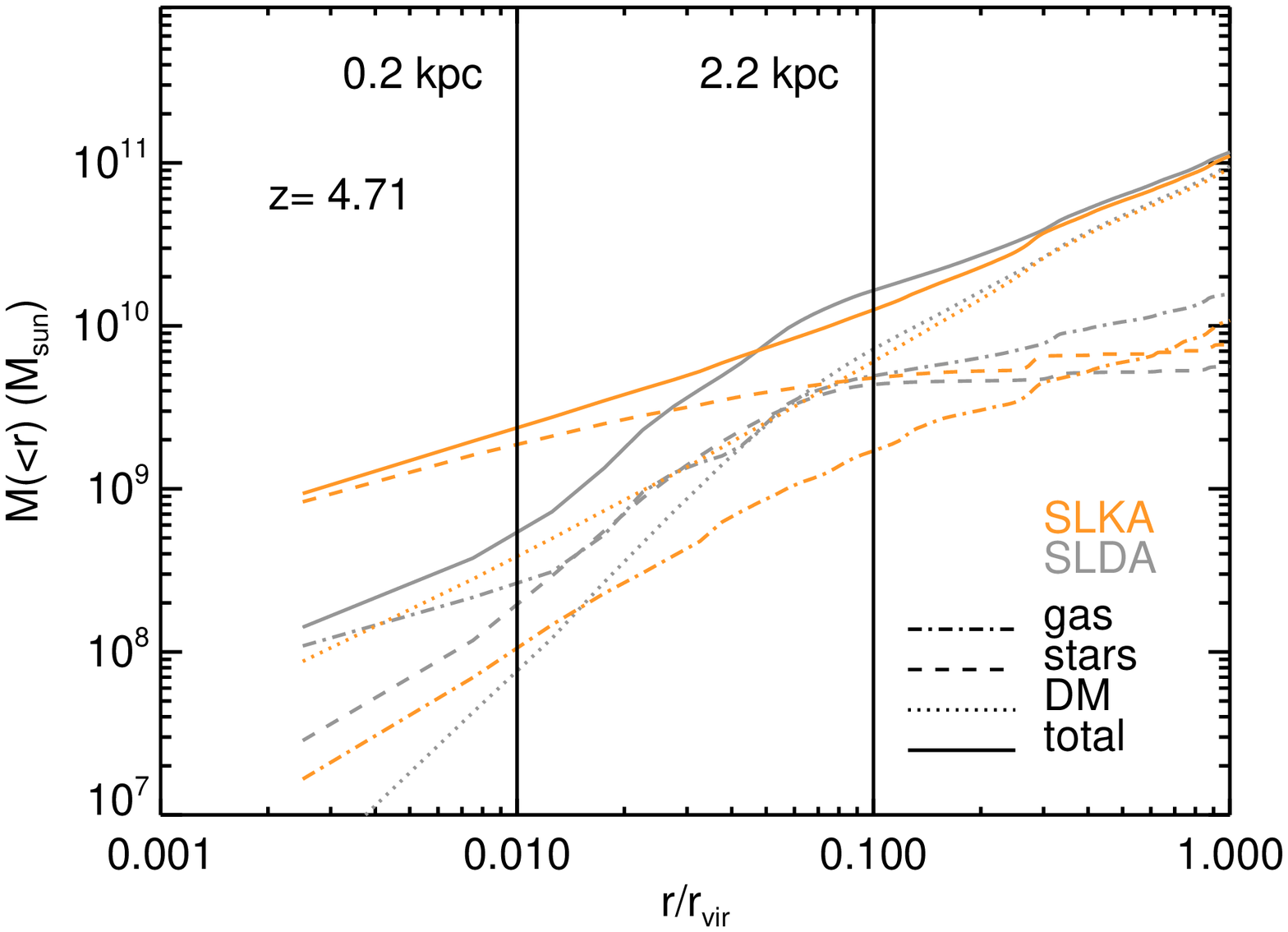}}}
  \centering{\resizebox*{!}{6.cm}{\includegraphics{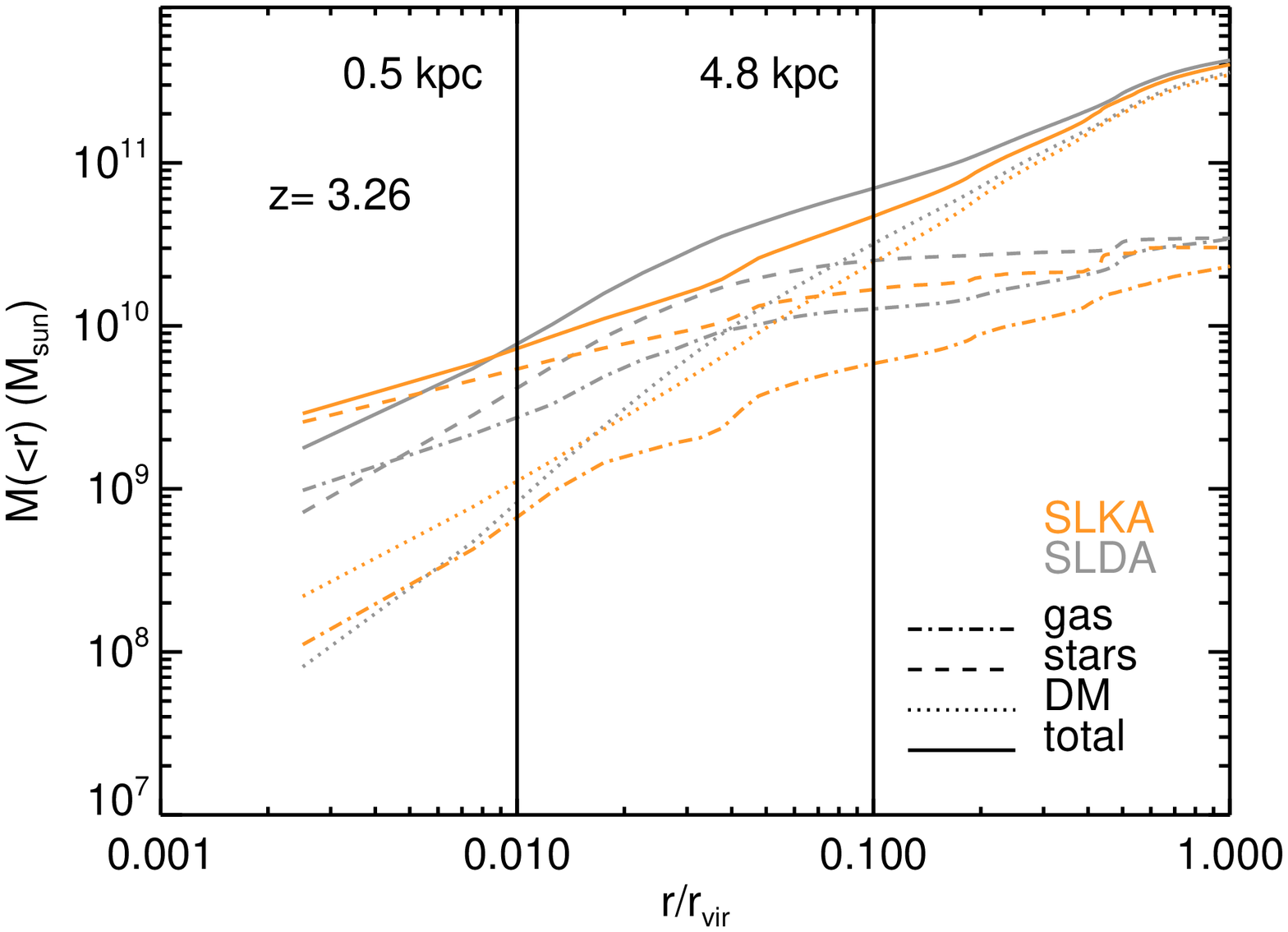}}}
  \caption{Integrated mass profiles as a function of radius normalised to the halo virial radius at $z=4.71$ (left panel) and $z=3.26$ (right panel). Grey lines are for the SLDA simulation and orange lines are for the SLKA simulation. Solid lines are the total masses, dotted lines the DM masses, dashed lines are the stellar masses and dot-dashed lines are the gas masses. At early times $z=4.71$, the simulation including SN feedback with delayed cooling (SLDA) has reduced the mass concentration within 10 per cent of the virial radius compared to the simulation with kinetic mode of SN feedback (SLKA).}
    \label{fig:mcomp}
\end{figure*}

From our set of simulations, we have seen that the central BH growth is initially inhibited by the presence of SNe when they are associated with a delayed cooling prescription that maximises the gas motion. Now, we want to compare the theoretical expectation of the velocity of SN-driven winds to the escape velocity of the galactic bulge. Therefore, if SN-driven winds are efficient, they should drive the gas out of the bulge (and star-forming clumps), until the bulge becomes sufficiently massive to retain any outflows in its sphere of influence.

The characteristic velocity of SN-driven winds in a star-forming clump is of the order of the Sedov velocity:
\begin{eqnarray}
u_{\rm SN}&\simeq&1.2 \sqrt{ m_{\rm new,s} f_{\rm blast} \eta_{\rm SN} e_{\rm SN} \over m_{\rm g}}\nonumber \\
&\simeq& 270 \sqrt{f_{\rm blast}\eta_{\rm SN}\over 0.1}\sqrt{ (m_{\rm new, s} / m_{\rm g})\over 0.1} \, \rm km\, s^{-1} \, ,
\end{eqnarray}
where $e_{\rm SN}=10^{50} \, \rm erg\, M_\odot^{-1}$ is the SN type II specific energy, $f_{\rm blast}\le 1$ is the fraction of SN energy input that is \emph{effectively} transferred into the gas (in the case of efficient energy losses through gas cooling $f_{\rm blast}$ can be much lower than 1), and $m_{\rm new, s}$ is the amount of new stars formed within the gas clump. $m_{\rm new, s}$ depends on the clump free-fall time (i.e. the clump gas density) and the time-scale before the ignition of the first SNe $t_{\rm SN}=10 \, \rm Myr$. Integrating the Schmidt law over time and volume, and assuming that the gas density is nearly constant over $t_{\rm SN}$, one obtains
\begin{equation}
{m_{\rm new, s}\over m_{\rm g}}\simeq \epsilon_* {t_{\rm SN} \over t_{\rm ff}}\, ,
\end{equation}
where the gas free-fall time $t_{\rm ff}$ for the central gas clump of mass $\sim10^8\, \rm M_\odot$ in the proto-bulge (see Fig.~\ref{fig:menc}) and of $\sim100 \, \rm pc$ radius\footnote{This radius is obtained doing a bulge-disc decomposition on the stellar density profile of SHDA using a fitting procedure with two exponentials. The bulge scale radius is of $\sim 25\, \rm pc$. We use a radius of 100 pc in order to account for 90 per cent of the stellar mass content in the bulge in that case.} is $\simeq1.6\, \rm Myr$. Therefore, the ratio of $m_{\rm new, s}/ m_{\rm g}\simeq 0.12$.
To summarize, the velocity of the blast propagating into the central gas clump (bulge) should be close to 270 $\rm km\, s^{-1}$ for the delayed cooling case, where $f_{\rm blast}~\sim 1$ is ensured by construction.

The escape velocity of a clump is 
\begin{equation}
u_{\rm esc}=\sqrt{ 2 G m_{\rm cl} \over r_{\rm cl} } \, ,
\end{equation}
where $m_{\rm cl}$ and $r_{\rm cl}$ are respectively the clump total mass and clump radius.
If we replace the clump radius $r_{\rm cl}$ by the bulge radius $r_{\rm b}\simeq 100 \, \rm pc$, and the clump mass $m_{\rm cl}$ by the bulge mass $M_{\rm b}\simeq10^9 \, \rm M_\odot$ that are the representative values when the BH starts growing at the Eddington rate.
The escape velocity is, therefore, $u_{\rm esc}\simeq300 \,\rm km\, s^{-1}$ and is comparable to the Sedov velocity produce by SN winds in the bulge.
Thus, below $M_{\rm b}\lesssim 10^9\,\rm M_\odot$, and provided that $f_{\rm blast}\sim 1$, the gas cannot accumulate in the bulge and is regularly ejected from the bulge gravitational potential well (see the top panel of  Fig.~\ref{fig:menc} for SHDA, SLDA and SLD), and above $M_{\rm b}\gtrsim 10^9\,\rm M_\odot$ the gas is locked into the bulge, in a fountain that ejects gas that quickly falls back, and provides the fuel to feed the central BH over a prolonged duration.

Fig.~\ref{fig:vesc} shows the escape velocity within 100 pc distance from the center of the galaxy.
For sake of simplicity, we call that region the bulge here, but note, however, that the bulge is not always well defined, especially at early times when only short-lived star forming clumps are present in the galaxy with delayed cooling SN feedback.
At early times ($z > 3.1$), since SNe in the delayed cooling cases prevent the gas from  accumulating too rapidly in the bulge, we observe that the escape velocity of the bulge increases more slowly than in the kinetic feedback cases.
At $z=3.1$, the escape velocity in the bulge in SLD and SLDA reaches 270 $\rm km\, s^{-1}$. 
This value is the predicted SN-driven wind velocity in the bulge for $f_{\rm blast}=1$ and marks the end of the SN-driven evolution of the bulge gas content. 
From this moment, the BH can grow close to the Eddington rate and settles into self-regulated growth $\sim 400$ Myr later at $z=2.6$.
Note that escape velocities in the bulge in both SLD and SLDA, respectively with and without AGN feedback, are similar during the SN-driven evolution since the AGN has as yet no impact on the surrounding gas as SN-driven winds prevent the BH from  accreting gas at  near-Eddington.
There is a difference in escape velocity between SLDA and SHDA (i.e. when changing resolution) at late times. 
This is attributed to the more compact bulge in the high resolution run ($\sim 25 \, \rm pc$ against $\sim 60 \, \rm pc$ bulge scale radius), although the mass of the bulge remains identical (see section~\ref{sec:inflow}).

For the kinetic cases, SLK and SLKA, the bulge endures a rapid mass growth already very early on due to the inefficiency of the coupling of the SN energy to the gas ($f_{\rm blast}\ll 1$).
Until the BH has reached its self-regulated phase in SLKA, the bulge mass can grow steadily, as it does in the case without any feedback from SNe nor AGN (SL run).
Finally, the growth of the bulge escape velocity (or equivalently the bulge mass) is suppressed once the BH is massive enough to have a significant impact on the surrounding gas, which is around $z\simeq6$ in SLKA.
Interestingly, the bulge escape velocities of SLDA and SLKA are similar $u_{\rm esc}\simeq 500 \, \rm km\, s^{-1}$ once both simulations have reached the BH self-regulated phase ($z<3$), and their value is smaller compared to the runs without AGN feedback $u_{\rm esc}\simeq 800 \, \rm km\, s^{-1}$ (SL, SLD and SLK).

A more in-depth investigation of the structure of the gravitational potential well, and, thus, mass distribution is shown in Fig.~\ref{fig:mcomp}, where we see how the gas mass accumulates differently in SLKA and SLDA at early epochs.
In the SLKA run, the escape velocity peaks at $\sim 400 \, \rm km\,s^{-1}$ at $z=4.71$ within the galactic bulge ($< 0.01 \, r_{\rm vir}=200\, \rm pc$) due to the high concentration of stars in the central region of the galaxy (and with a low concentration of gas and DM).
Vice versa, in the SLDA run at the same redshift, the peak of the escape velocity is $250 \, \rm km\, s^{-1}$ at a distance $\simeq 0.05\, r_{\rm vir}=1 \, \rm kpc$ that roughly corresponds to the galaxy radius. 
The escape velocity within the galactic bulge reaches lower values around $150\, \rm km\, s^{-1}$ a factor of two below the SN wind velocity.
In this simulation, the dominant mass component in the bulge is the gas with a fraction above $50\%$. 
Though the amount of gas within the bulge is larger by a factor of 2 in SLDA than in SLKA, the amount of total mass is lower in SLDA because of the frequent expulsions of gas, that translate into a much lower amount of stellar mass.
Note also the strong enhancement of the DM concentration within the central region in the SLKA run due to adiabatic contraction~\citep{gnedinetal04}.

At the near-Eddington growth of the BH in SLDA ($z=3.26$), the escape velocity profile in the galaxy has been boosted by a factor $\sim 2,$ reaching $\sim 400 \, \rm km\, s^{-1}$ at $r\simeq0.03\, r_{\rm vir}$ ($350 \, \rm km\, s^{-1}$ at the galaxy center). 
These new values of the escape velocity in SLDA create a gravitational barrier large enough for the SN winds to be trapped within the core of the galaxy, and, therefore, allow  maintenance of a steady reservoir of cold gas for the BH to grow from.
The SLKA escape profile shows a small  variation with the $z=4.71$ profile, it is simply rescaled by a $\gtrsim 1$ factor.
Although the stellar mass and DM mass content within the galaxy bulge are now similar in SLKA and SLDA, the amount of gas is still a factor 5 different, and the gas fraction in SLDA is close to $30\%$ while it is below $10\%$ in the SLKA.

\subsection{The triggering of nuclear inflows by a major merger}
\label{sec:inflow}

\begin{figure}
  \centering{\resizebox*{!}{6cm}{\includegraphics{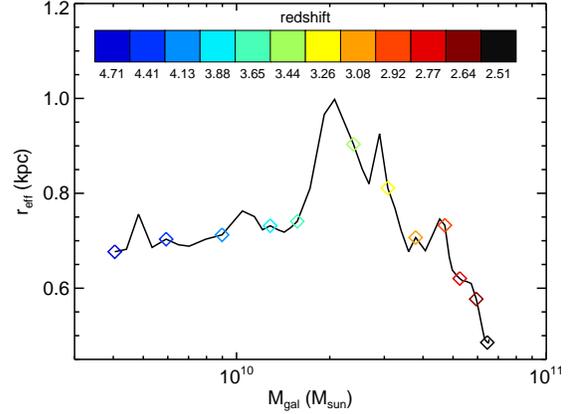}}}
  \caption{Evolution with redshift of the effective radius and the stellar mass of the central galaxy in SHDA at different redshifts (indicated by the colors). The 1:3 merger is happening from $z=3.7$ to $3.5$ when the effective radius quickly increases and as quickly contracts.}
    \label{fig:rmvsz}
\end{figure}

\begin{figure*}
  \centering{\resizebox*{!}{4.2cm}{\includegraphics{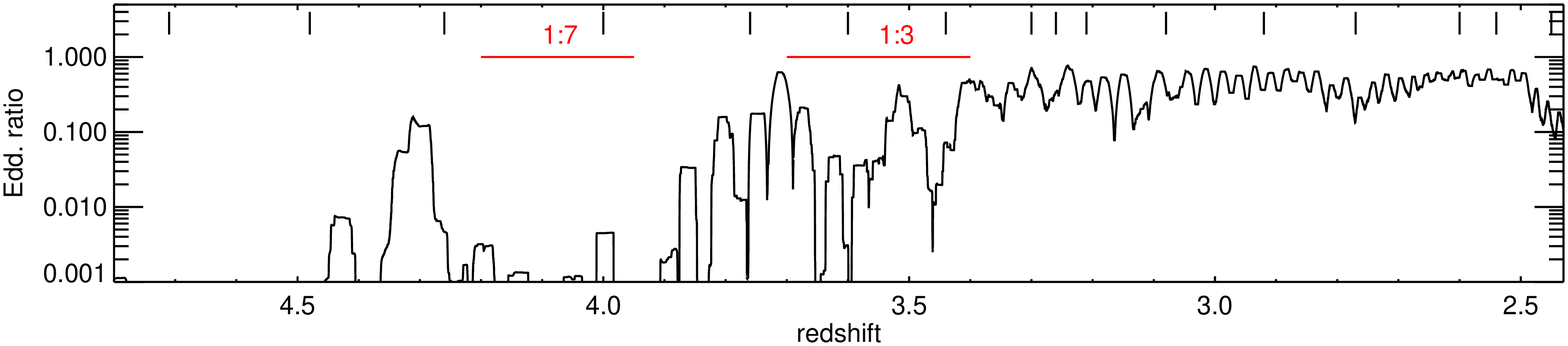}}}
  \centering{\resizebox*{!}{4.2cm}{\includegraphics{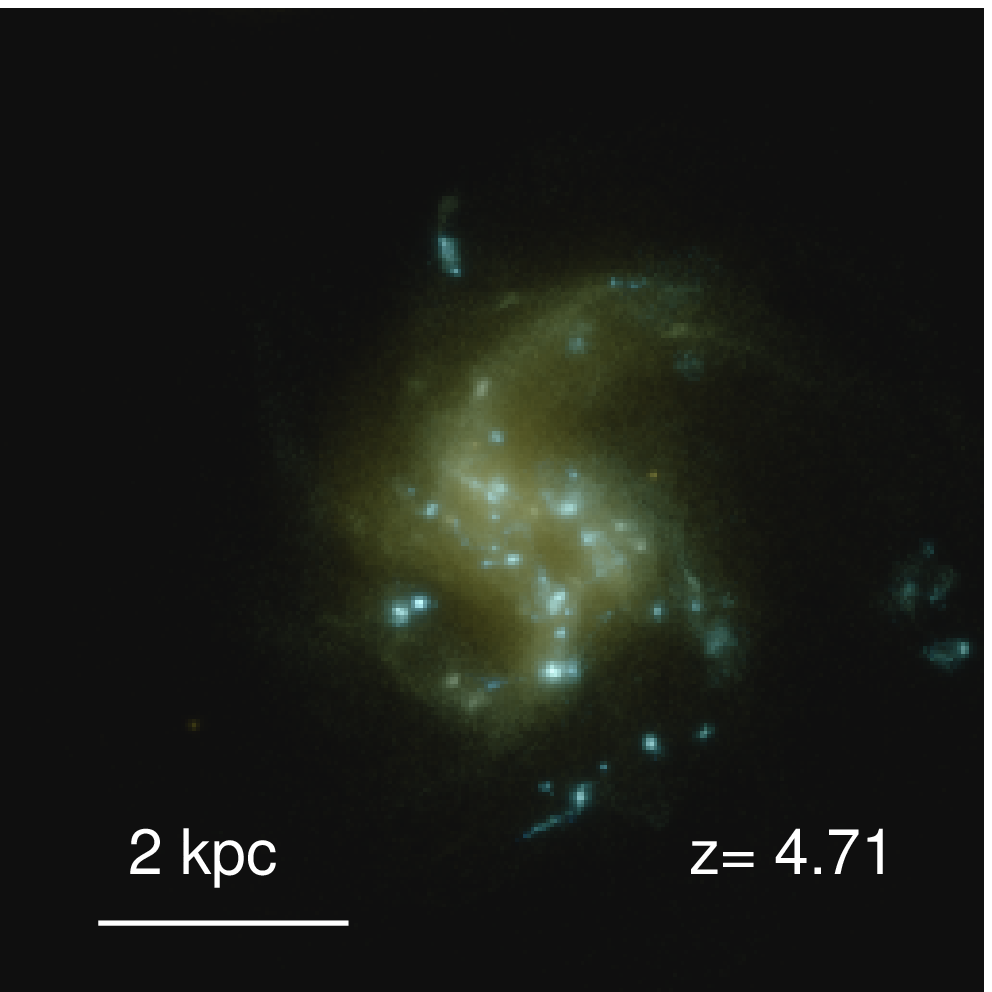}}}
  \centering{\resizebox*{!}{4.2cm}{\includegraphics{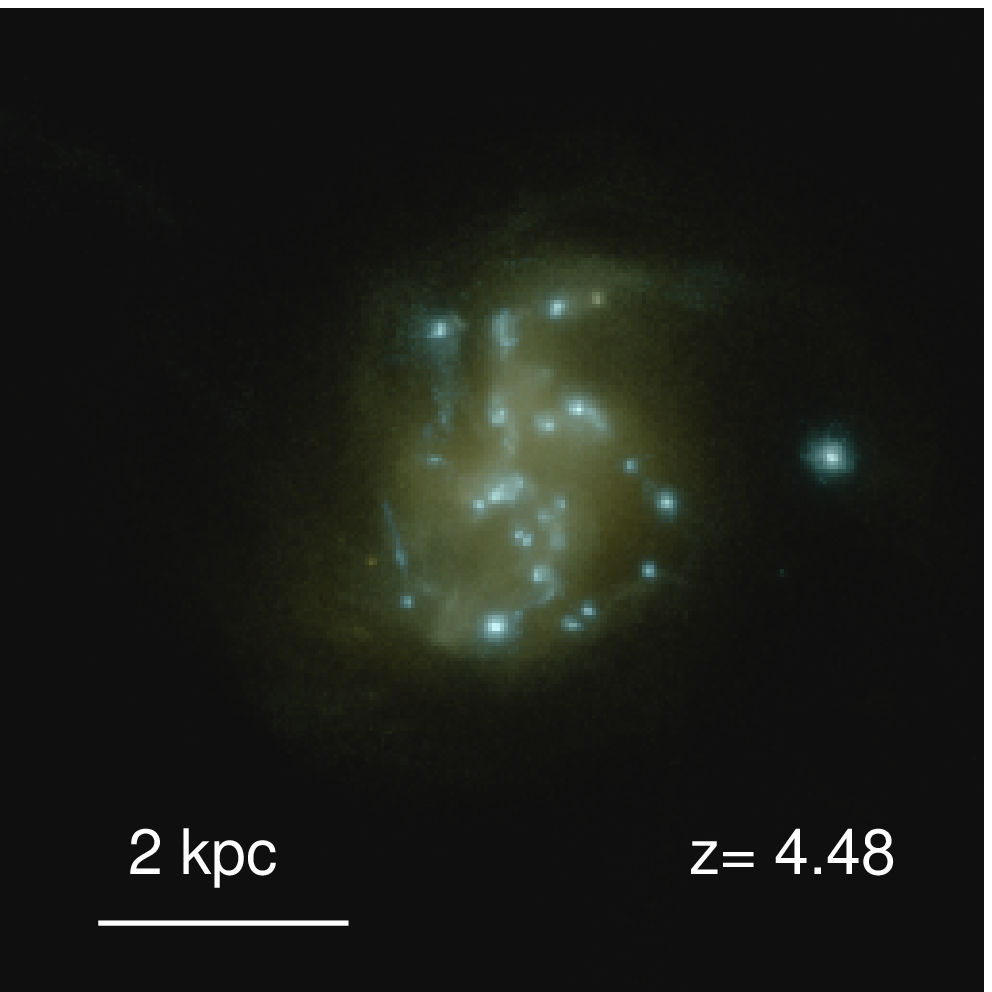}}}
  \centering{\resizebox*{!}{4.2cm}{\includegraphics{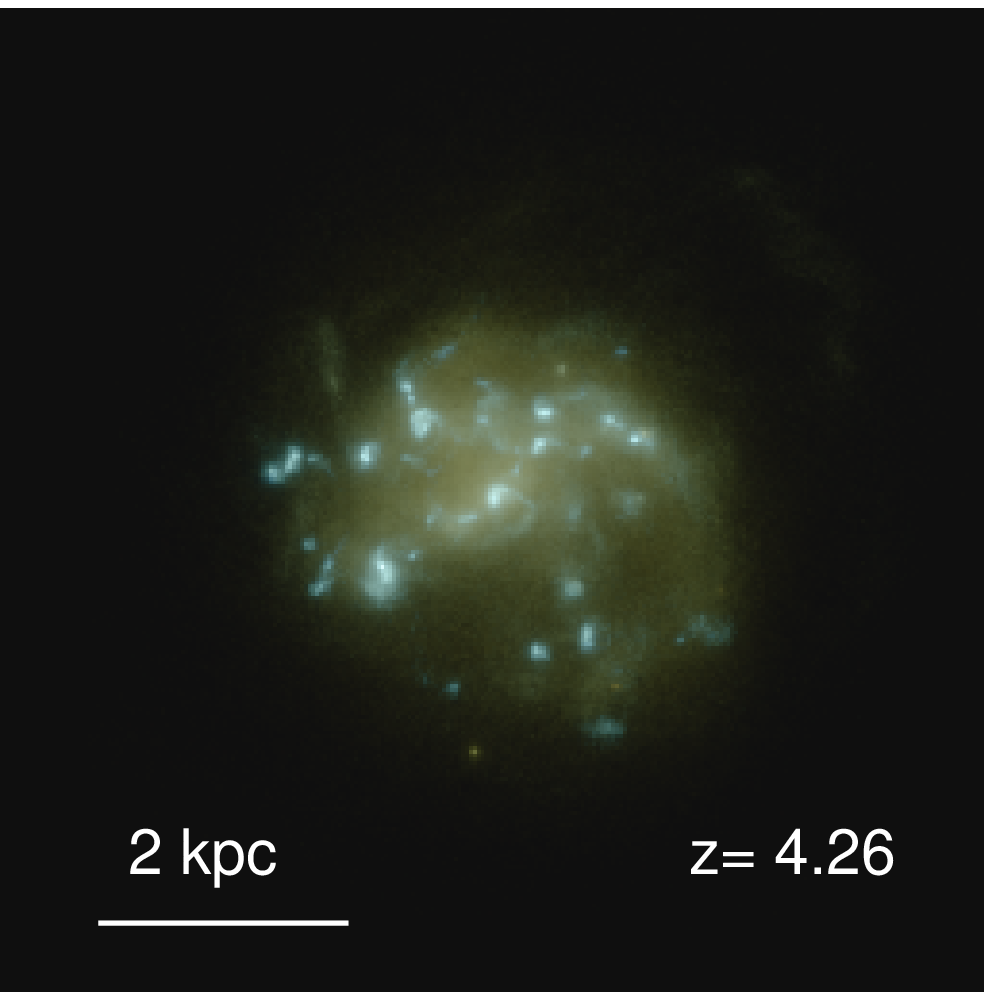}}}
  \centering{\resizebox*{!}{4.2cm}{\includegraphics{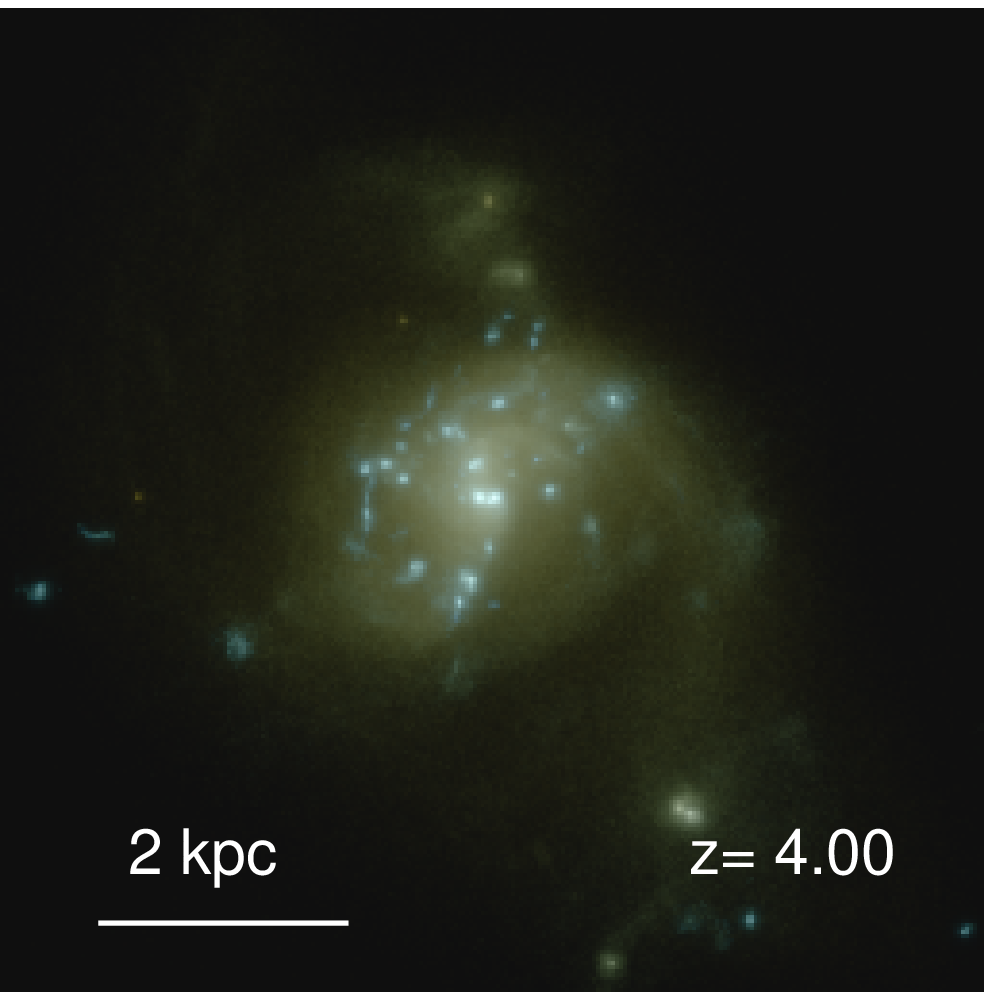}}}
  \centering{\resizebox*{!}{4.2cm}{\includegraphics{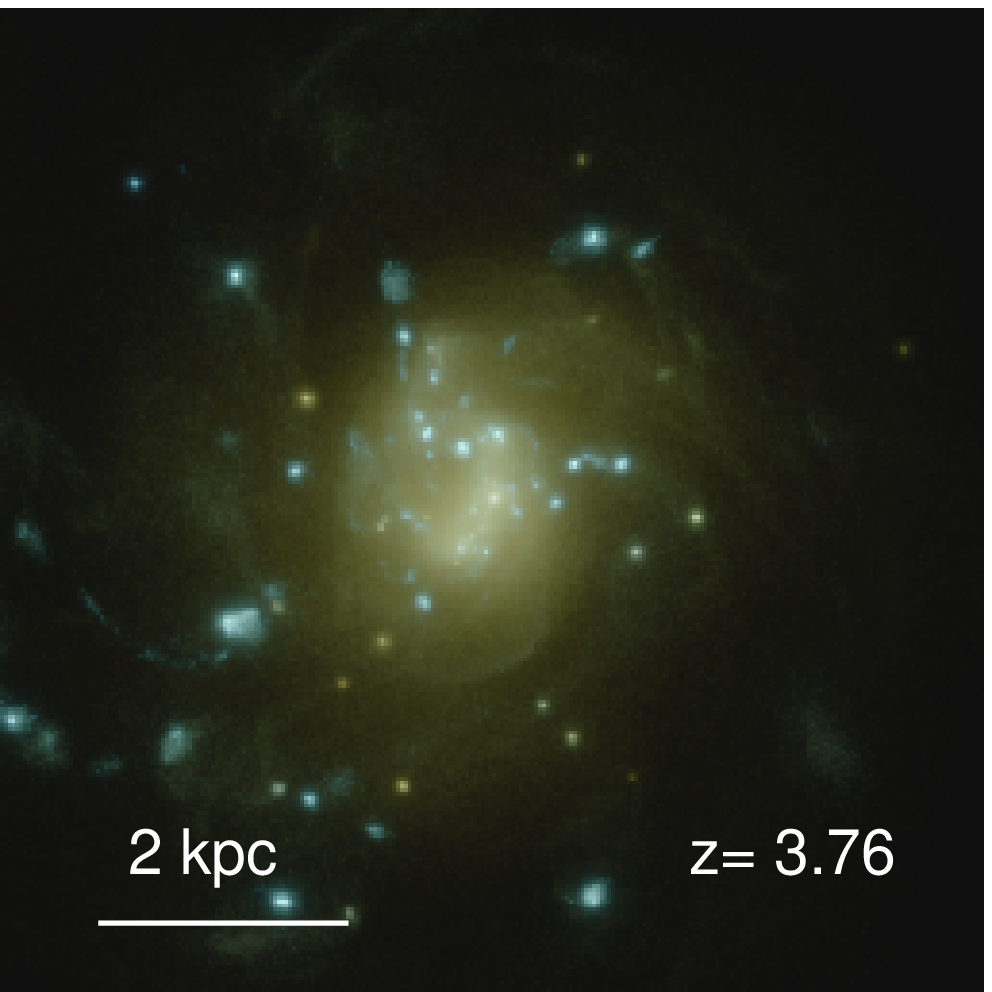}}}
  \centering{\resizebox*{!}{4.2cm}{\includegraphics{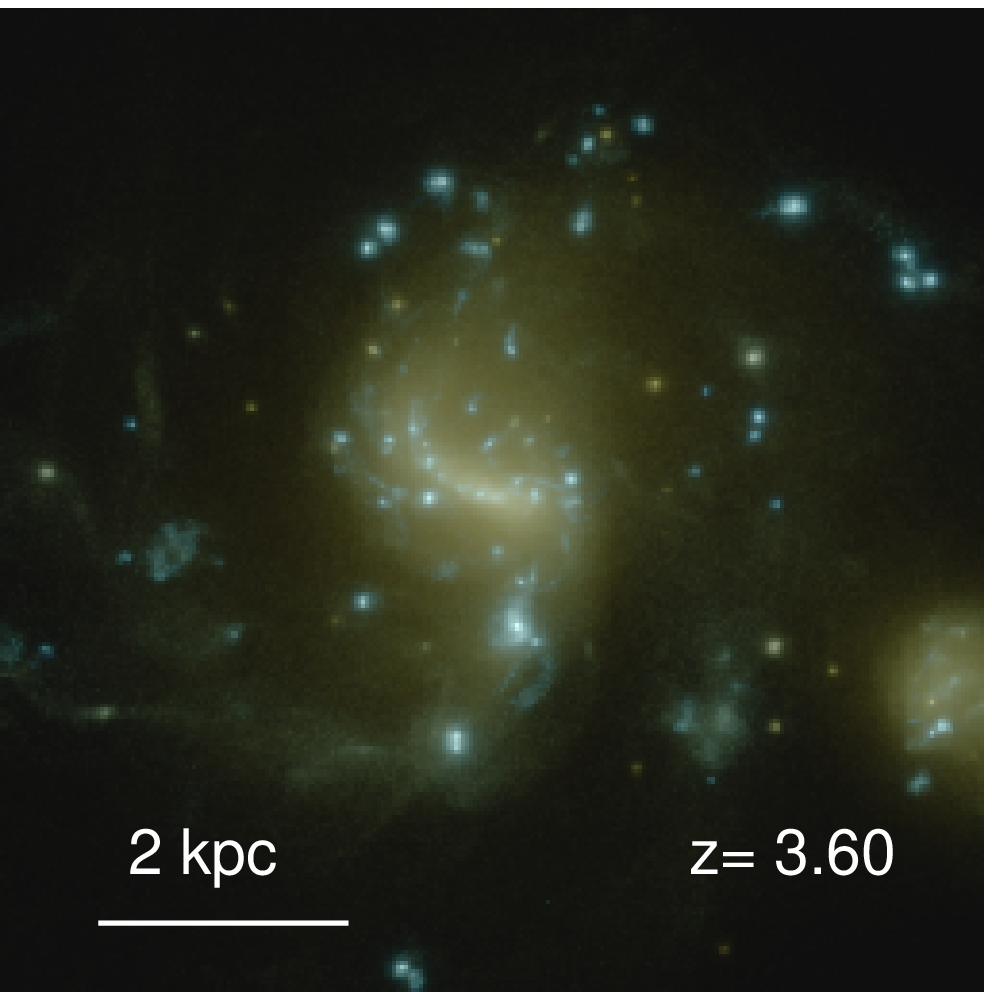}}}
  \centering{\resizebox*{!}{4.2cm}{\includegraphics{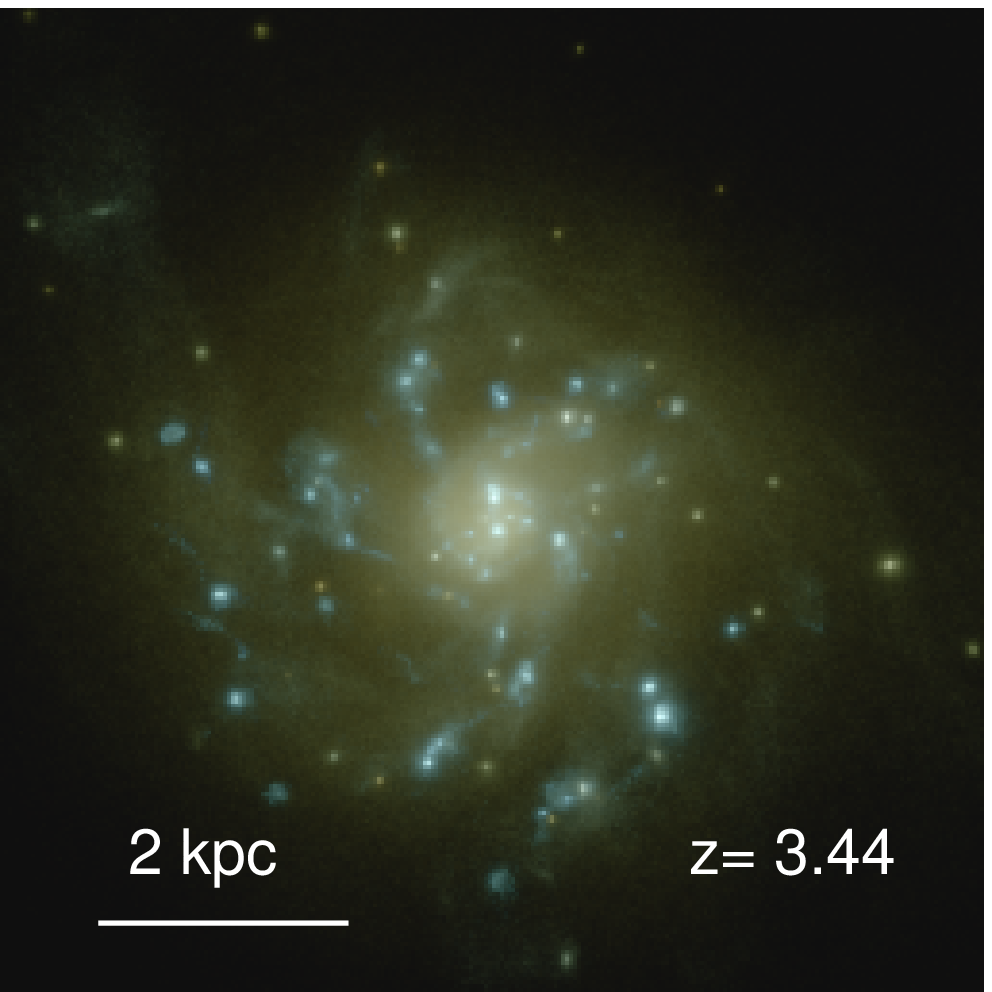}}}
  \centering{\resizebox*{!}{4.2cm}{\includegraphics{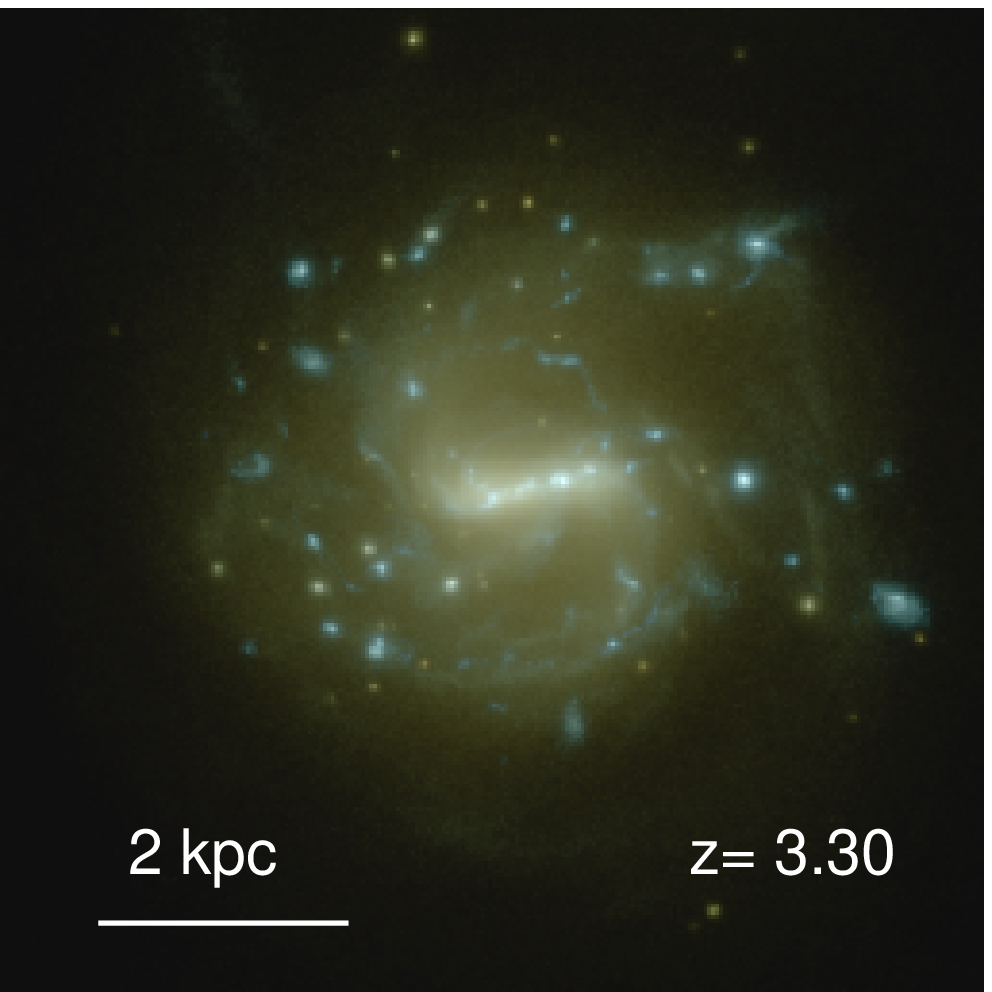}}}
  \centering{\resizebox*{!}{4.2cm}{\includegraphics{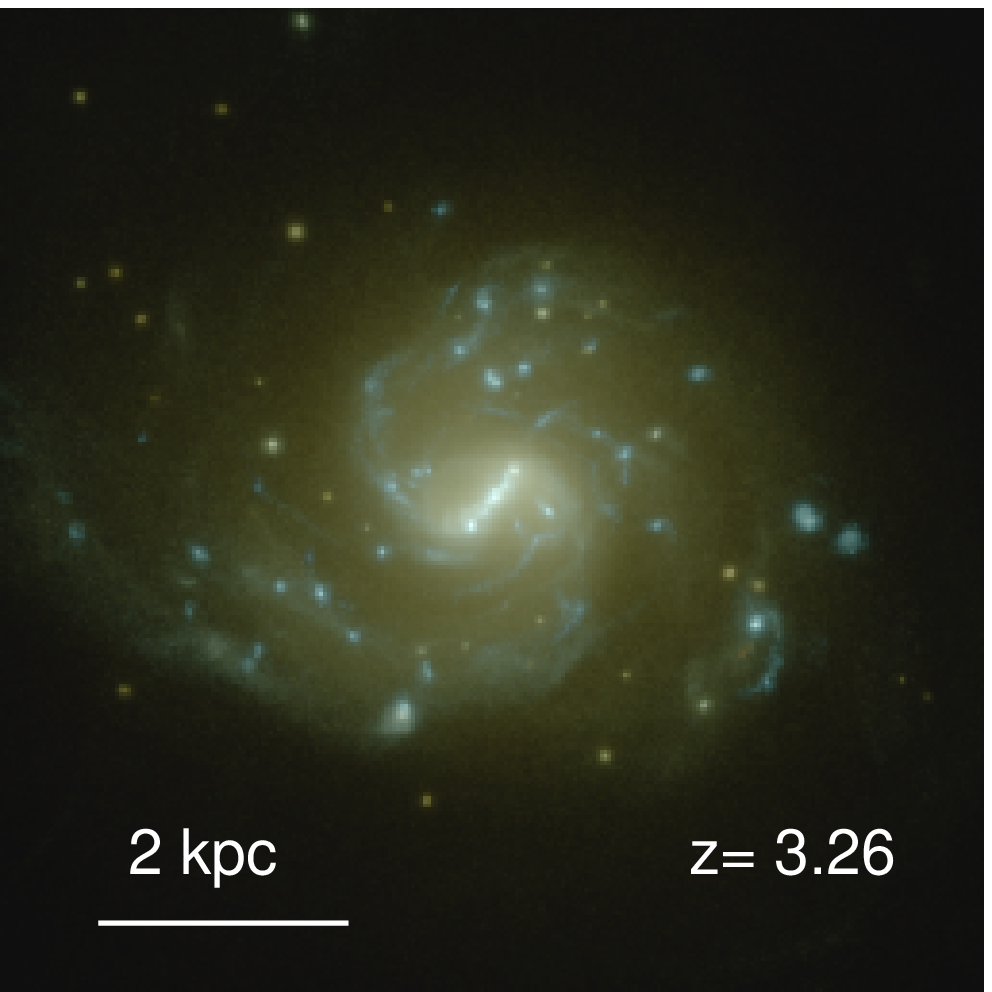}}}
  \centering{\resizebox*{!}{4.2cm}{\includegraphics{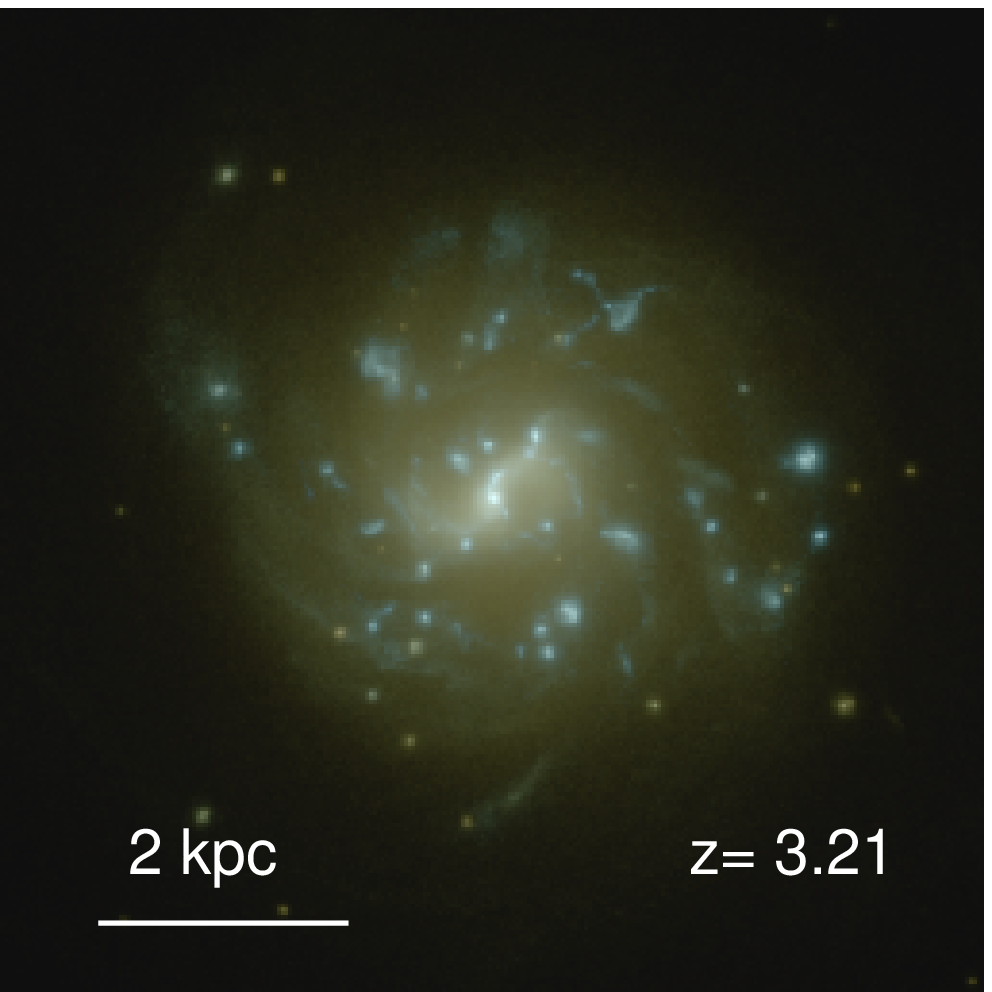}}}
  \centering{\resizebox*{!}{4.2cm}{\includegraphics{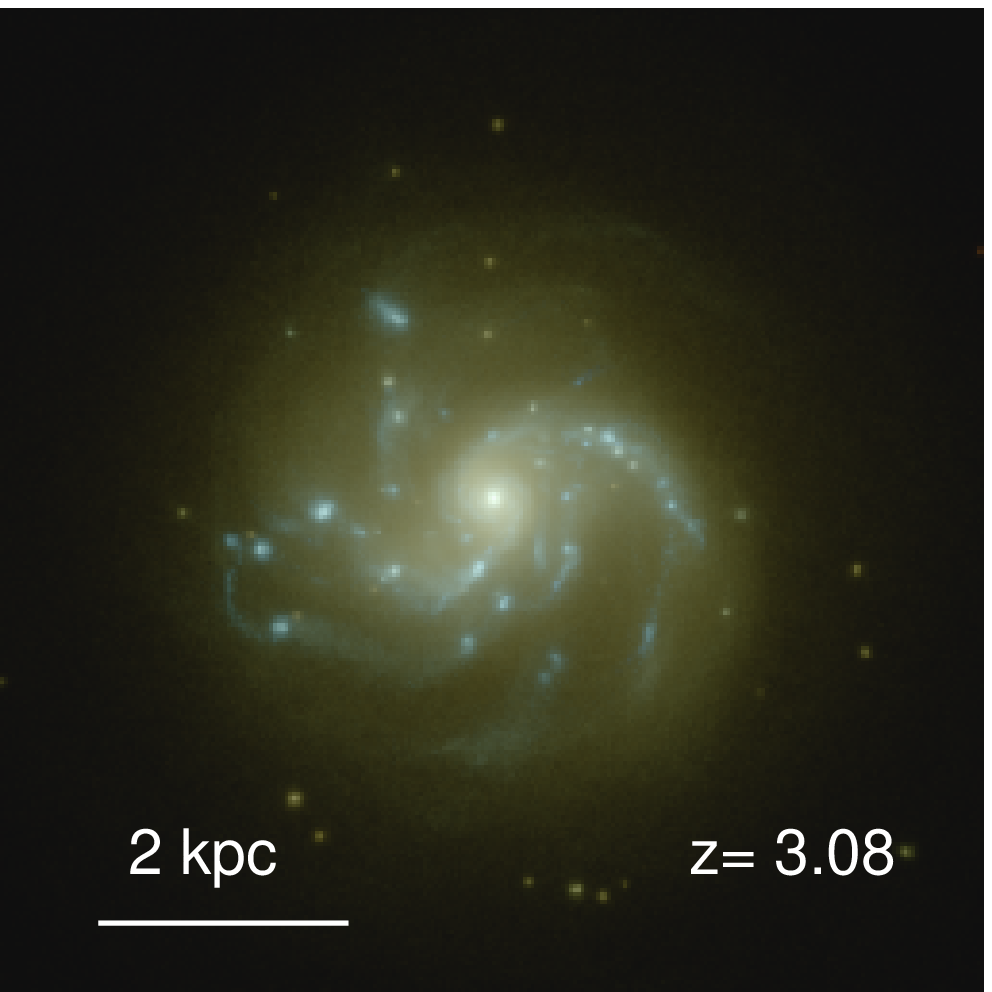}}}
  \centering{\resizebox*{!}{4.2cm}{\includegraphics{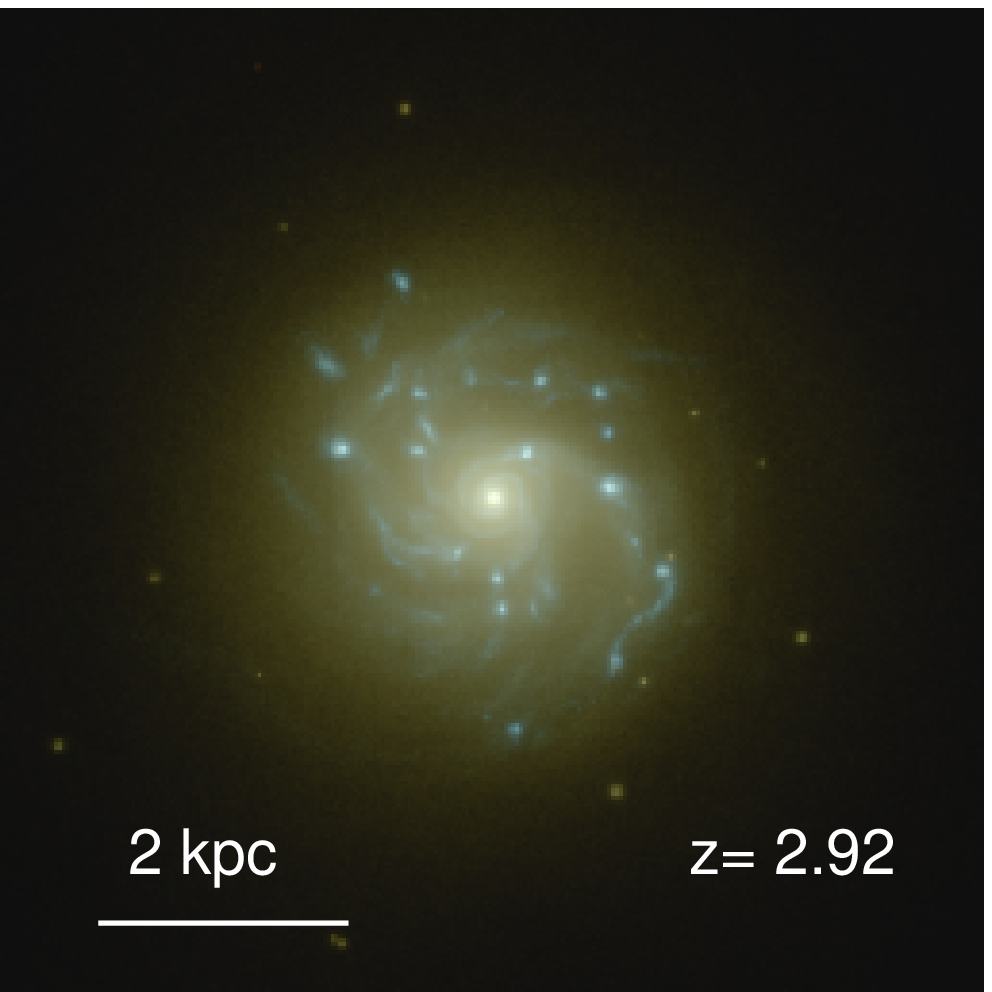}}}
  \centering{\resizebox*{!}{4.2cm}{\includegraphics{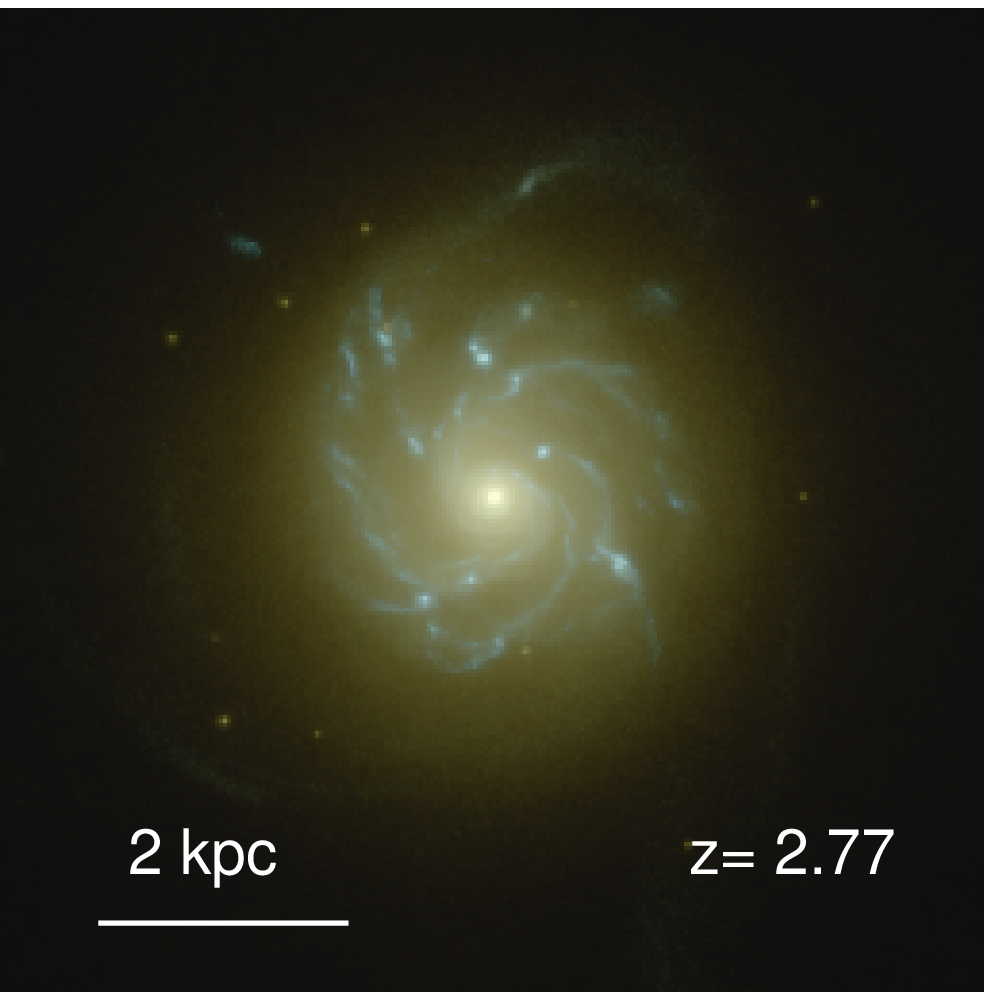}}}
  \centering{\resizebox*{!}{4.2cm}{\includegraphics{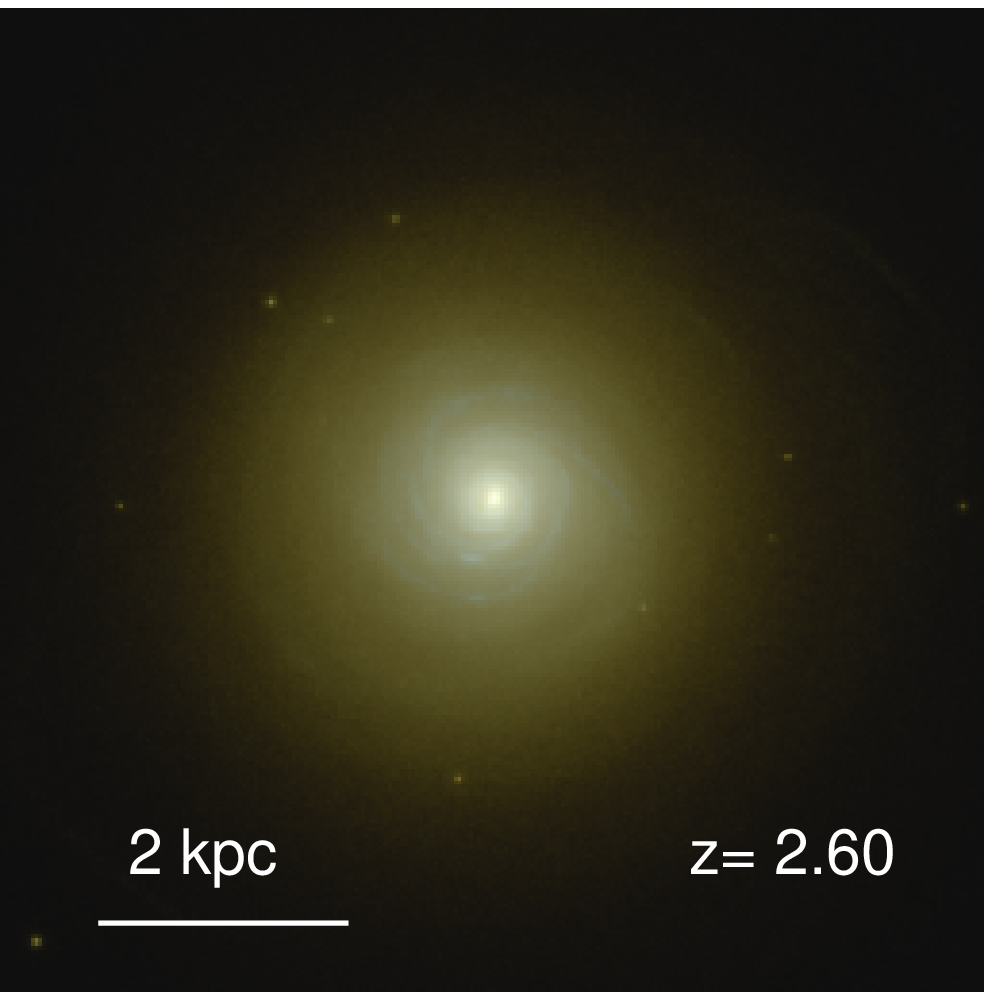}}}
  \centering{\resizebox*{!}{4.2cm}{\includegraphics{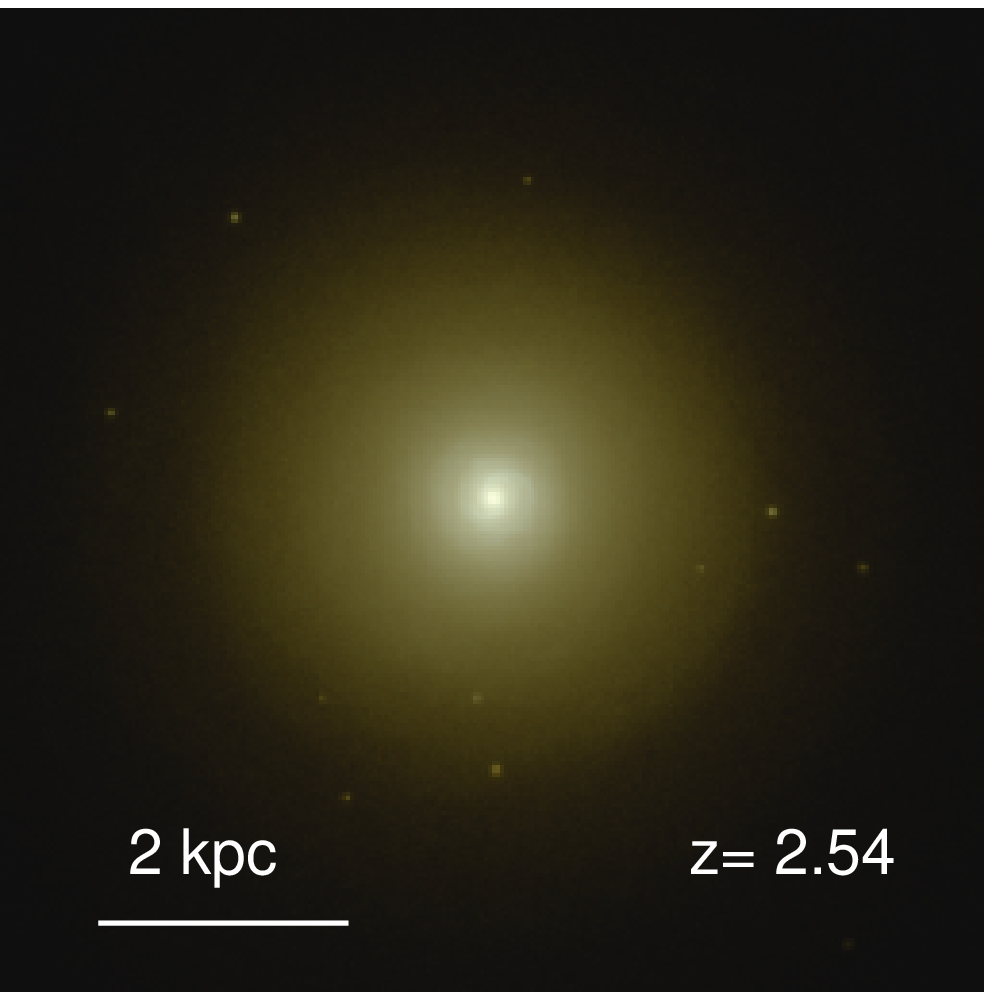}}}
  \centering{\resizebox*{!}{4.2cm}{\includegraphics{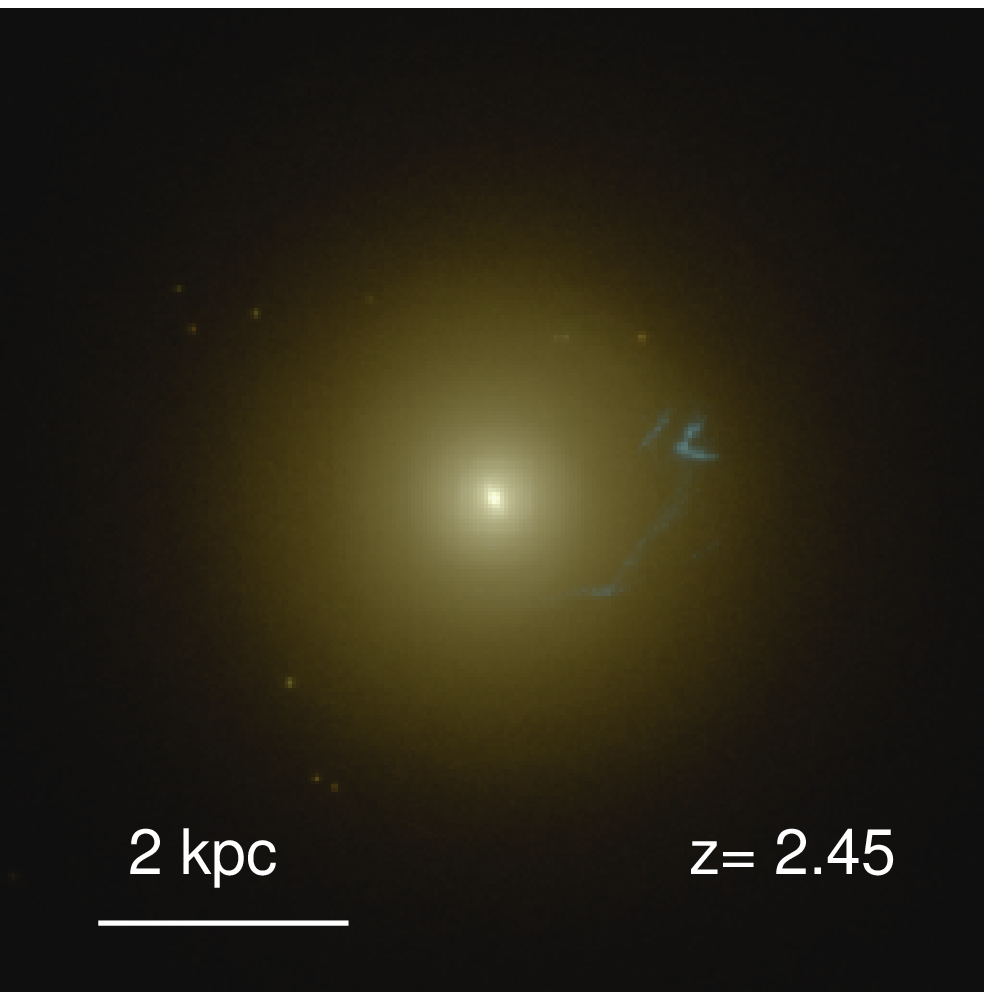}}}
  \caption{Top: Eddington ratio of the central BH as a function of redshift for the SHDA. Important mergers for the BH growth are indicated as the red horizontal lines with their stellar mass ratio. Small vertical bars pinpoints the redshifts of the stellar images. Bottom: Stellar emission of the central galaxy at different redshifts for SHDA as they would be observed face-on through u-g-i filter bands in the rest-frame. Extinction by dust is not taken into account. We see three main phases of galaxy evolution: the morphologically disturbed phase ($z> 3.6$), the disc settling phase with a bar ($2.77\le z\le 3.6$), and the red nugget ($z\le 2.6$).}
    \label{fig:shda}
\end{figure*}

\begin{figure}
  \centering{\resizebox*{!}{6cm}{\includegraphics{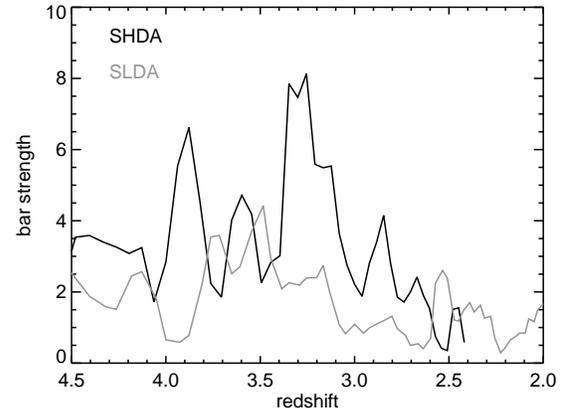}}}
  \caption{Evolution with redshift of the bar strength of the central galaxy in SHDA (black) and SLDA (grey). The bar becomes stronger between $3.1<z<3.4$ after the major merger (occurring at $3.4<z<3.7$) in SHDA, and is the signature of a nuclear inflow of gas that triggers the steady Eddington-limited growth of the central BH at this redshift.}
    \label{fig:barvsz}
\end{figure}

In the simulations with delayed-cooling SN feedback, the BH enters its Eddington-limited growth phase once the galaxy encounters a significant merger with a stellar mass ratio of $\sim$1:3.
Fig.~\ref{fig:rmvsz} shows the galaxy effective radius in SHDA as a function of its stellar mass for different redshifts. 
The stellar mass of a galaxy is defined by the total amount of stars which belong to the same structure using a galaxy finder algorithm. We employ the {\sc AdaptaHOP} finder~\citep{aubertetal04, tweedetal09} onto the distribution of all stars in the simulation.
For computing the effective radius, we project the stellar mass along the 3 cartesian axes of the box, measure the half-mass projected radius of each projection, and take the geometrical average of the three components.
In the early stage of the galaxy evolution ($z>4$) its effective radius stay nearly constant even though the galaxy stellar mass increases.
Once the 1:3 merger happens from $z=3.7$ to $z=3.5$ the galaxy radius quickly increases, but then contracts further.

Fig.~\ref{fig:shda} illustrates further the connection between the BH feeding and the morphological transformation of the galaxy in SHDA.
This figure shows (at the  top) the BH growth relative to Eddington as a function of redshift with the two largest merger mass ratios highlighted in red, and the panels show the stellar emission in rest-frame u, g, i filters at various redshifts.
We see that above $z>3.6$ the galaxy is a very disturbed object with no bright galaxy bulge, plenty of blue stellar clusters and a disc scale radius almost constant.
Between $3.21<z<3.6$, a stellar disc forms with a central bar that trigger the inflow of gas in the center of the galaxy resulting in a compact bulge later on at $z<3.21$.
In this redshift range, the galaxy settles into a disc-like shape and has increased its radius (see also Fig.~\ref{fig:rmvsz}) at the onset of the merger with its massive companion (that one can see at the east/south-east of the panel corresponding to $z=3.6$).
We suggest that the central bar and the resulting nuclear inflow is induced by the new gravitational torques exerted by the merger, which explain the rise in the BH growth at the near-Eddington rate at this redshift.
Finally, this merger has allowed  sufficient amounts of mass to concentrate around the  BH through new gas inflows.
In return, the gravitational barrier (or escape velocity) of the bulge has become large enough to prevent the SN winds from depleting the cold gas content in the center of the galaxy.
The consequence for the BH is that there is now plenty of gas residing in the bulge, and the BH can, from $z=3.6$, grow at the near-Eddington rate until it reaches self-regulation $z\simeq2.5$.
At the end of the merger ($z=3.4$), the galaxy starts to shrink and later on ($z=2.6$) form a red nugget with a smooth red disc component and a compact stellar bulge.
In Fig.~\ref{fig:slda} of the Appendix~\ref{app:galmorphlowres}, we show the morphological evolution of the central galaxy in the SLDA run extended to lower redshift $z=2$.
The morphological evolution is remarkably comparable to the high-resolution run SHDA, and shows the formation of a spiral blue disc galaxy below the redshift of $z<2.5$.

In Fig.~\ref{fig:barvsz}, we show the bar strength of simulations SHDA and SLDA. 
The bar strength is obtained by measuring the inertia tensor of young stars (younger than 10 Myr) within 1 kpc radius from the center of the galaxy.
The inertia tensor is then diagonalized to get the three principal axis of the ellipsoid $e_1 < e_2 <e_3$.
We, then, use those three axis to evaluate the strength of the asymmetric distribution of young stars (i.e. the so-called ``bar'') with the bar strength being equal to $(e_3/e_2-1)e_3/e_1$.
We see that the bar strength in SHDA has two main peaks.
The first one at $z\simeq 3.9$, which correspond to the triggering of accretion bursts onto the central BH, though they are discontinuous and not at Eddington (see top panel of Fig.~\ref{fig:shda} between $3.6<z<3.9$).
The second more extended peak at $3.1<z<3.4$, and that is the result of the major merger occurring before at $3.4<z<3.7$ and which is responsible for the global destabilization of the disc and the central nuclear inflow.
This nuclear inflow takes the form of a bar that can be seen in the stellar light distribution in Fig.~\ref{fig:shda}.
Note that the same bar strength increase is observed for the lower resolution run SLDA, except that it appears slightly earlier in time ($3.1<z<3.8$) with a lower amplitude (that can probably be attributed to resolution effects).
It also explains why the Eddington-limited growth in SLDA appears slightly earlier than in SHDA (see top panel of Fig.~\ref{fig:mbhevol}).

\begin{figure}
  \centering{\resizebox*{!}{4.0cm}{\includegraphics{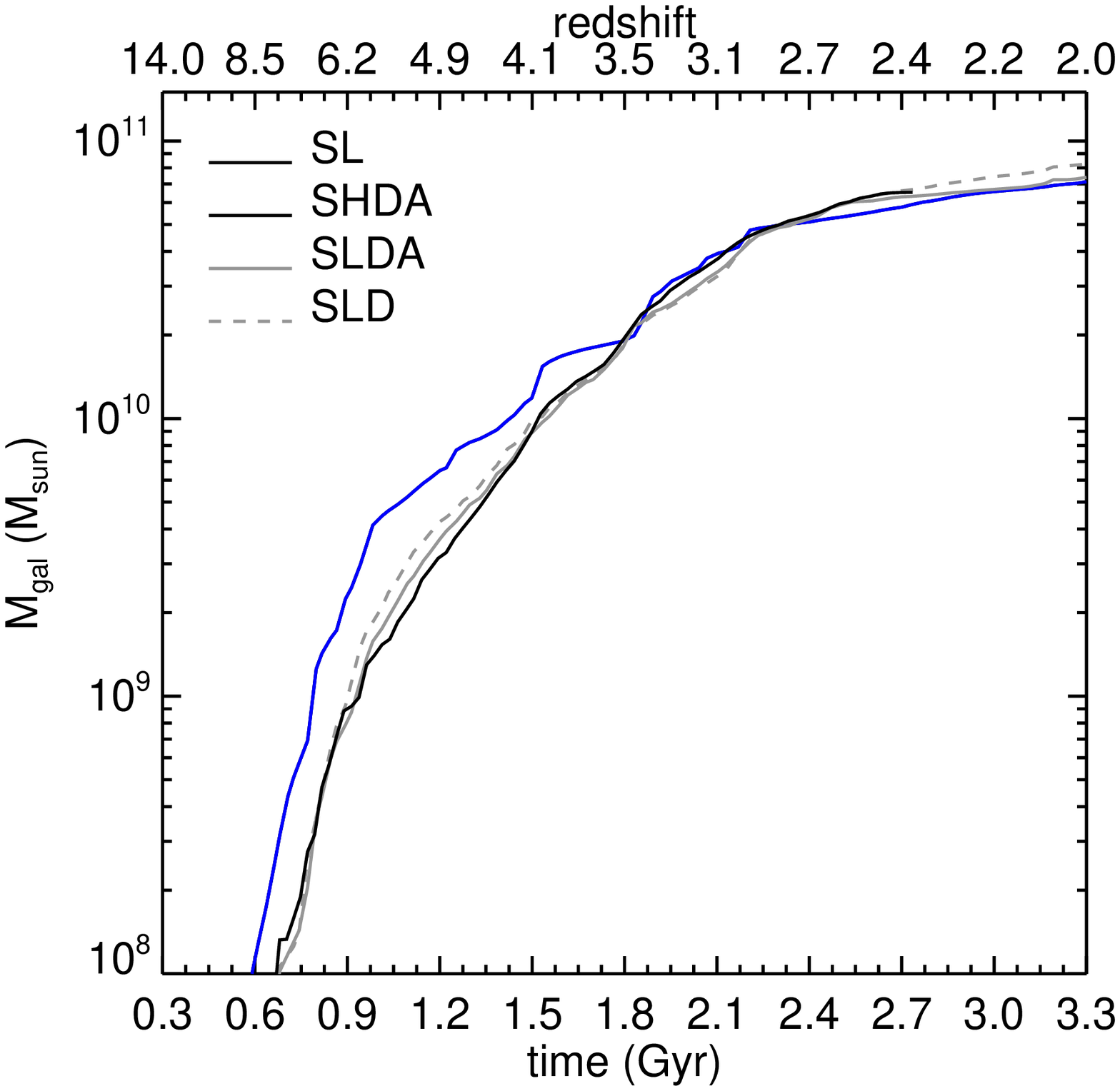}}}\hspace{-0.7cm}
  \centering{\resizebox*{!}{4.0cm}{\includegraphics{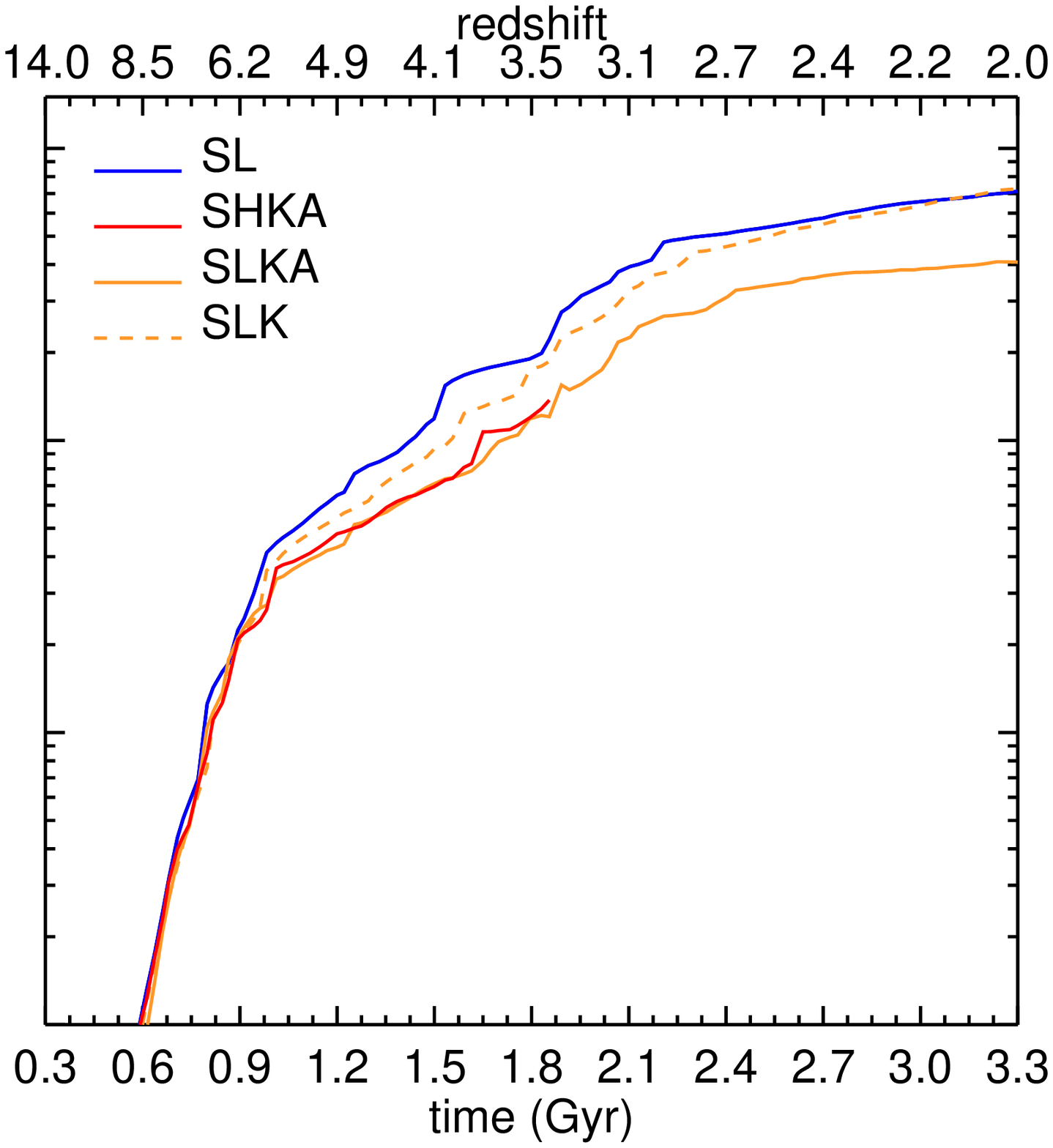}}}\\
  \centering{\resizebox*{!}{4.0cm}{\includegraphics{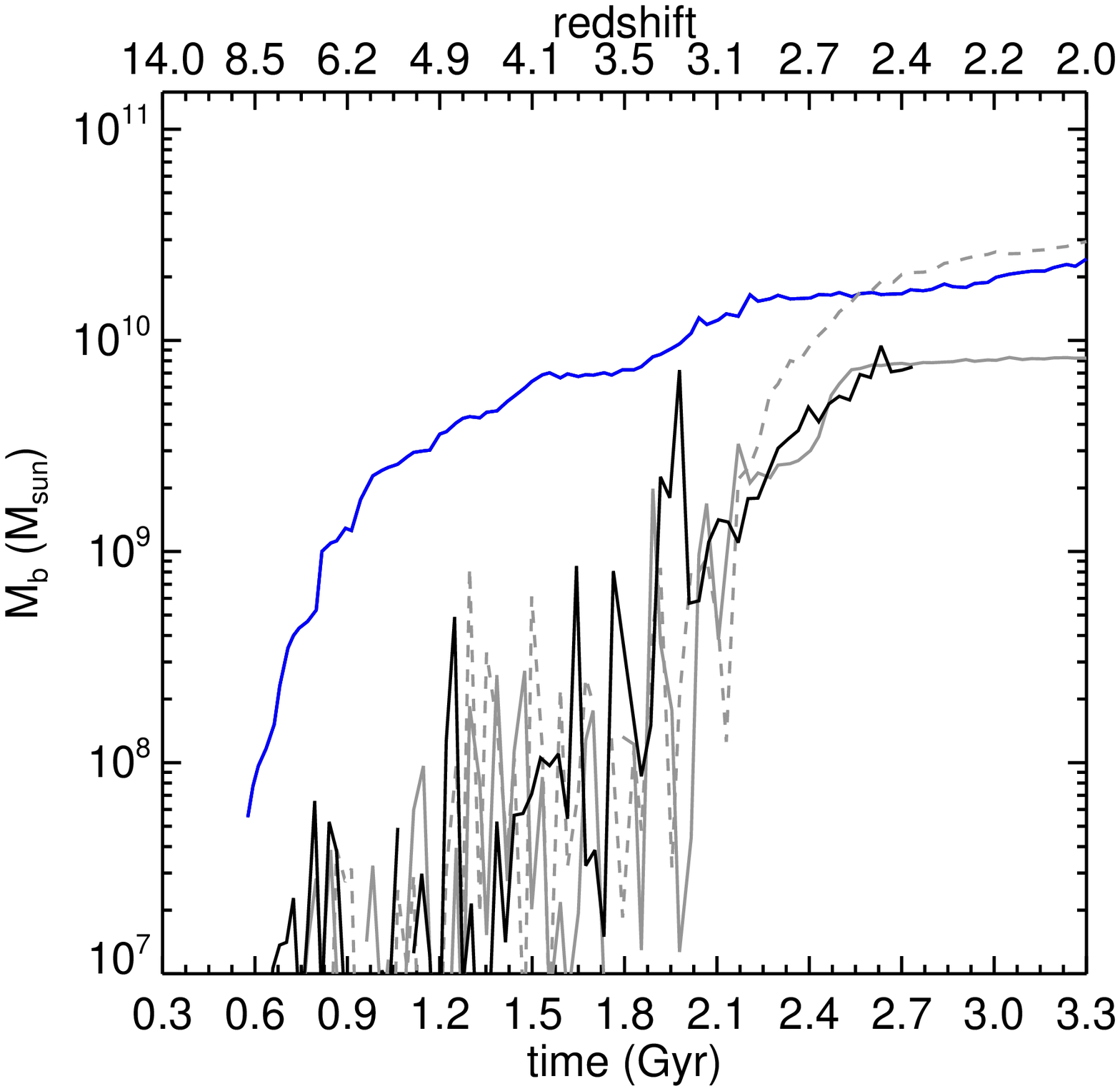}}}\hspace{-0.7cm}
  \centering{\resizebox*{!}{4.0cm}{\includegraphics{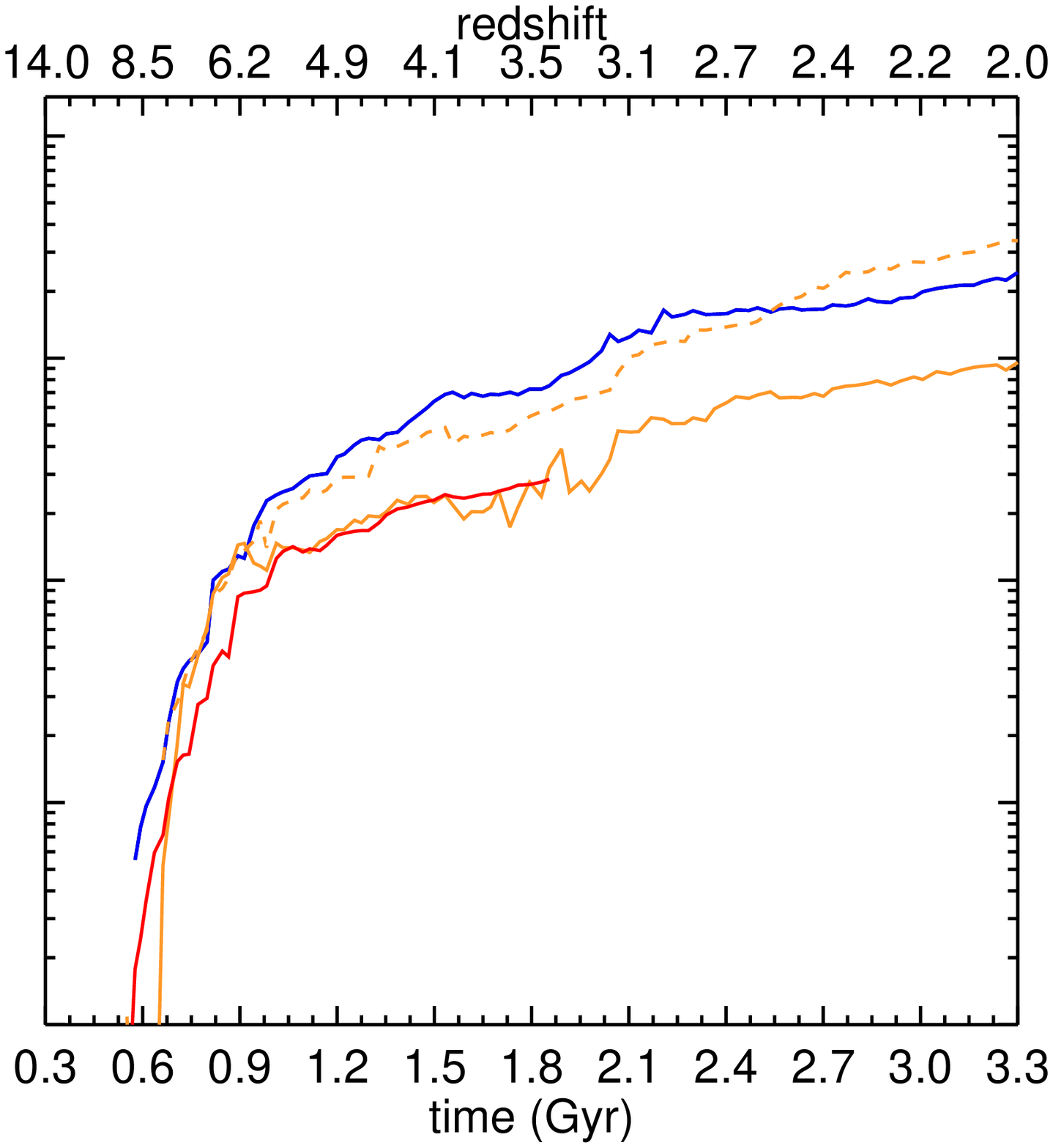}}}\\
  \centering{\resizebox*{!}{4.0cm}{\includegraphics{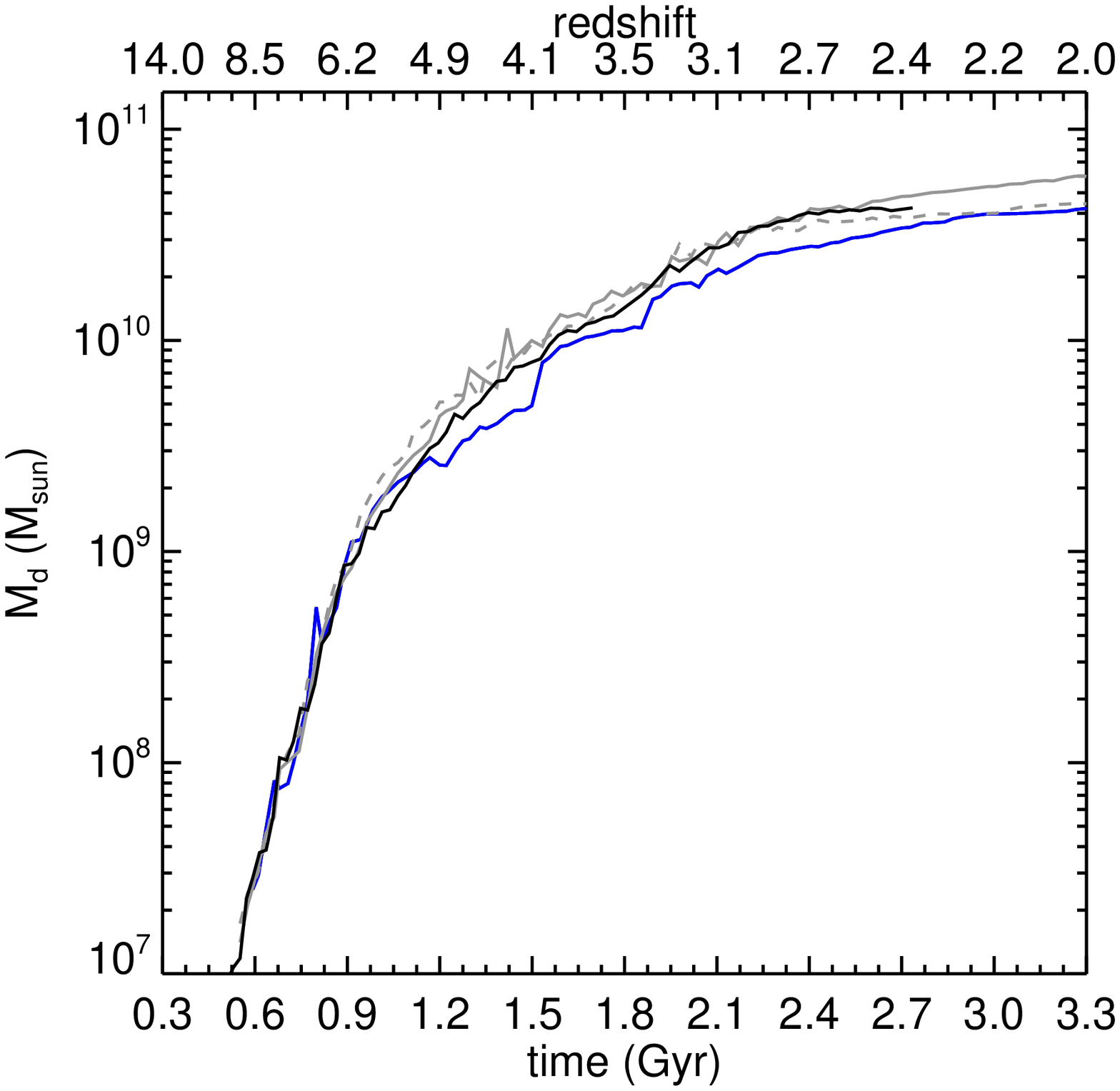}}}\hspace{-0.7cm}
  \centering{\resizebox*{!}{4.0cm}{\includegraphics{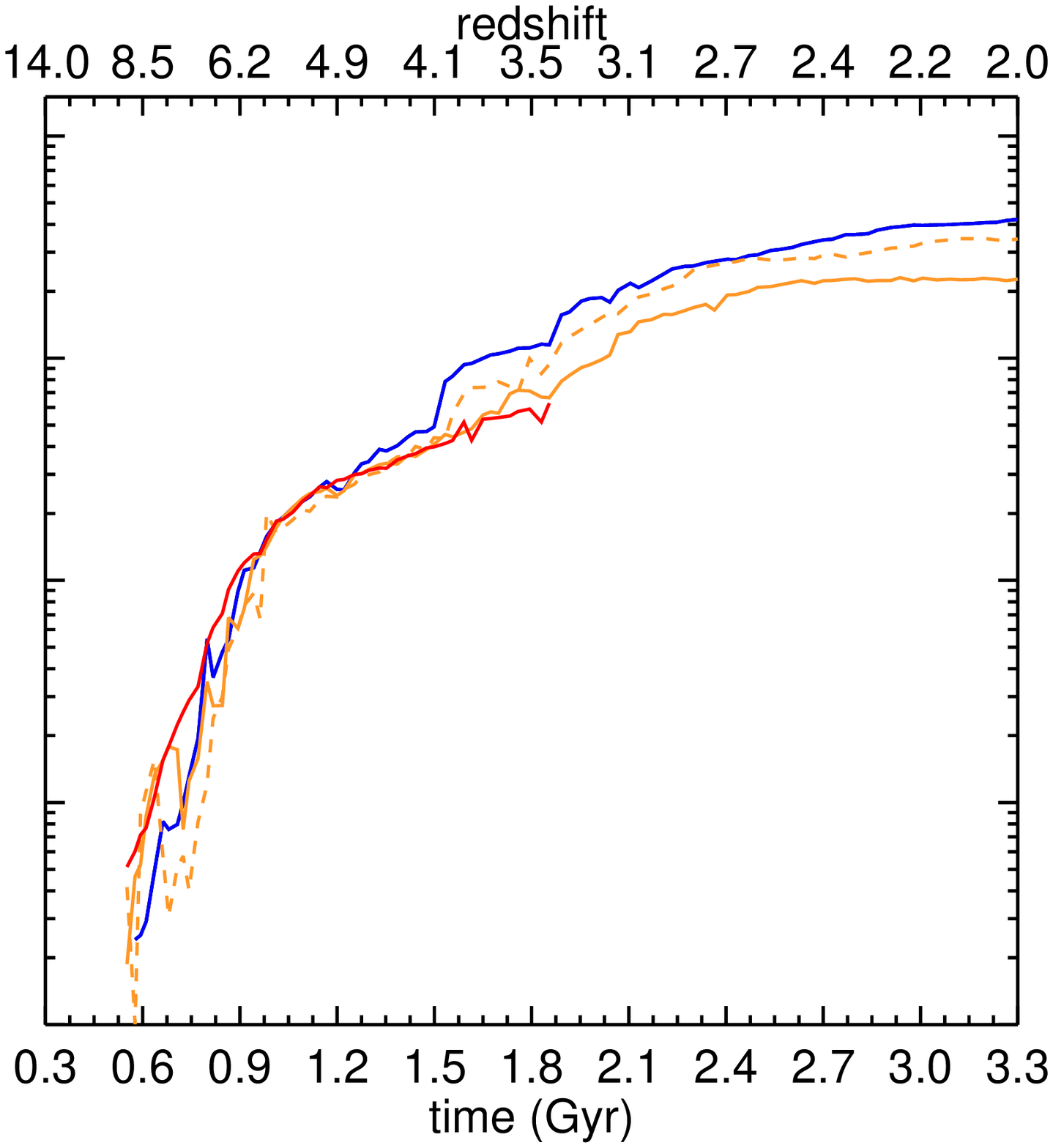}}}\\
  \caption{Galaxy mass (top panels), bulge mass (middle panels), and disc mass (bottom panels) as a function of time (or redshift). In the presence of delayed cooling SN feedback (SLD, SLDA and SHDA) the bulge is suppressed at early times (above $z>3$). The presence of kinetic SN feedback has little impact on galaxy mass, both bulge and disc (SLK compared to SL). However the presence of the AGN greatly reduces the bulge mass content  independently both of the SN implementation and resolution. }
    \label{fig:mgalvsz}
\end{figure}

Fig.~\ref{fig:mgalvsz} shows the galaxy stellar mass as a function of time for the different simulations together with the bulge and disc mass content. 
To compute the bulge and disc mass, we build the one-dimensional stellar surface density profile extracted from the galaxy finder (in order to remove stellar sub-structures using {\sc AdaptaHOP}), and fit two exponentials $\Sigma_{\rm b,d}=\Sigma_{0, b,d}\exp (-r/r_{\rm b,d})$, where $\Sigma_{0, b,d}$ and $r_{\rm b,d}$ are the free parameters of the fit. 
The bulge $M_{\rm b}$ and disc $M_{\rm d}$ masses are simply $M_{\rm b, d}=\Sigma_{\rm b, d} 2\pi r_{\rm b, d}^2$, where the bulge is represented by the profile with the smaller $r_{\rm b}$.
The galaxy mass is smaller in all delayed cooling SN feedback runs (SLD, SLDA, SHDA) at early times compared to the no feedback run (SL), until $z\simeq 3$, but this is essentially due to the suppression of the bulge component.
The disc component in SL, SLDA and SHDA is comparable and somewhat larger than the disc component of the SL run.

The kinetic mode of SN feedback, instead, has very little impact on the stellar mass content, both for the bulge and disc: the mass evolution of the SLK and SL are very close.
We see that the bulge mass in the SLDA and SHDA runs finally settles at $z=3$ once the major merger has happened at $z=3.5$.
Before the settling of the bulge, its mass shows extreme variations with time.
This effect is the consequence of the quick removal of gas due to SN explosions, creating rapid variations in the bulge gravitational potential well and leading to the flattening of the stellar mass distribution~\citep{governatoetal10} with the destruction of the central stellar clump.
Note that the presence of the AGN feedback has a non-negligible effect on the mass content of the bulge.
Once the BH is self-regulating ($z\simeq 6$ for SLKA and SHKA, and $z\simeq 2.5$ for SLDA and SHDA), the bulge mass growth is regulated by the presence of the AGN feedback with more than a factor two difference at $z=2$ compared to the simulations without AGN.

At $z=2$, the halo mass is $M_{\rm h}=10^{12}\, \rm M_\odot$, and the expected stellar mass at this redshift obtained from abundance matching is $M_{\rm gal}\simeq 2\times 10^{10} \, \rm M_\odot$~\citep{mosteretal13}.
In all our simulations, the central galaxy ends up being more massive than expected from semi-empirical predictions by a factor 2 ($M_{\rm gal}\simeq 4\times 10^{10} \, \rm M_\odot$) for SLKA and by a factor $3.5$ ($M_{\rm gal}\simeq 7\times 10^{10} \, \rm M_\odot$) for SLDA.
This tension is possibly due to the missing extra sources of stellar feedback including stellar winds and radiation pressure from young stars~\citep{hopkinsetal11}.
Including more stellar feedback and having a larger star formation efficiency should reduce the stellar mass content of the galaxy and delay the formation of the bulge~\citep{governatoetal10, roskaretal14, agertz&kravtsov14}, and, as consequence, further inhibits the growth of the central BH.
It is worth noting that the SN-quenching of BH growth in SLDA occurs until the halo reaches a mass of $M_{\rm h}\simeq 3\times 10^{11}\, \rm M_\odot$ at $z\simeq3$, which is smaller than the expected peak of the stellar conversion efficiency at $z=3$ ($M_{\rm h, peak}\simeq 3\times 10^{12}\, \rm M_\odot$, \citealp{mosteretal13}).
Therefore, if more realistic stellar feedback can produce lower stellar mass at a corresponding halo mass before that peak, we can assume that the halo mass at which this SN-quenching of BH growth mechanism becomes ineffective, appears for more massive halos, probably around the peak of stellar conversion efficiency.

\begin{figure}
  \centering{\resizebox*{!}{6.0cm}{\includegraphics{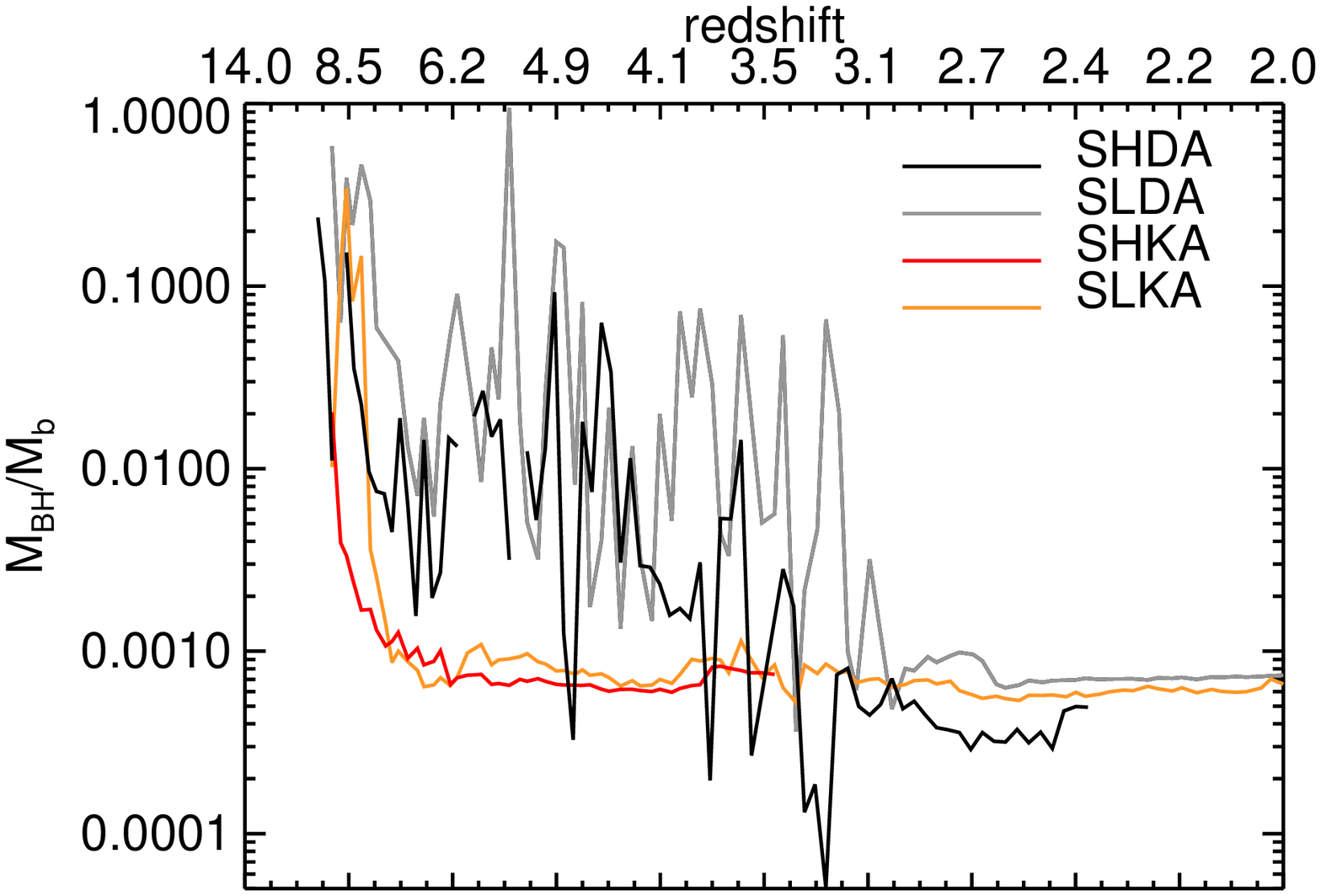}}}\vspace{-1.25cm}
  \centering{\resizebox*{!}{6.0cm}{\includegraphics{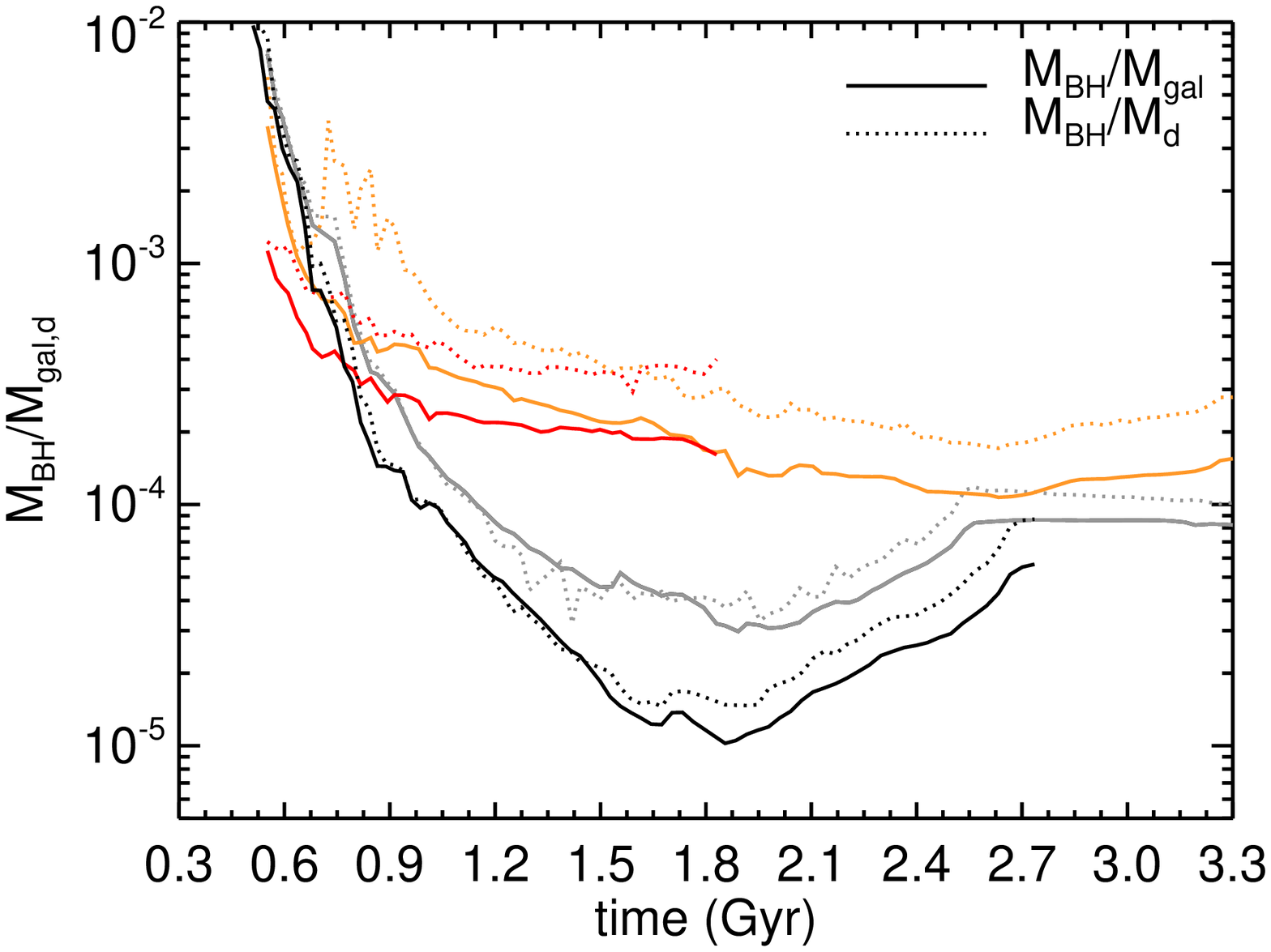}}}
  \caption{BH mass over bulge mass (top panel), BH mass over galaxy mass (solid) or disc mass (dotted, bottom panel) as a function of time (or redshift) for the SHDA (black), SLDA (grey), SHKA (red), and SLKA (orange) simulations. The BH over bulge mass ratio is $\simeq7\times 10^{-4}$ for all simulations when the BH has reached the self-regulation and stays approximately constant. }
    \label{fig:mbhmgal}
\end{figure}

In the case with delayed cooling SN feedback, the BH starts to grow (see Fig.~\ref{fig:mbhevol}) when the bulge definitively settles at $z\sim3.5$, while in the kinetic feedback case bulge and BH start growing earlier on. In Fig.~\ref{fig:mbhmgal} we compare, in the top panel, the ratio between BH and bulge mass as a function of time. In SLKA and SHKA a roughly constant ratio, $\simeq7\times 10^{-4}$, is maintained from early times, $z\sim 6$ all the way to the end of the simulations. This value is only slightly lower than the ratio found at $z=0$, $1-3\times10^{-3}$. SHDA and SLDA present instead a very different evolution: the BH's growth is very slow until $z\sim3.5$, and the bulge has not settled yet. Therefore, before that time, the ratio between BH and bulge mass varies rapidly, despite the BH mass remaining roughly constant. In this case a measure of the BH-to-bulge ratio would indicate a value typically higher than $\sim10^{-3}$. The bulge is able to grow, and to let the BH grow, only after the potential of the bulge has reached a large enough value (cf. section 3.2).  We have also evaluated the ratio of the BH to total galaxy mass (bottom panel of Fig.~\ref{fig:mbhevol}). In fact, except for a few exceptions \citep{2006ApJ...649..616P,2013ApJ...767...13S} high-z measurements of the masses of quasar and AGN hosts cannot perform a robust bulge-to-disc decomposition, and often the total stellar mass is measured (but  compared to the BH-bulge relationship at $z=0$). 
In SLKA and SHKA the ratio gently decreases with cosmic time, highlighting that the total stellar mass grows faster than the BH. 
Therefore, in the SLKA and SHKA cases, the BH bulge-mass relation would be roughly independent of redshift below $z<6$ while the BH-galaxy mass relation would decrease smoothly with redshift.
In SHDA and SLDA, the situation is more complex: initially the galaxy grows faster than the BH ($z>3.5$), then the BH grows faster than the galaxy ($3.5<z<2.5$), and eventually they start tracking each other.  In this case, low mass BHs at high-redshift would be considered {\it overmassive} with respect to their bulge, but {\it undermassive} if the measured quantity is the total galaxy mass, and the comparison is made to today's BH-bulge relationship (i.e., an expectation value of $\sim10^{-3}$).

\begin{figure*}
  \centering{\resizebox*{!}{4.5cm}{\includegraphics{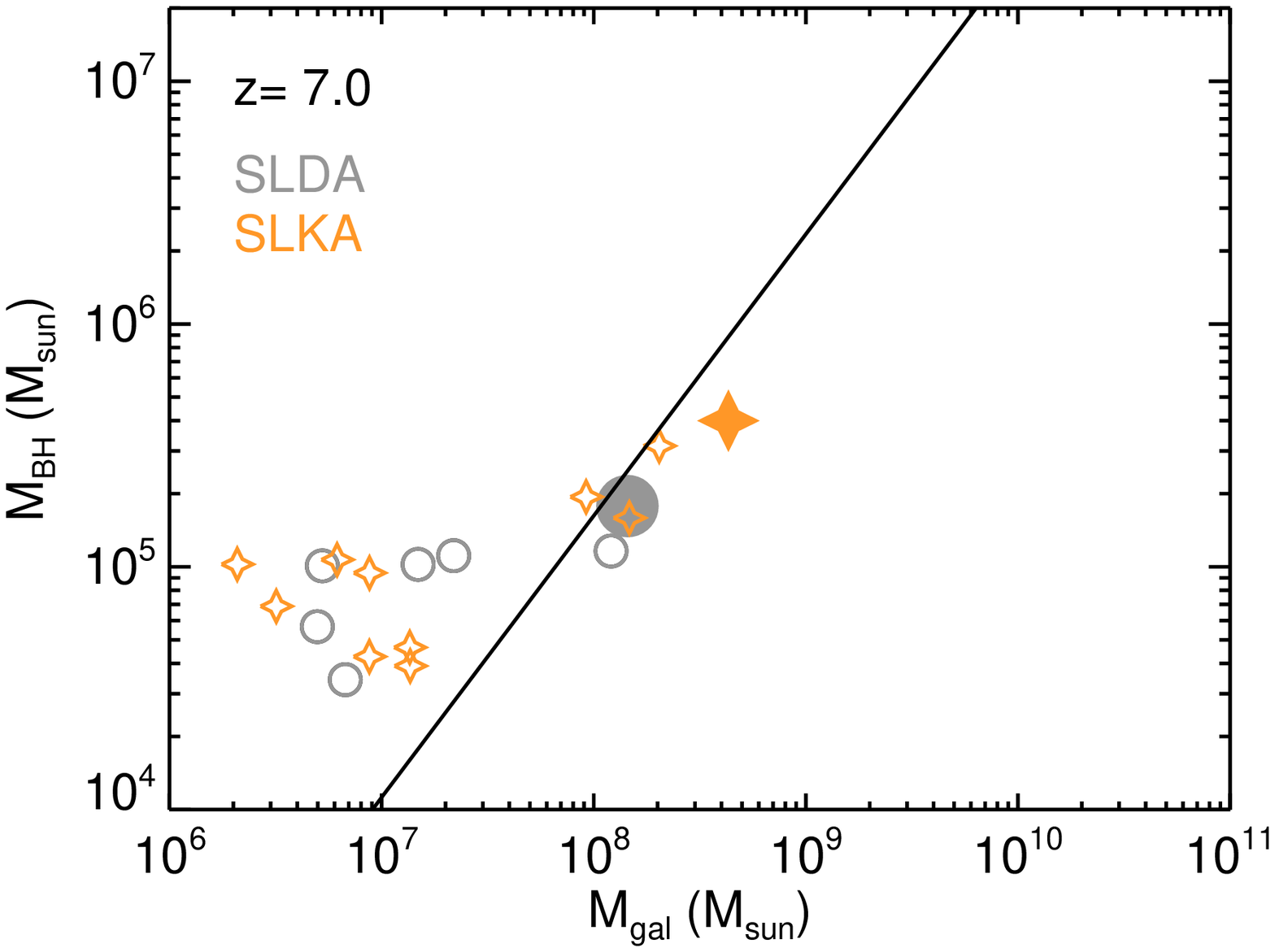}}}\hspace{-0.5cm}
  \centering{\resizebox*{!}{4.5cm}{\includegraphics{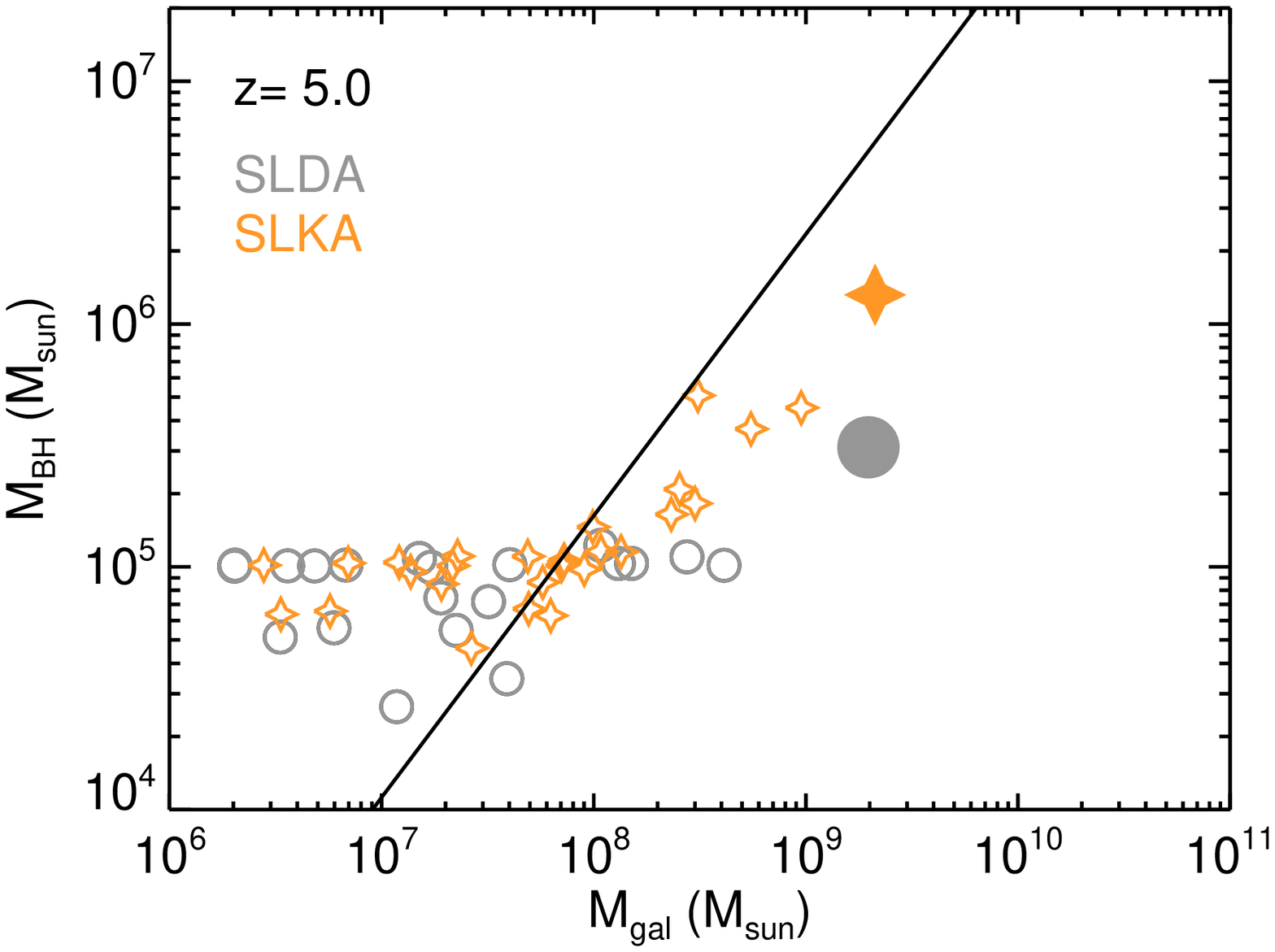}}}\hspace{-0.5cm}
  \centering{\resizebox*{!}{4.5cm}{\includegraphics{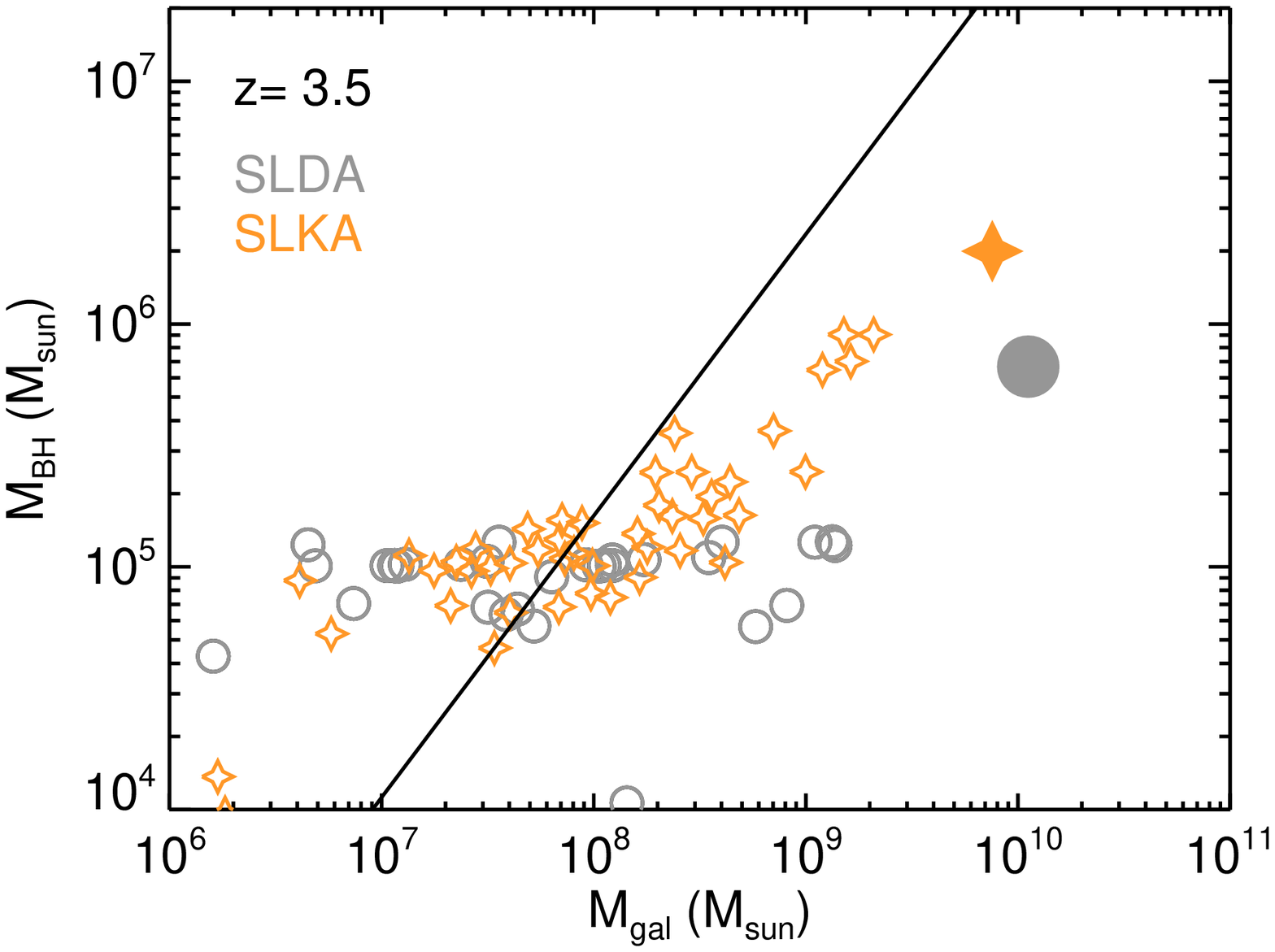}}}\vspace{-0.25cm}
  \centering{\resizebox*{!}{4.5cm}{\includegraphics{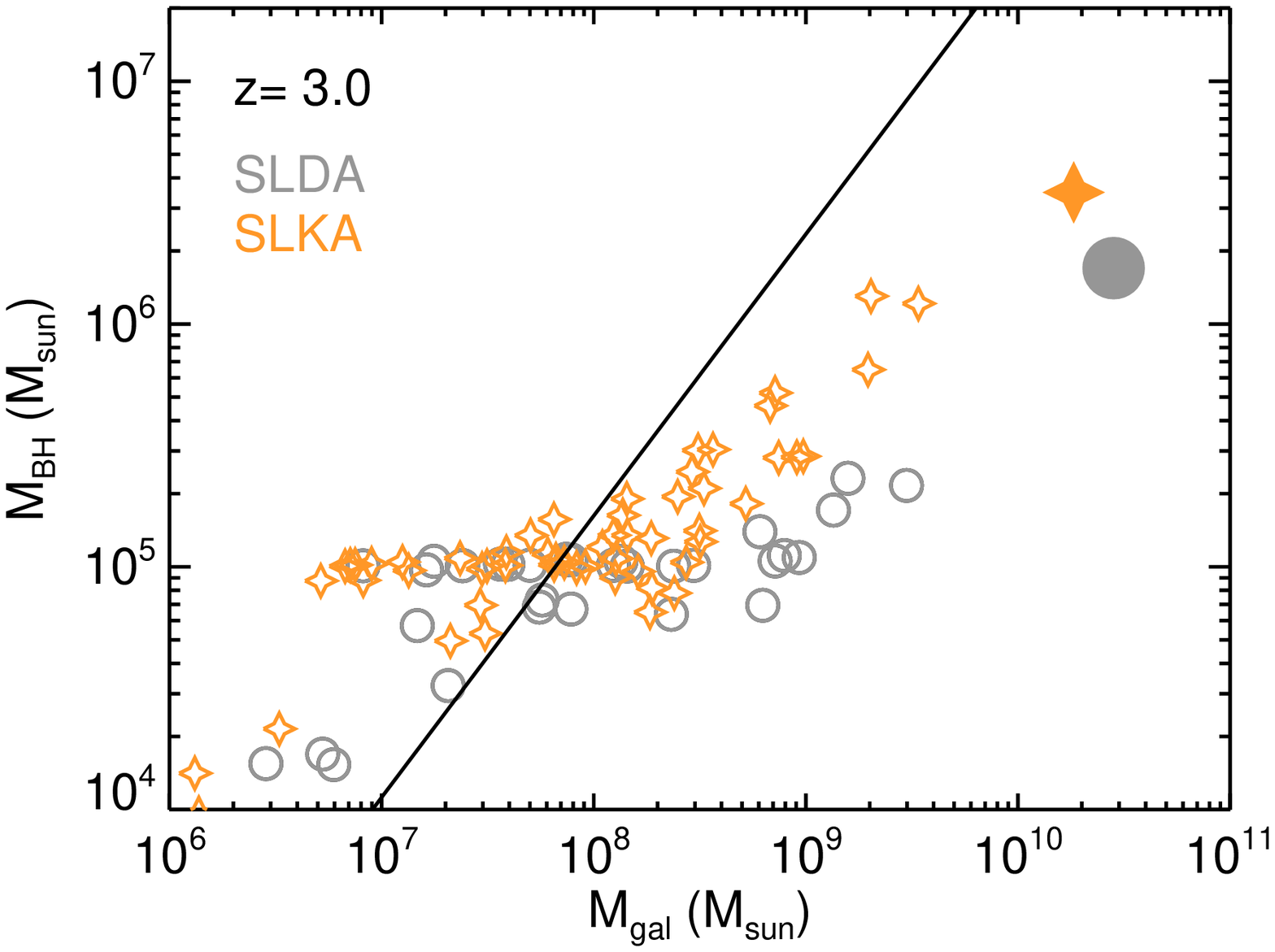}}}\hspace{-0.5cm}
  \centering{\resizebox*{!}{4.5cm}{\includegraphics{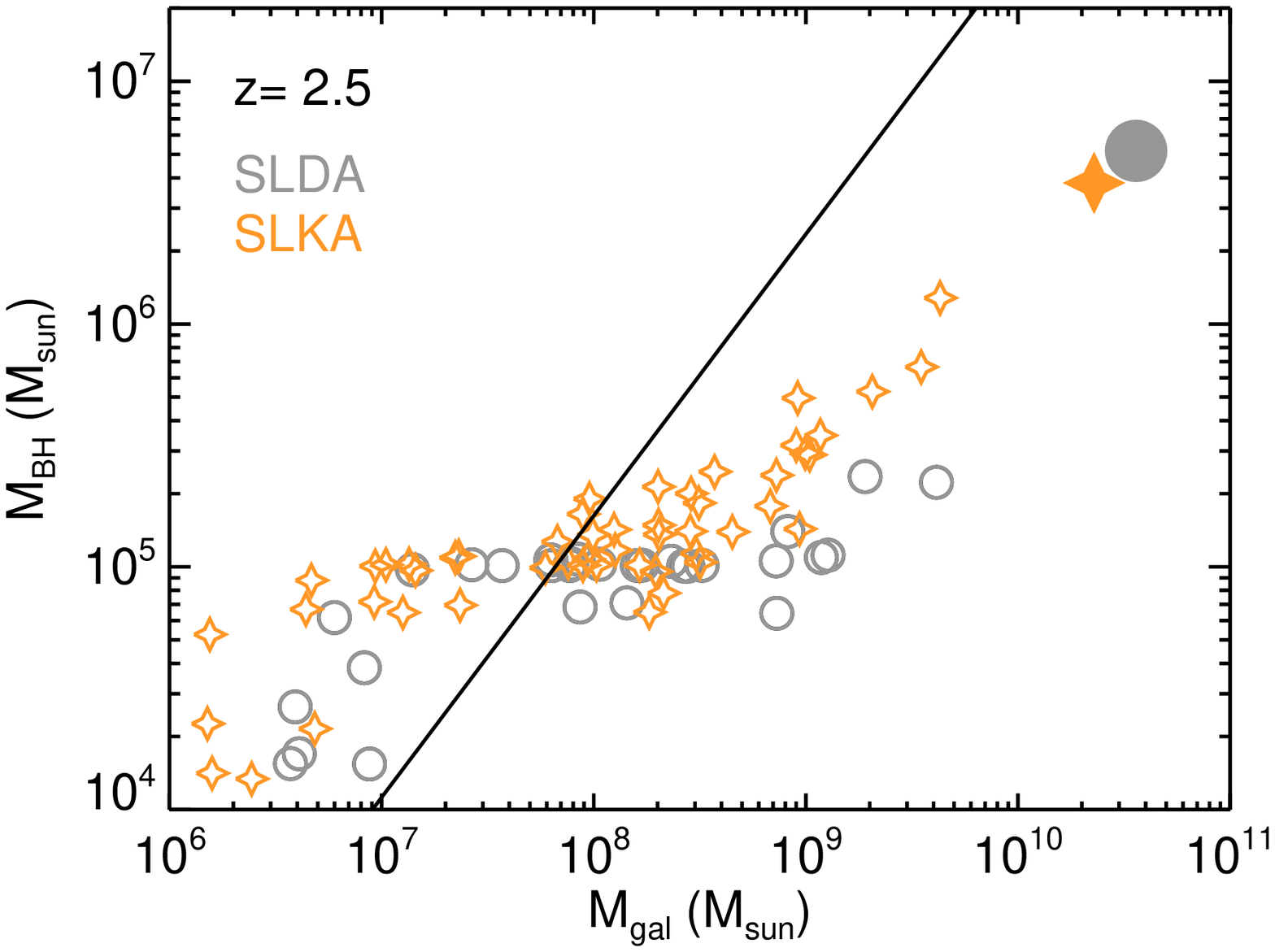}}}\hspace{-0.5cm}
  \centering{\resizebox*{!}{4.5cm}{\includegraphics{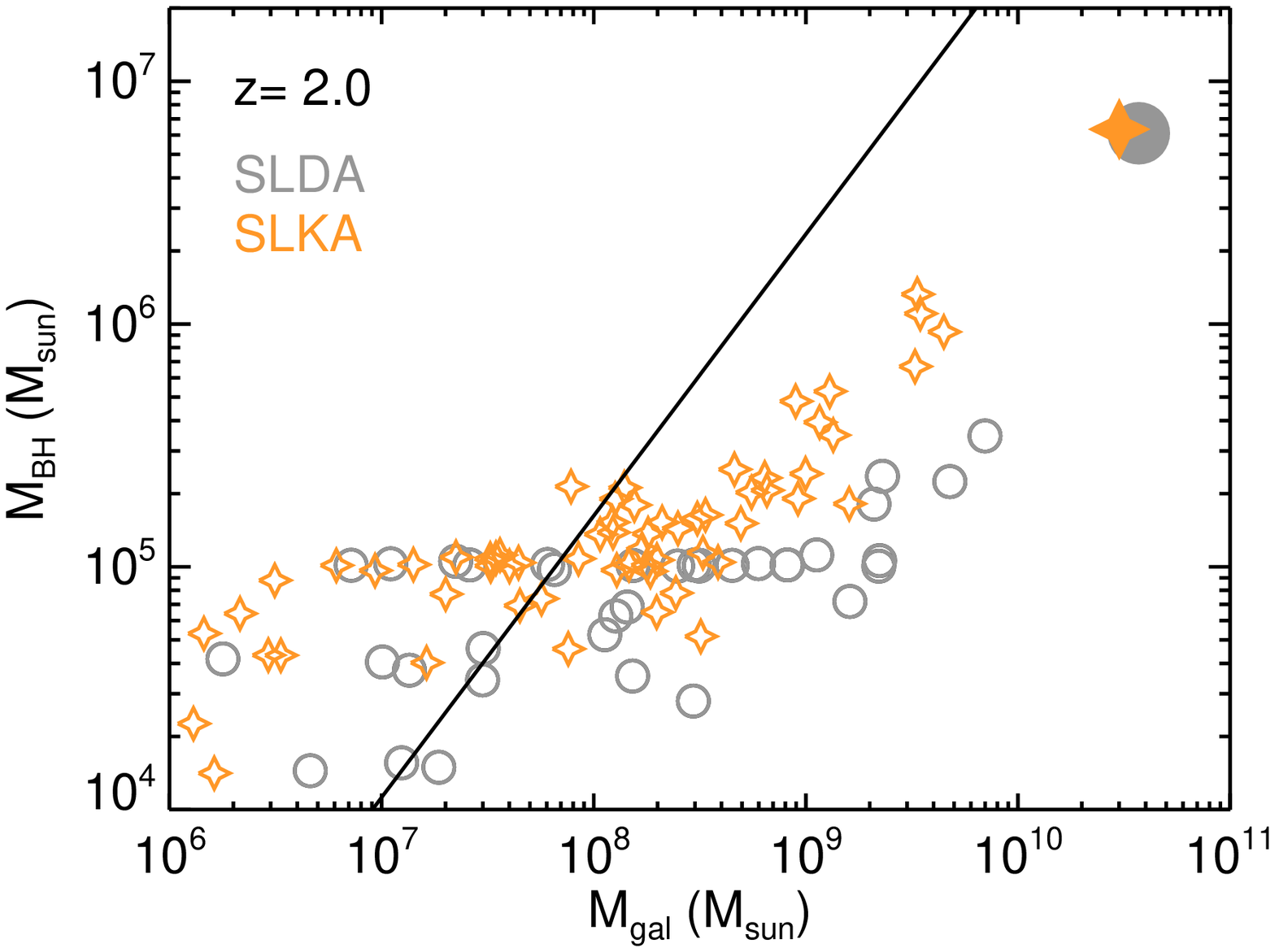}}}
  \caption{BH mass versus stellar mass at different redshifts as indicated in the different panels for SLDA (grey circles) and SLKA (orange stars). The main central galaxy is indicated with the filled symbol. The solid line is for the $z=0$ observed relation from~\citet{kormendy&ho13}. In the SLKA simulation, BHs reach the self-regulated state very early-on, while the presence of delayed-cooling SN feedback in SLDA prevents the BHs in the less gravitationally bound bulges from growing rapidly.}
    \label{fig:mbhvsmstar}
\end{figure*}

\subsection{Extension to the surrounding galaxies}

To check whether the impact of efficient SN explosions applies to low-mass galaxies in a general fashion, we measure the BH mass and galaxy mass relation for the other galaxies within the zoom region. We only consider galaxies that are at a distance smaller than $5\, r_{\rm vir}$ of the main galaxy. Fig.~\ref{fig:mbhvsmstar} shows the values of $M_{\rm BH}$-$M_{\rm gal}$  for these galaxies at various redshifts in the SLDA and SLKA runs, with the observed $M_{\rm BH}$-$M_{\rm gal}$ relation at $z=0$ from~\cite{haring&rix04}.
As already highlighted in the previous sections for the main galaxy, we recognize the slowing down of the growth of the most massive BH (the BH of the main galaxy) at high redshift in the run with delayed cooling SN feedback (SLDA), while the most massive BH grows quickly together with its host galaxy mass in the kinetic SN run (SLKA).
Other smaller mass galaxies behave in a similar fashion: in SLKA the BH mass follows the $M_{\rm BH}$-$M_{\rm gal}$ relation (though with a larger normalization factor) and BHs starts growing in galaxies of $M_{\rm gal}\simeq 10^8 \, \rm M_\odot$, in SLDA the BH growth is delayed in low-mass galaxies, i.e. BHs start growing in more massive galaxies of $M_{\rm gal}\simeq 2\times 10^9 \, \rm M_\odot$, the normalization of $M_{\rm BH}$-$M_{\rm gal}$ relation is further increased.
Note that there is a plateau in our $M_{\rm BH}$-$M_{\rm gal}$ relation at low galaxy masses which is due to our choice of initial seed BH mass $10^5\, \rm M_\odot$ (there are BHs with smaller values because we cannot take more mass within a gas cell that is actually available), therefore, those BHs are already overly massive for their galaxy mass.

\begin{figure}
  \centering{\resizebox*{!}{1.5cm}{\includegraphics{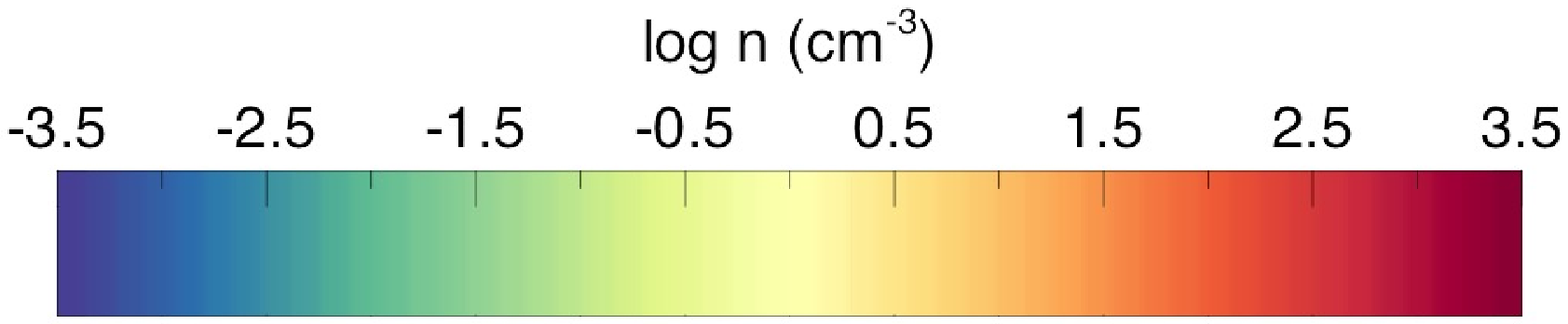}}}
  \centering{\resizebox*{!}{4.0cm}{\includegraphics{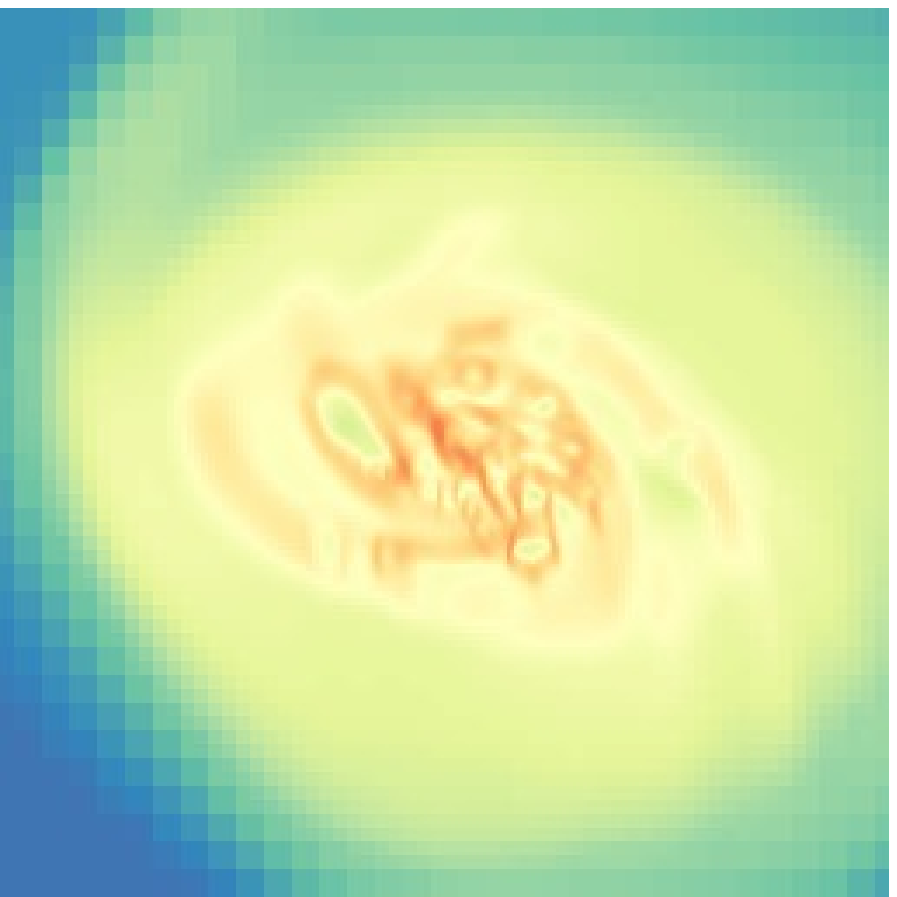}}}
  \centering{\resizebox*{!}{4.0cm}{\includegraphics{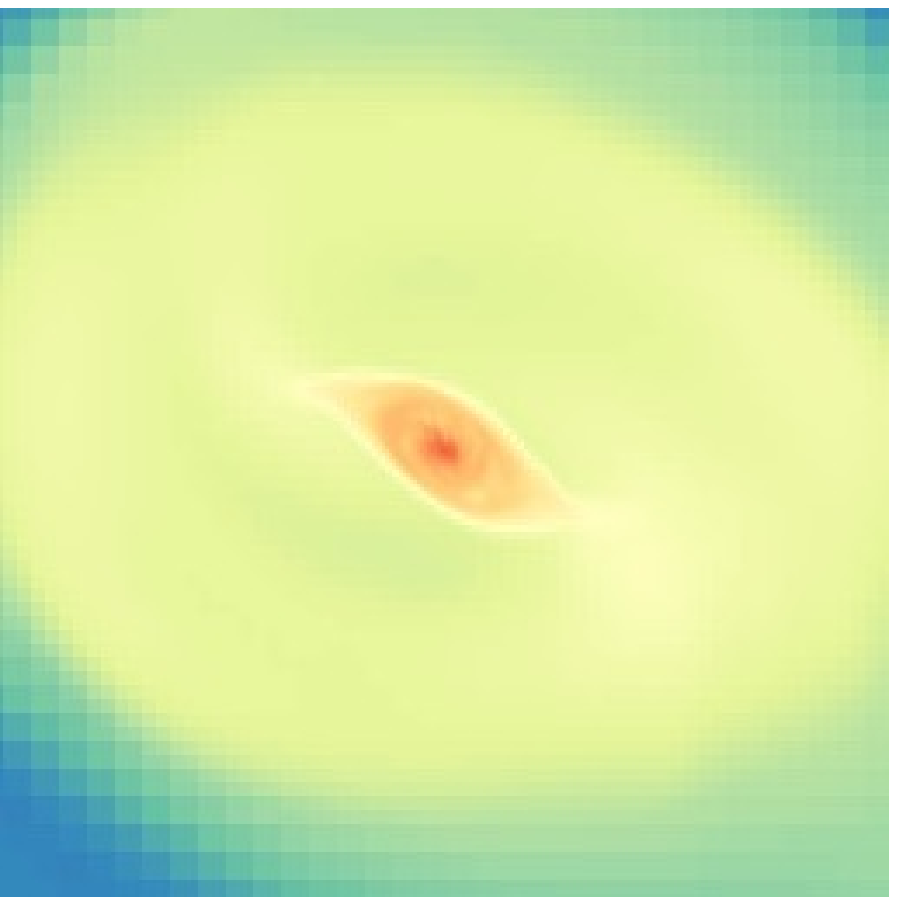}}}
  \centering{\resizebox*{!}{4.0cm}{\includegraphics{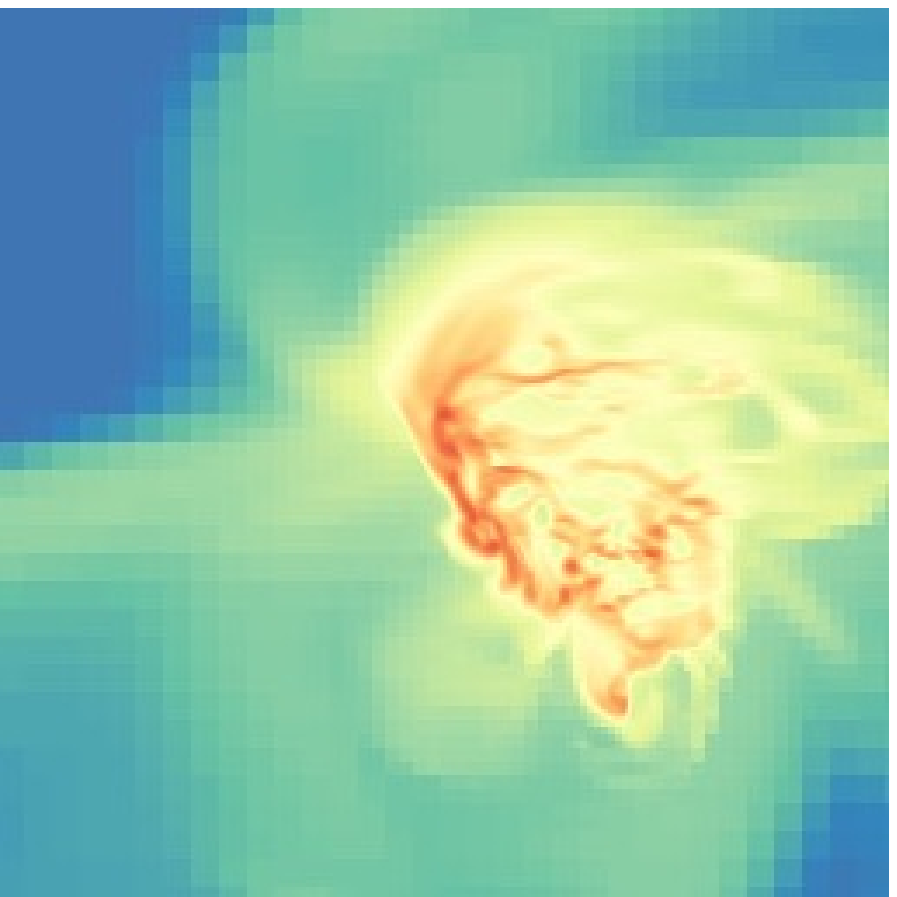}}}
  \centering{\resizebox*{!}{4.0cm}{\includegraphics{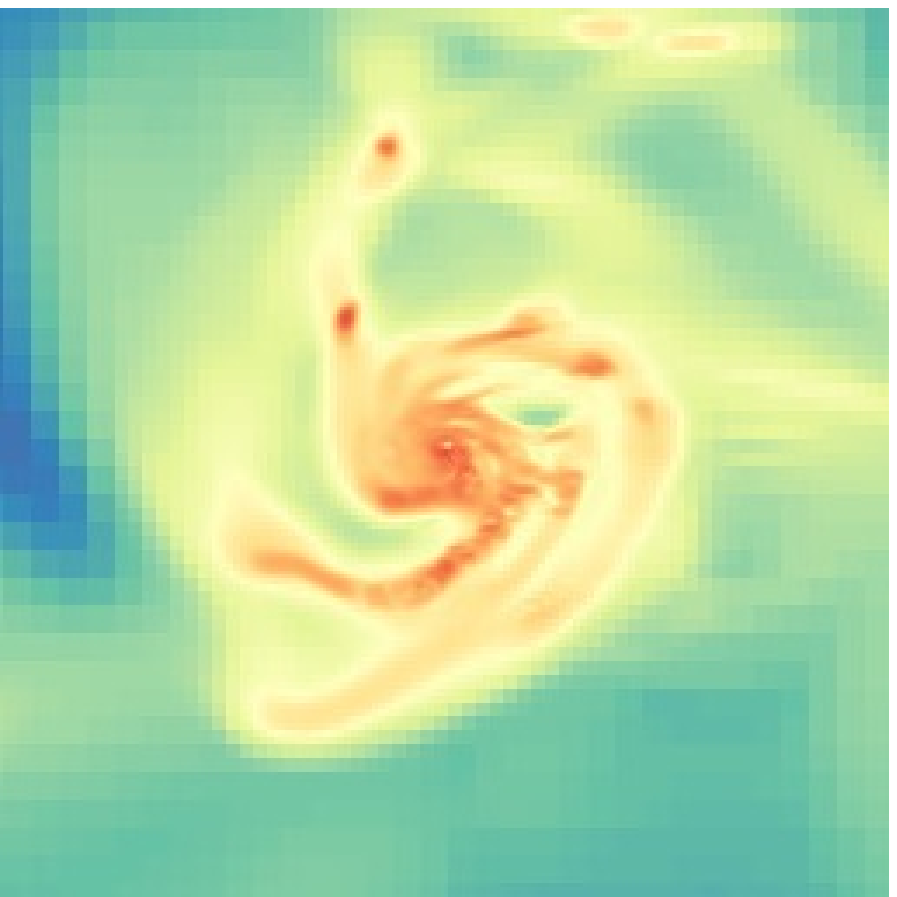}}}
  \centering{\resizebox*{!}{4.0cm}{\includegraphics{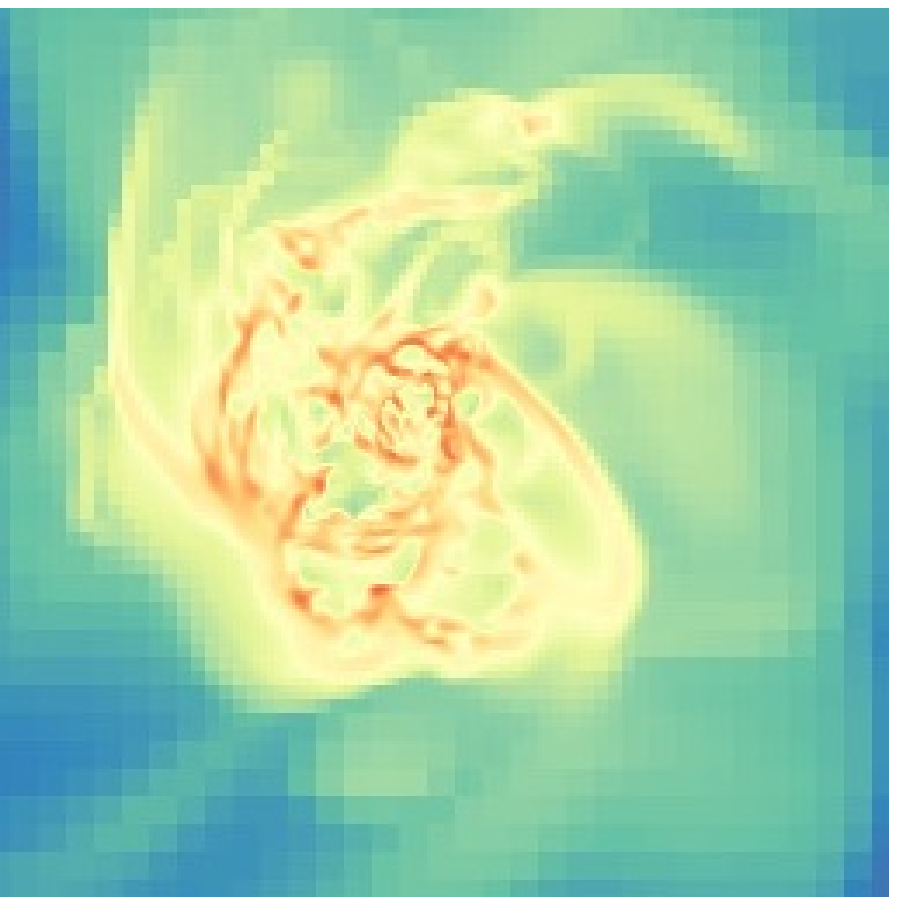}}}
  \centering{\resizebox*{!}{4.0cm}{\includegraphics{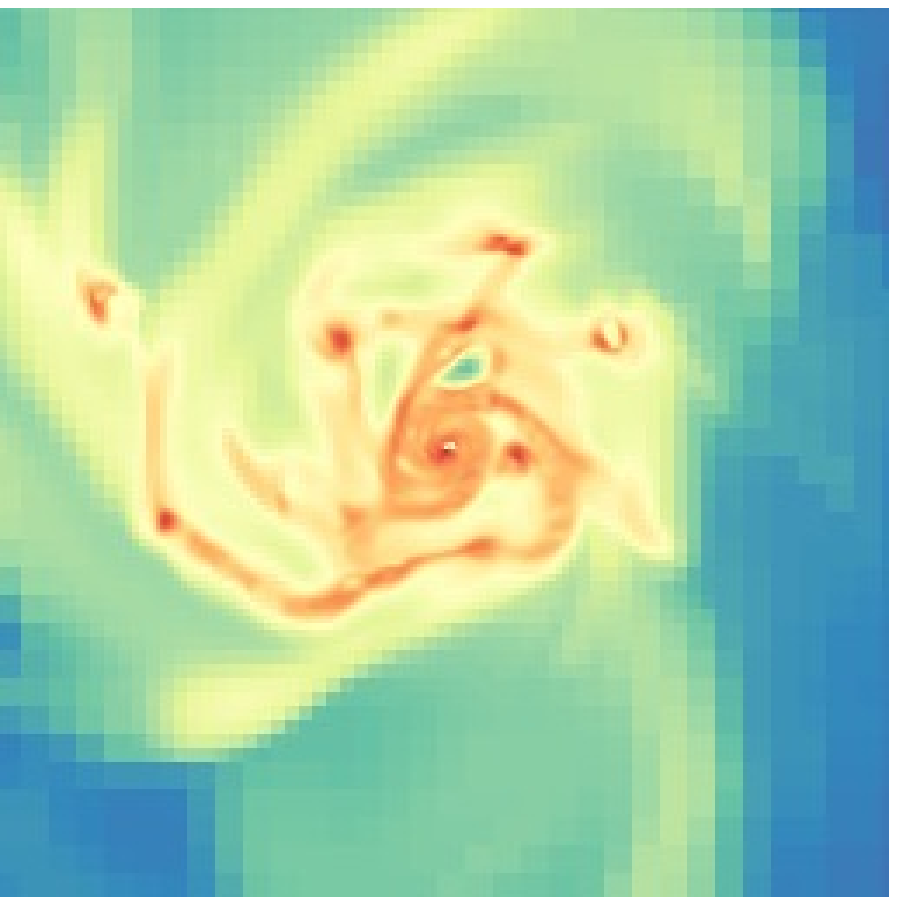}}}
  \caption{Projected gas densities for three other galaxies than the central one at $z=2$ for SLDA (left column) and SLKA (right row). The images are 10 kpc across. These galaxies are not sub-structures of the main halo, and each of this image in SLDA correspond to its equivalent in SLKA (i.e. same halo host). Galaxies simulated with delayed cooling SN feedback appear more disturbed and turbulent than galaxies simulated with kinetic SN feedback, explaining the lack of early BH growth that is a generic effect in low-mass galaxies.}
    \label{fig:mmgal}
\end{figure}

In Fig.~\ref{fig:mmgal}, we compare the projected gas densities of a few galaxies in SLDA and SLKA at $z=2$.
Galaxies in SLDA exhibit a porous and turbulent ISM that is driven by SN explosions. In comparison, galaxies in SLKA systematically show a more uniform gas distribution with spiral arms encompassing star-forming gas clumps but the ISM is not as porous as in the SLDA run.
As discussed earlier for the main galaxy, this extra turbulence (or porosity) driven by efficient SN explosions, is responsible for the inhibition of BH growth in the low-mass galaxies.
Therefore, it seems plausible that this inhibition mechanism should hold on average for all galaxies with galaxy masses below a few $10^9 \, \rm M_\odot$.

\section{Conclusions}
\label{section:conclusion}

Our set of cosmological hydrodynamical simulations (the Seth suit of simulations) including either SN feedback with a kinetic model from~\cite{dubois&teyssier08winds} or a delayed cooling prescription from~\cite{teyssieretal13} shows that the strength of SN feedback can dramatically alter the BH evolution at early epochs of galaxy formation, or conversely in low-mass galaxies.
In the former  (weak SN feedback)  case, a cold gas reservoir is maintained in the center of the galaxy sustained by inflows of dense star-forming clumps of gas, which provides an efficient gas fuelling of the central BH, driving an Eddington-limited growth early on.
In the later  (strong SN feedback) case, the energy release from SNe is maximised, which allows for the destruction of the dense gas clumps within the core of the galaxy.
In return, the early growth of the BH is prevented due to the lack of a sustained cold gas reservoir, and this effect is systematically observed for all galaxies within the simulated region.
We have shown that the characteristic SN wind velocity is of the order of $270\, \rm km \, s^{-1}$, a value larger than the escape velocity created by the gravitational potential of a galactic bulge of $M_{\rm b}\lesssim10^9\, \rm M_\odot$ (with a fixed radius of $100 \, \rm pc$).
This efficient mode of SN feedback prevents the rapid growth of the bulge although the disc component continues growing almost unimpeded.
Once enough mass has accumulated in the center of the galaxy, in our case driven by a major merger, the gravitational barrier prevents the removal of the cold gas and the BH can grow at a near-Eddington rate until it reaches self-regulation.

For the efficient SN feedback scenario, the bulge mass $\sim$ a few $10^9\, \rm M_\odot$ above which the accretion at the Eddington rate turns on is a factor $\sim 10$ smaller than the bulge mass $\sim 10^{10}\, \rm M_\odot$ below which abnormally light BHs are observed and exhibit a larger scatter in the BH-bulge correlation~\citep{greeneetal10, jiangetal11, graham&scott15}.
In our simulations, only feedback from type II SNe has been included.
An extra source of energy or momentum input such has radiation pressure from young stars could enhance the wind velocities, and, thus, increase the bulge mass above which the gas ejected by stellar feedback is trapped in the bulge.
Therefore, the fact that BHs undergo less growth in low-mass galaxies could be the signature of a strong stellar feedback activity, which is able to efficiently destroy the cold clumps of star-forming gas, establishing an indirect evidence of short-lived clusters of gas below a given mass.

We have tested those two scenarios with respect to spatial resolution effects at $\Delta x= 34.8\, \rm pc$ and $\Delta x=8.7$, and our results are qualitatively independent on the adopted resolution.
The Eddington growth phases appear at the same epochs: as soon as the galaxy forms for the case of kinetic SN feedback runs, and once the major merger happens in the delayed-cooling SN feedback model.
Even the quantitative aspects show relatively good agreement: the bulge and disc mass are in remarkable good agreement, and the BH mass at self-regulation end up being the same within a factor 2.

In the Seth simulation, the 1:3 major gas-rich merger happening at $z=3.5$ is the culprit for the triggering of the BH growth in the runs including the delayed cooling prescription for SN feedback.
It does not imply that all galaxies would necessarily start growing their BH every time they endure a major merger, because they probably need to reach a mass sufficiently large enough ($M_{\rm b}>10^9\, \rm M_\odot$, therefore $M_{\rm gal}>10^9\, \rm M_\odot$).
Once the galaxy is massive enough, a persistent bulge needs to be built, and the major merger is the leading mechanism for the Seth simulation because of the induced torques and consequent nuclear inflows.
Though major mergers are the favourite candidates for creating the strongest torques, one can imagine that a massive flow of cold streams misaligned with the galactic spin would have a similar impact on the galaxy.
Only extra simulations with different accretion histories, and in particular more quiescent objects, could help us to fully address that question and verify whether major mergers are systematically responsible for the ignition of BH growth in the case of efficient SN feedback.

The detailed interaction between star formation, SN feedback, BH growth and AGN feedback is probably the key to understanding the establishment of the scaling relation between BHs and their host galaxies, their redshift evolution, the overall co-evolution of BHs and galaxies, and as a consequence the properties of the galaxy population.  At high redshift, estimates of BH and galaxy stellar masses  of AGN samples typically find that the BH mass is larger than expected from this correlation at a given stellar mass\footnote{High-redshift stellar masses are not equivalent to bulge masses, nor are measured in a self-consistent way throughout different samples and at different redshift.}, except for submillimeter-selected galaxy \citep{2008AJ....135.1968A}. \cite{2010AJ....140..546W}, however, based on the comparison between the BH and galaxy mass functions at $z\sim 6$, suggest that most high-z galaxies host BHs that are smaller than expected from this correlation at a given stellar mass (or host no BH at all).  Similar conclusions are reached with X--ray searches of $z>6$ galaxies: current data are sensitive to $10^5-10^6$M$_\odot$ BHs accreting at $1-10$\% Eddington \citep{2011ApJ...742L...8W,2012A&A...537A..16F,2012ApJ...748...50C,2013ApJ...778..130T,2015arXiv150202562G,2015arXiv150106580W}, but very few, if any are found, implying strong limits on the early growth of BHs in relatively small galaxies. The observational evidence, therefore, seems to suggest that high-redshift BHs in the most massive galaxies are actively accreting and are on or above the correlation with their hosts, while at lower galaxy mass the behaviour is opposite \citep{2011MNRAS.417.2085V}. This new SN-regulated BH growth mechanism is a promising tool to explain why BHs in relatively low-mass galaxies at high-redshift are unable to grow.

Though our method for modelling SN feedback is still very crude, it highlights two extreme regimes.
Future improvements should model SNe with more physically motivated assumptions using the correct amount of momentum imparted by the explosions~\citep{agertzetal13, kimmetal15} and study the interplay between circumnuclear star formation and black hole growth.
We should also consider the steady input of energy from stellar winds and radiation from bright OB associations~\citep[e.g.][]{hopkinsetal11} with dedicated simulations to probe their effect on the BH growth.
Note that, even though there will be extra energy delivered by these new stellar feedback processes, it is unclear how they will interact with the gas content and star formation.
As a matter of fact, stellar winds and UV radiation from young stars initiate earlier than type II SNe, and could result in an earlier impact on the gas content of the star-forming cloud but could also limit the total amount of stars formed in one star-formation cycle, thus, limiting the impact of feedback~\citep{rosdahletal15}.
However, a scenario with efficient SN (or stellar) feedback mechanism is probably favoured over a less extreme mode in order to explain the low stellar mass content~\citep[e.g.][]{hopkinsetal14} and the absence of bulge in dwarf galaxies~\citep{governatoetal10}.
We dedicate these investigations to future work.

We have limited our simulations to the high-redshift ($z>2$) environment of the progenitor of a group of galaxies and have shown that all low-mass galaxies ($M_{\rm gal}\lesssim 10^9\, \rm M_\odot$) within this region exhibit un-grown BHs.
Though, we have not explicitly demonstrated that the same mechanism is at play in the low-redshift dwarf galaxies, we argue that they should behave similarly, particularly given that galaxies at low redshift are less compact than their high-redshift equivalent, and, thus, it is easier for SN explosions to drive outflows from them.
This epoch of galaxy evolution remains to be investigated and our proposed mechanism to be validated by new numerical experiments of low-redshift dwarf galaxies.

\section*{Acknowledgments}
We thank the anonymous referee for useful comments, which improved the clarity of the paper.
The simulations presented here were run on the DiRAC facility jointly funded by STFC, the Large Facilities Capital Fund of BIS and the University of Oxford. 
This research is part of the Horizon-UK project. 
YD and JS acknowledge support by the ERC  by  ERC project 267117 (DARK) hosted by Universit\'e Pierre et Marie Curie - Paris 6, and JS for support at JHU by National Science Foundation grant OIA-1124403 and by the Templeton Foundation. 
This work is partially supported by the Spin(e) grants ANR-13- BS05-0005 of the French Agence Nationale de la Recherche.
MV acknowledges funding support from NASA, through award ATP NNX10AC84G, and from a Marie Curie Career Integration grant (PCIG10-GA-2011-303609). 
YD thanks Jonathan Patterson for his technical support during the course of this work.

\bibliographystyle{mn2e}
\bibliography{author}

\appendix

\section{Setting up parameters for the delayed cooling mode of SN feedback}
\label{app:delcool}

The time variation of the non-thermal energy component in star-forming regions follows
\begin{equation}
{D e_{\rm NT}\over D t}= \eta_{\rm SN} \dot \rho_{\rm s} {10^{51} \rm \, erg\over 10 \rm \, M_\odot }- {e_{\rm NT}\over t_{\rm diss}}
\label{enernt}
\end{equation}
with the first term of the right-hand side being the injection of energy by SN explosions, and the second term is the decay rate due to the dissipation of the energy by sub-grid processes (decay of turbulence, cooling of cosmic rays, etc.). 
At equilibrium the two terms exactly compensate and lead to 
\begin{equation}
 e_{\rm NT} = \eta_{\rm SN} \dot \rho_{\rm s} {10^{51} \rm \, erg\over 10 \rm \, M_\odot } t_{\rm diss}\, ,
\end{equation}
or
\begin{equation}
 {e_{\rm NT}\over \rho} = \eta_{\rm SN} {10^{51} \rm \, erg\over 10 \rm \, M_\odot } {t_{\rm diss}\over t_{\rm ff}(\rho)}\epsilon_{*}
\end{equation}
for a Schmidt law.
It comes that the corresponding non-thermal velocity dispersion is ($e_{\rm NT}=0.5\rho \sigma_{\rm NT}^2$):
\begin{equation}
 \sigma_{\rm NT}^2 = 2 \eta_{\rm SN} {10^{51} \rm \, erg\over 10 \rm \, M_\odot } {t_{\rm diss}\over t_{\rm ff}(\rho)}\epsilon_{*}\, .
 \label{eq:snt}
\end{equation}
The injected energy from SNe is transferred from unresolved small scales, with high gas density where it is deposited, up to resolved larger scales, with low gas density.
Explicit resolved turbulence will dissipate over its own time-scale which is provided by the dynamics of shocks in the ISM.
Note that, here, we derive a model for sub-grid unresolved turbulence, and the time-scale to dissipate this energy component (non-thermal energy) is turbulent crossing-time over a few resolution elements (typically one Jeans length) $l_{\rm J}=N_{\rm cell} \Delta x$ (with $N_{\rm cell}\ge4$), so that $t_{\rm diss}\simeq l_{\rm J}/\sigma_{\rm NT}$.
Therefore, eq.~(\ref{eq:snt}) is replaced by 
\begin{equation}
\sigma_{\rm NT} = \left ( 2 \eta_{\rm SN} {10^{51} \rm \, erg\over 10 \rm \, M_\odot } {l_{\rm J}\over t_{\rm ff}(\rho)}\epsilon_{*}\right )^{1/3}\, .
\end{equation}
We can simplify the expression further by imposing that the gas free-fall time is that at the threshold of star formation. Thus  
\begin{equation}
t_*(\rho)= { t_{\rm ff}(\rho) \over \epsilon_{*}} = {1\over \epsilon_*}\sqrt{3\pi \over {32 G \rho_0}} \, .
\end{equation}
Finally, we can write $\sigma_{\rm NT}$ as a simple function of all known parameters for star formation and SN feedback efficiency, and resolution:
\begin{eqnarray}
\label{sigmant}
\sigma_{\rm NT}&\simeq& 48 \left( \eta_{\rm SN} \over 0.1\right)^{1/3} \times \left( \epsilon_* \over 0.01\right)^{1/3} \times \left( N_{\rm cell} \Delta x \over 4\times 10 \, {\rm pc} \right)^{1/3}\nonumber\\
	&\times & \left( n_0 \over 200 \, {\rm cm^{-3}}\right)^{1/6} \rm \, km\, s^{-1} \, ,
\end{eqnarray}
where $n_0$ is the number density threshold for star formation ($n_0=n_{\rm H,0}/X_{\rm H}$, $X_{\rm H}=0.76$).
The dissipative time-scale is also fully determined by the choice of $\eta_{\rm SN}$, $\epsilon_*$, $\Delta x$ and $n_0$:
\begin{eqnarray}
t_{\rm diss}&\simeq& 0.82 \left( \eta_{\rm SN} \over 0.1\right)^{-1/3} \times \left( \epsilon_* \over 0.01\right)^{-1/3} \times \left( N_{\rm cell} \Delta x \over 4\times 10 \, {\rm pc} \right)^{2/3}\nonumber\\
	&\times & \left( n_0 \over 200 \, {\rm cm^{-3}}\right)^{-1/6} \rm \, Myr \, .
\end{eqnarray}
For this set of simulations we have taken $N_{\rm cell}=4$ corresponding to the typical Jeans length  at high gas densities (pressure-supported by the polytropic equation of state, see section~\ref{section:subgridgal}).
Note that $\Sigma_{\rm NT}$ becomes resolution independent for our given choice of polytropic equation of state.
Since $4\times \Delta x=l_J\propto \sqrt{T_0/n_0}$, the density dependancy in equation~(\ref{sigmant}) vanishes.
This simple model ensures that SN energy deposit can support the propagation of the blast wave up to $N_{\rm cell}\Delta x$ without any radiative loss since cooling is blocked up to that distance.
Beyond that distance, we consider that the blast wave is resolved and radiative losses can proceed, which is ensured by the calibration of the dissipation time-scale $t_{\rm diss}$ to that of the sound-crossing time up to $N_{\rm cell}\Delta x$ for an energy deposit of $0.5\rho \sigma_{\rm NT}^2$.

\section{Galaxy morphology at lower resolution}
\label{app:galmorphlowres}

\begin{figure*}
  \centering{\resizebox*{!}{4.2cm}{\includegraphics{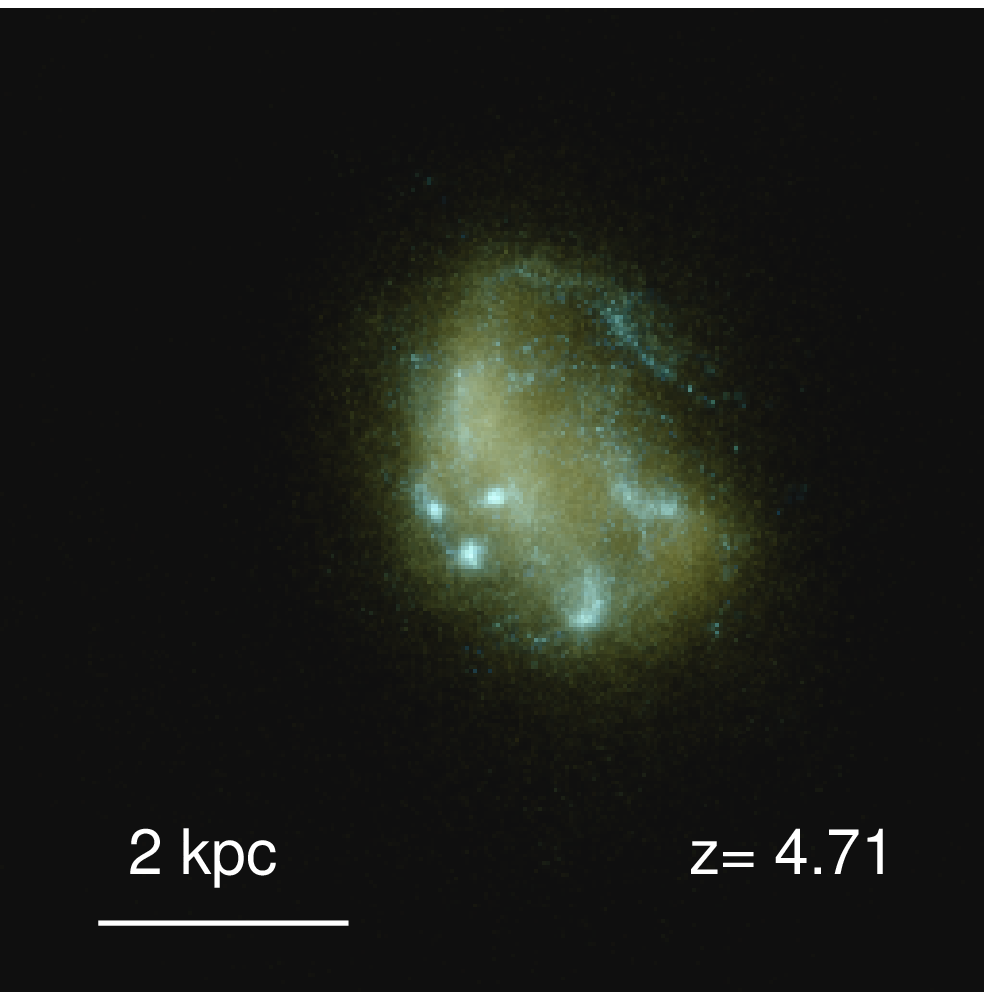}}}
  \centering{\resizebox*{!}{4.2cm}{\includegraphics{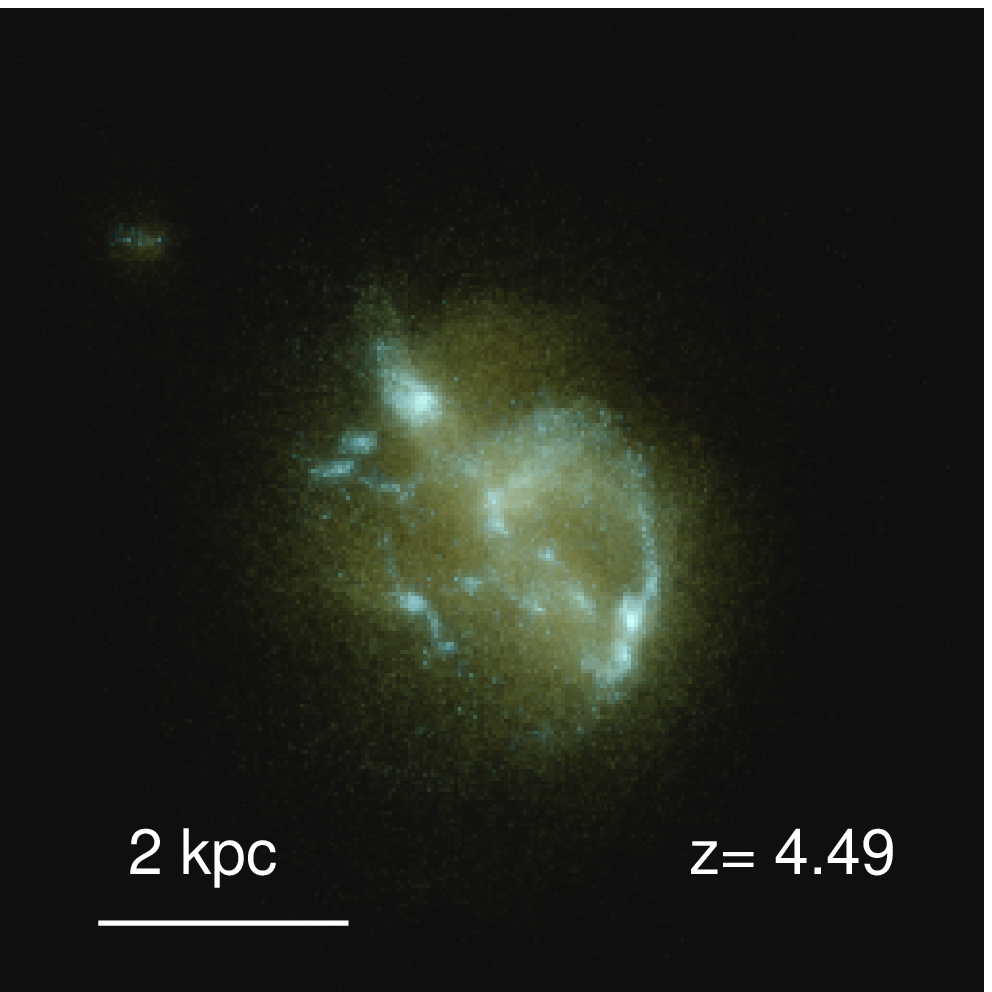}}}
  \centering{\resizebox*{!}{4.2cm}{\includegraphics{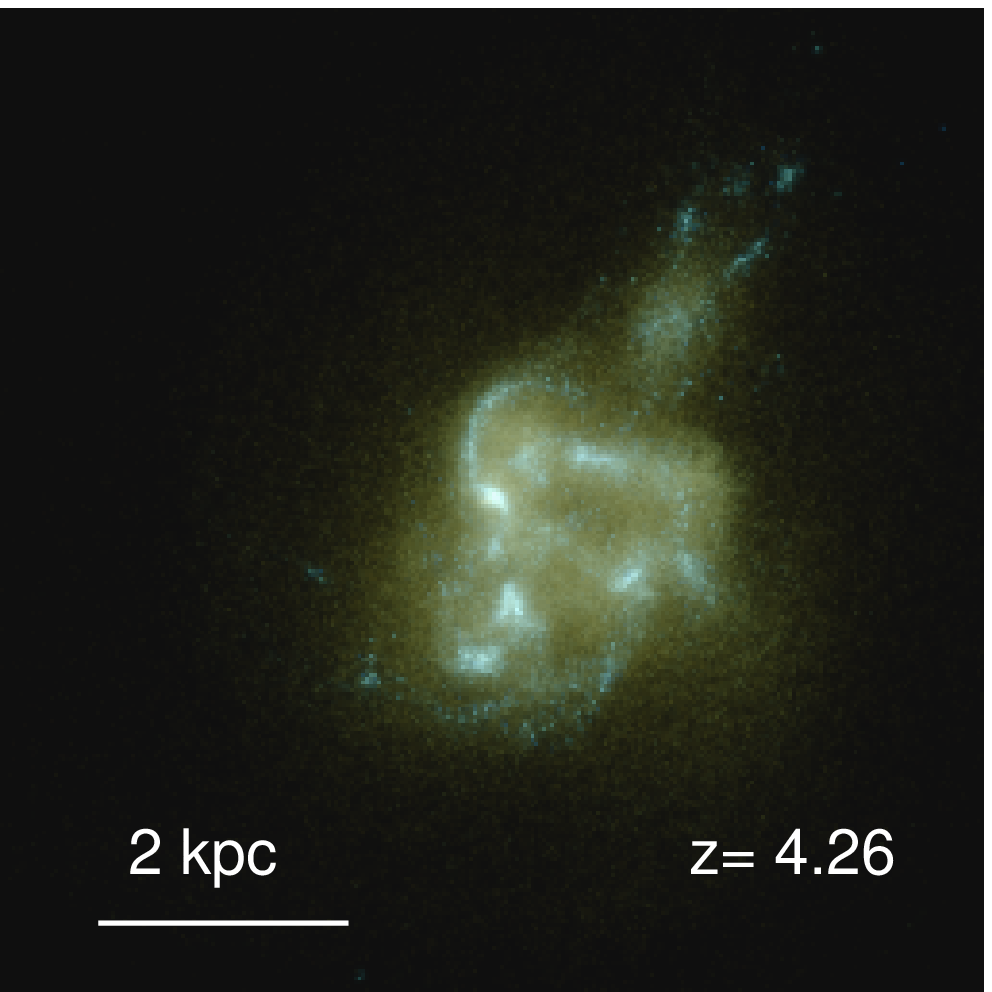}}}
  \centering{\resizebox*{!}{4.2cm}{\includegraphics{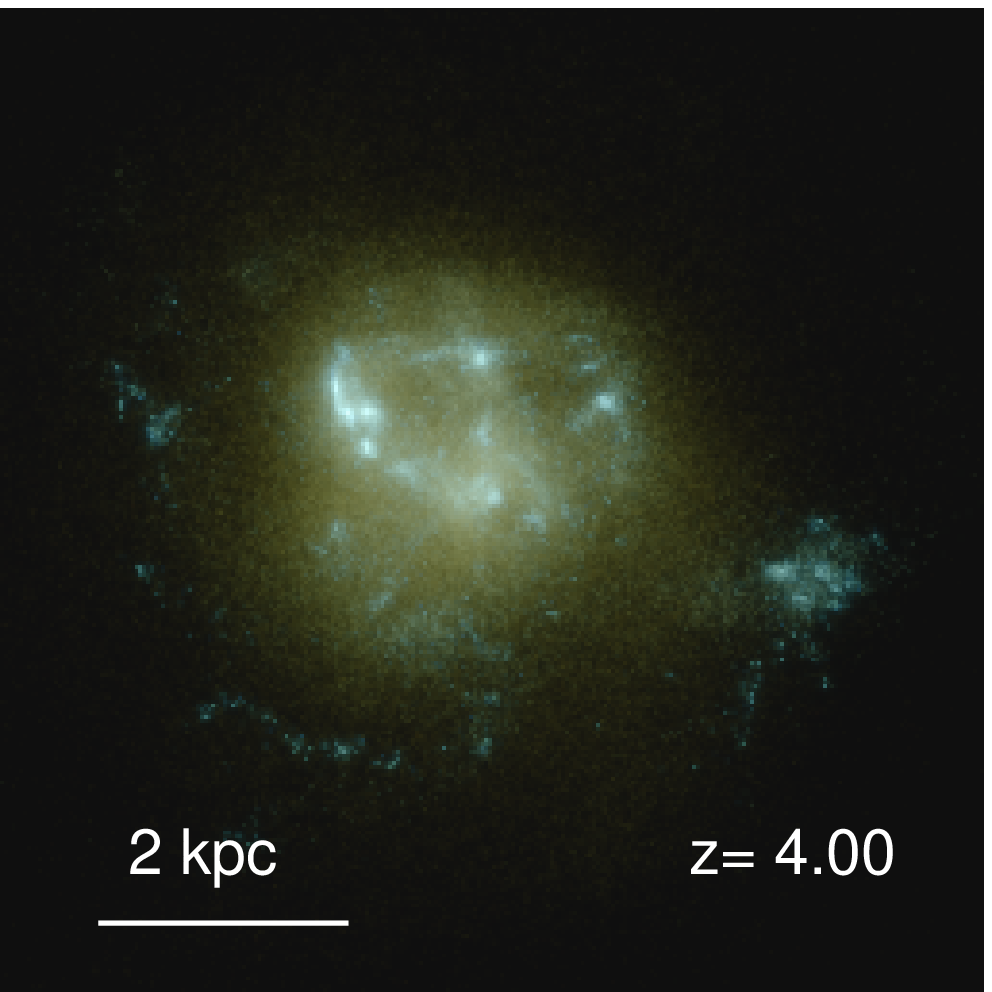}}}
  \centering{\resizebox*{!}{4.2cm}{\includegraphics{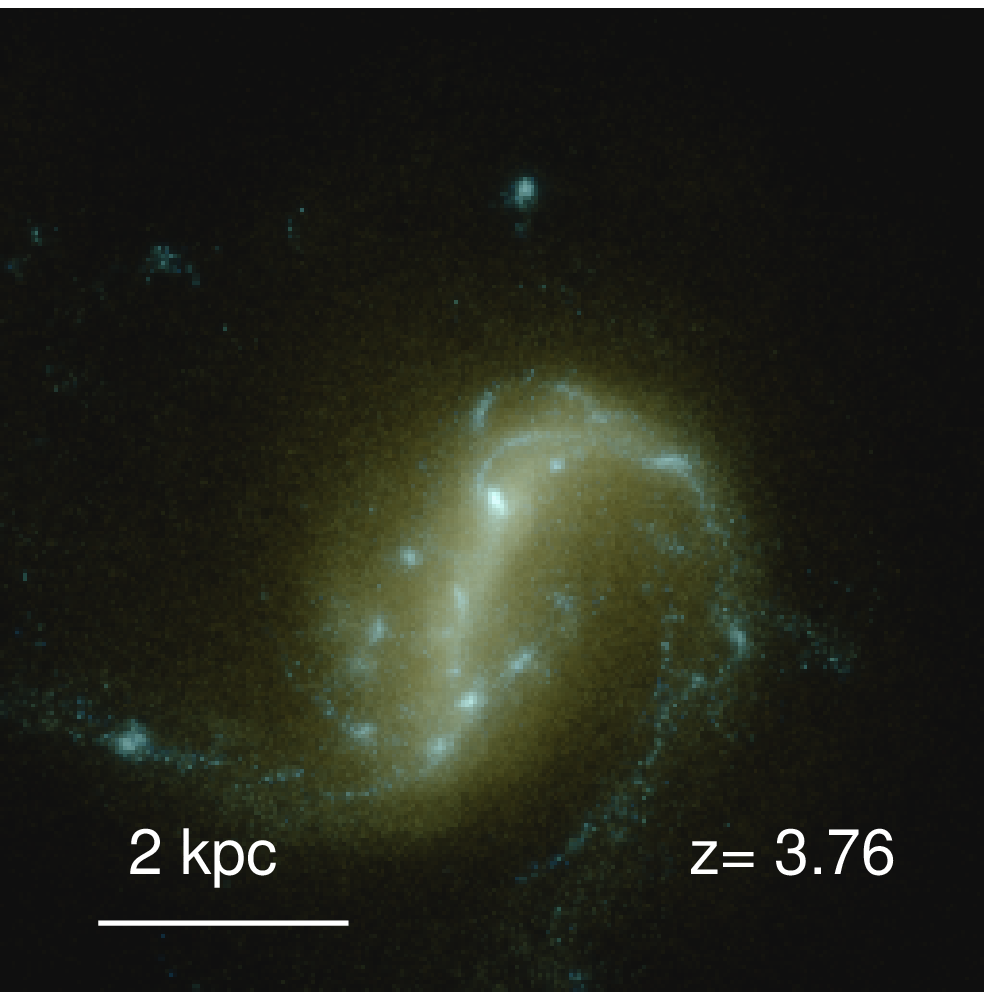}}}
  \centering{\resizebox*{!}{4.2cm}{\includegraphics{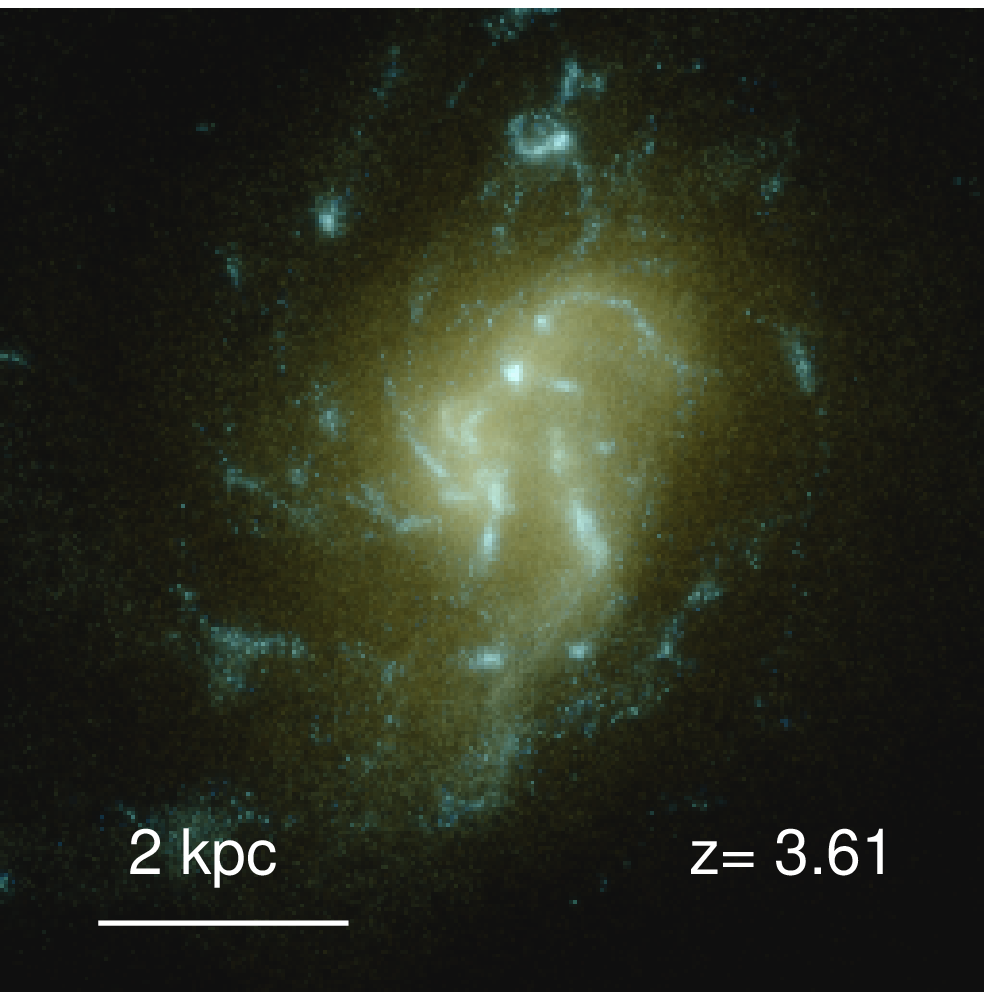}}}
  \centering{\resizebox*{!}{4.2cm}{\includegraphics{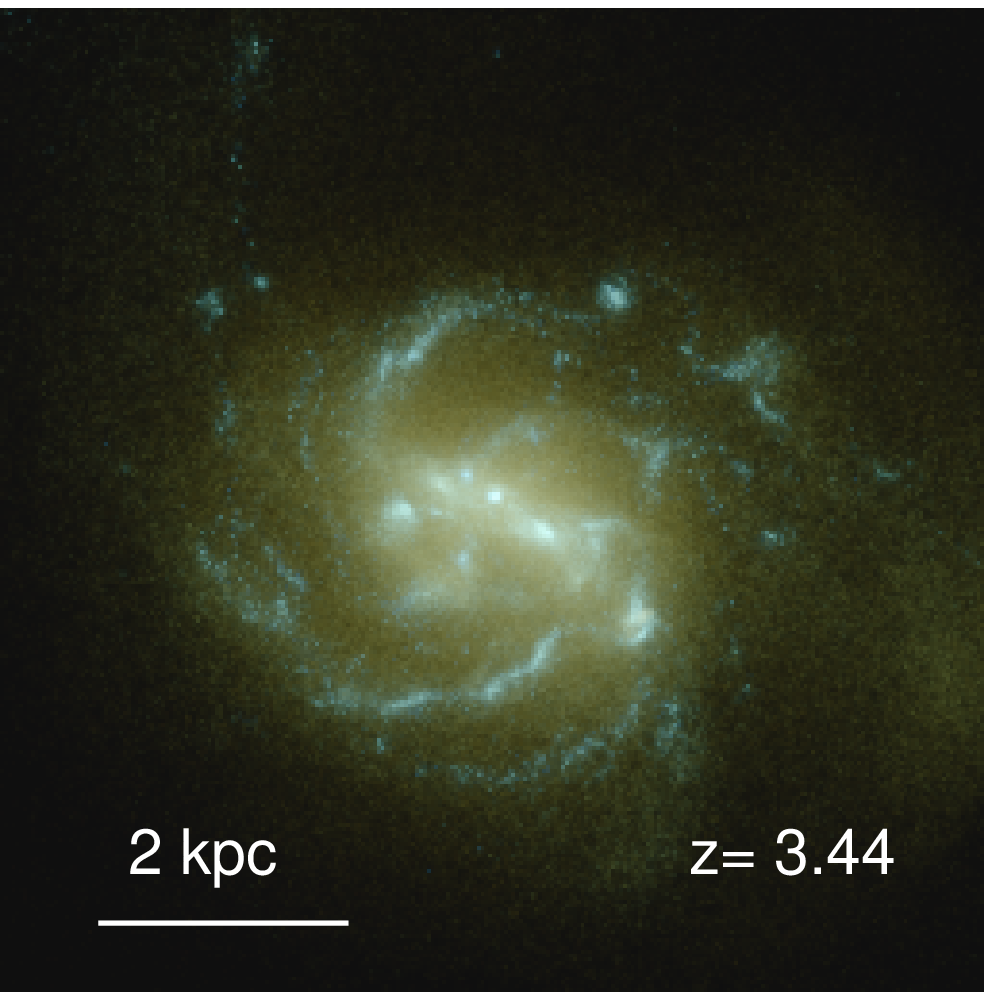}}}
  \centering{\resizebox*{!}{4.2cm}{\includegraphics{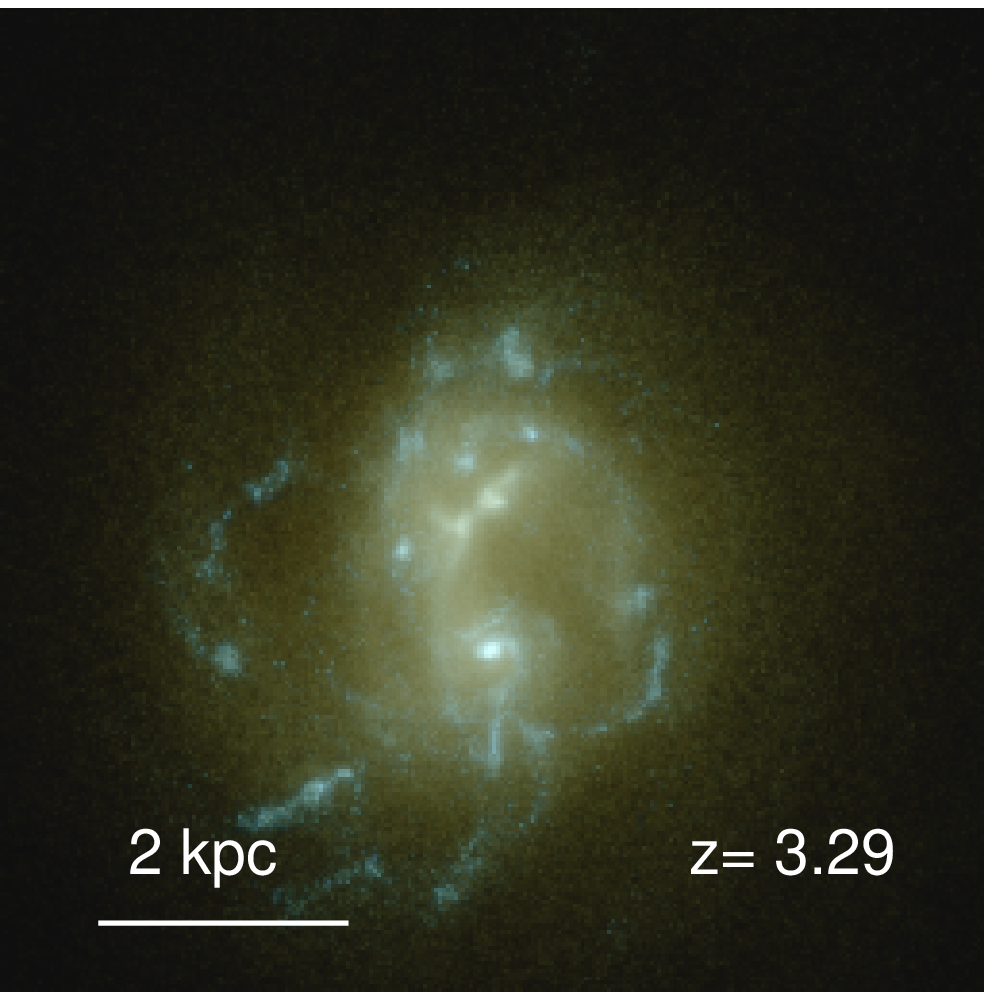}}}
  \centering{\resizebox*{!}{4.2cm}{\includegraphics{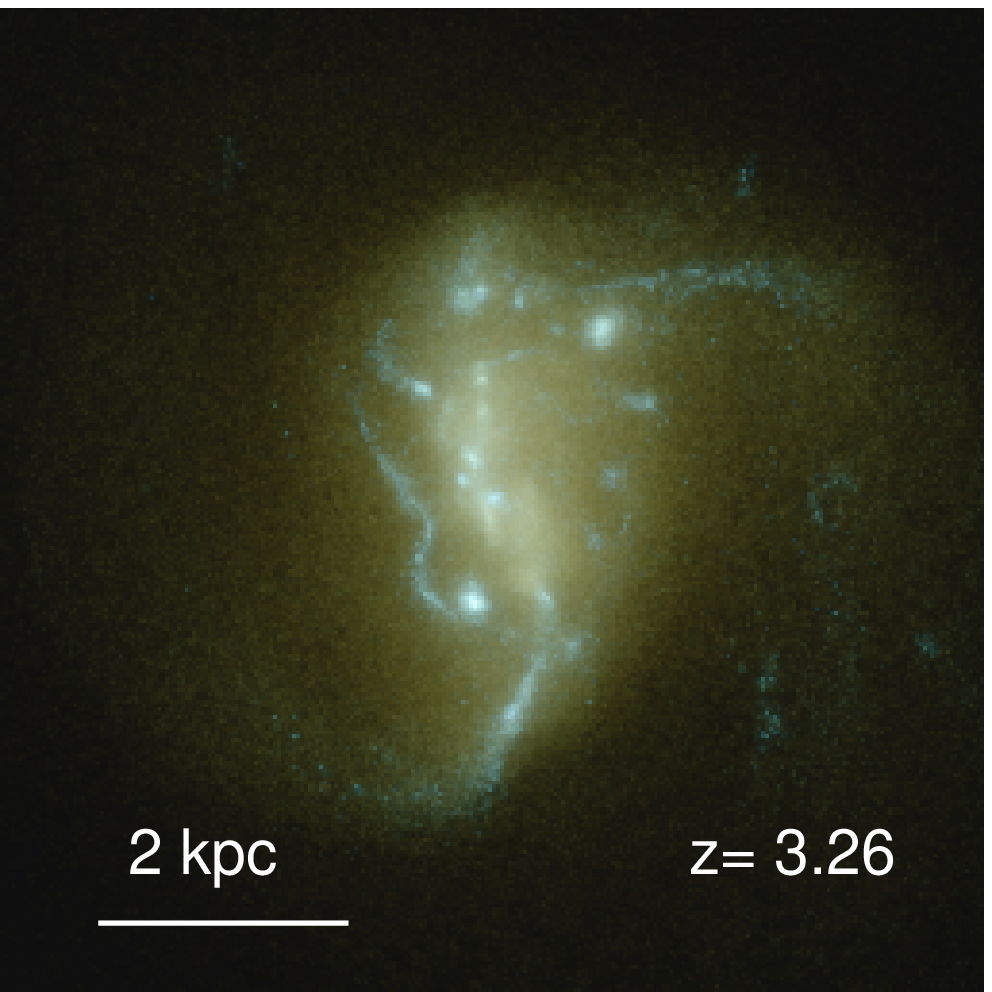}}}
  \centering{\resizebox*{!}{4.2cm}{\includegraphics{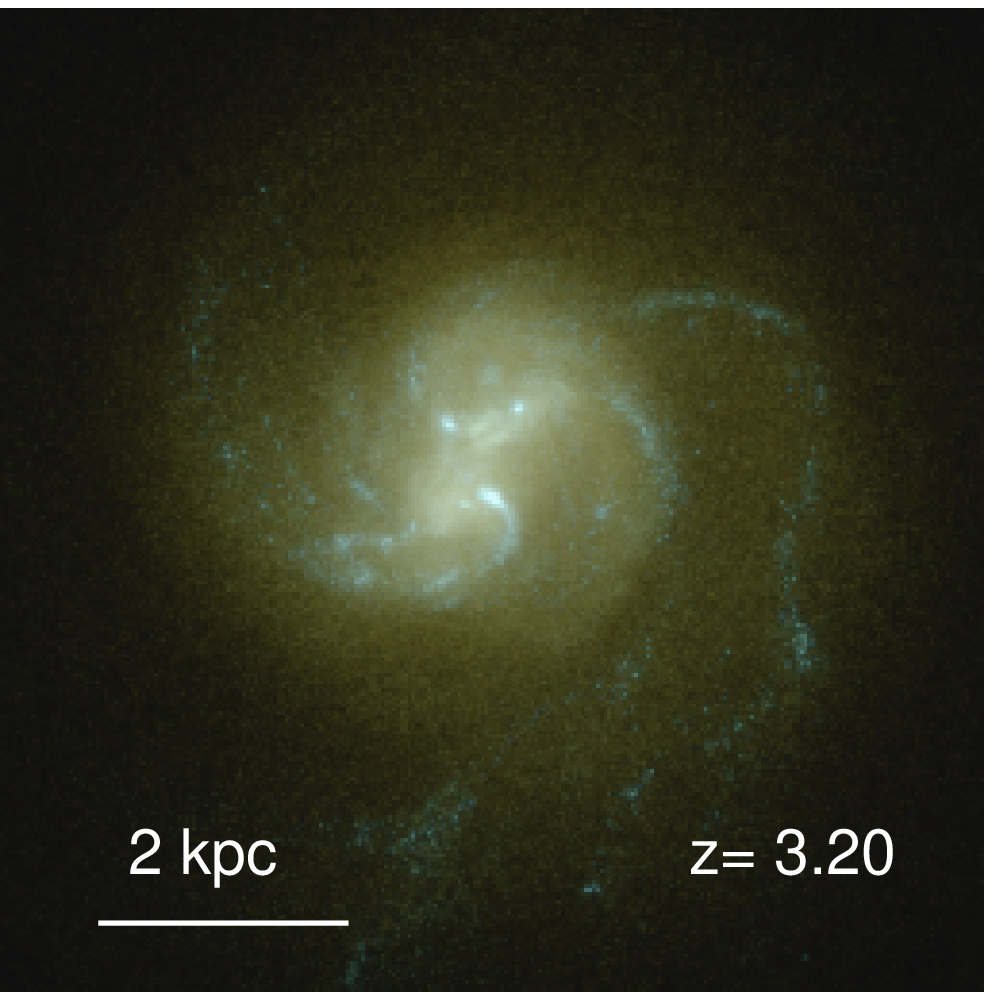}}}
  \centering{\resizebox*{!}{4.2cm}{\includegraphics{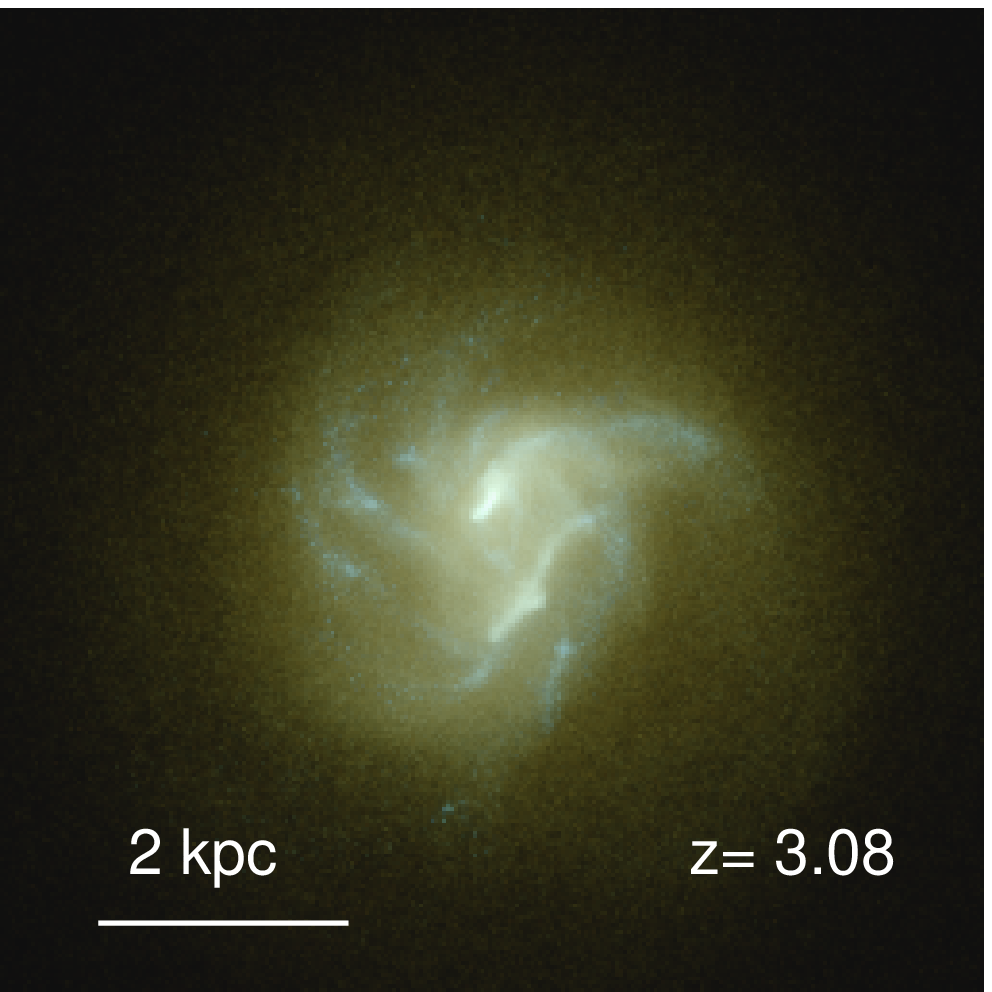}}}
  \centering{\resizebox*{!}{4.2cm}{\includegraphics{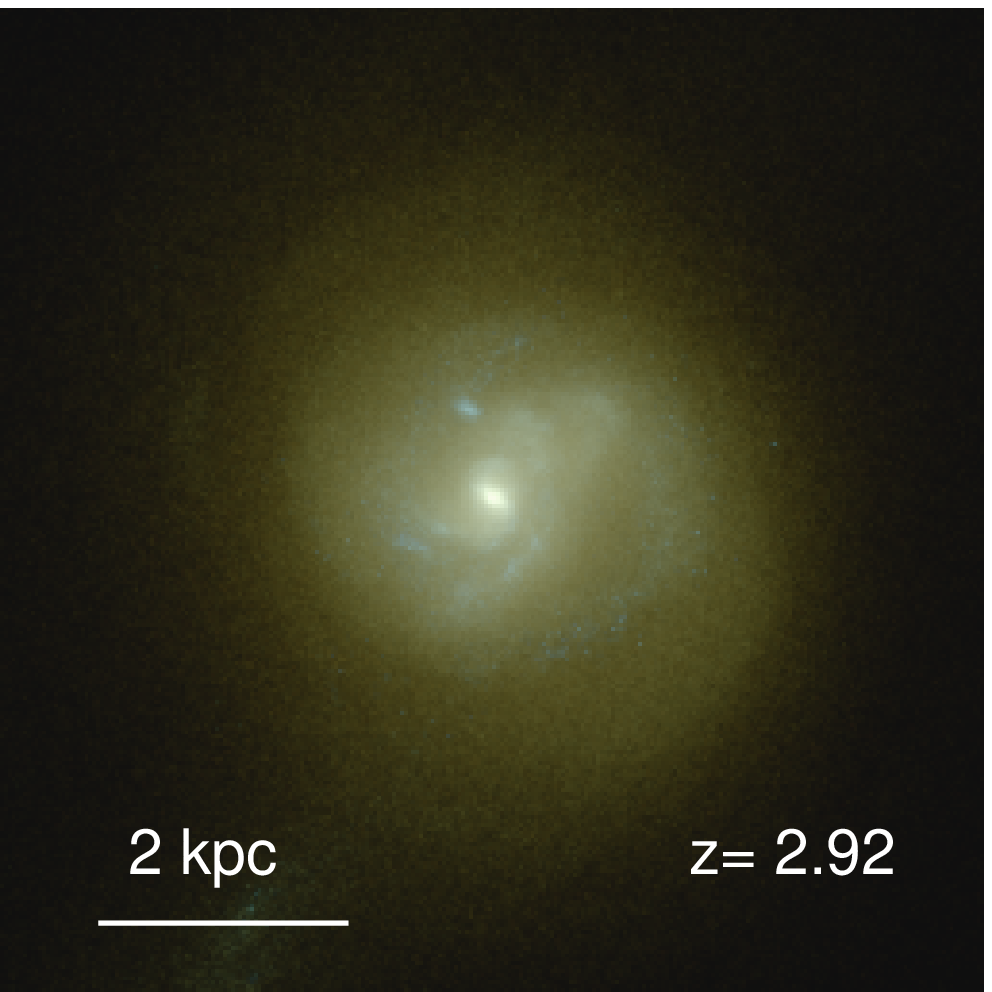}}}
  \centering{\resizebox*{!}{4.2cm}{\includegraphics{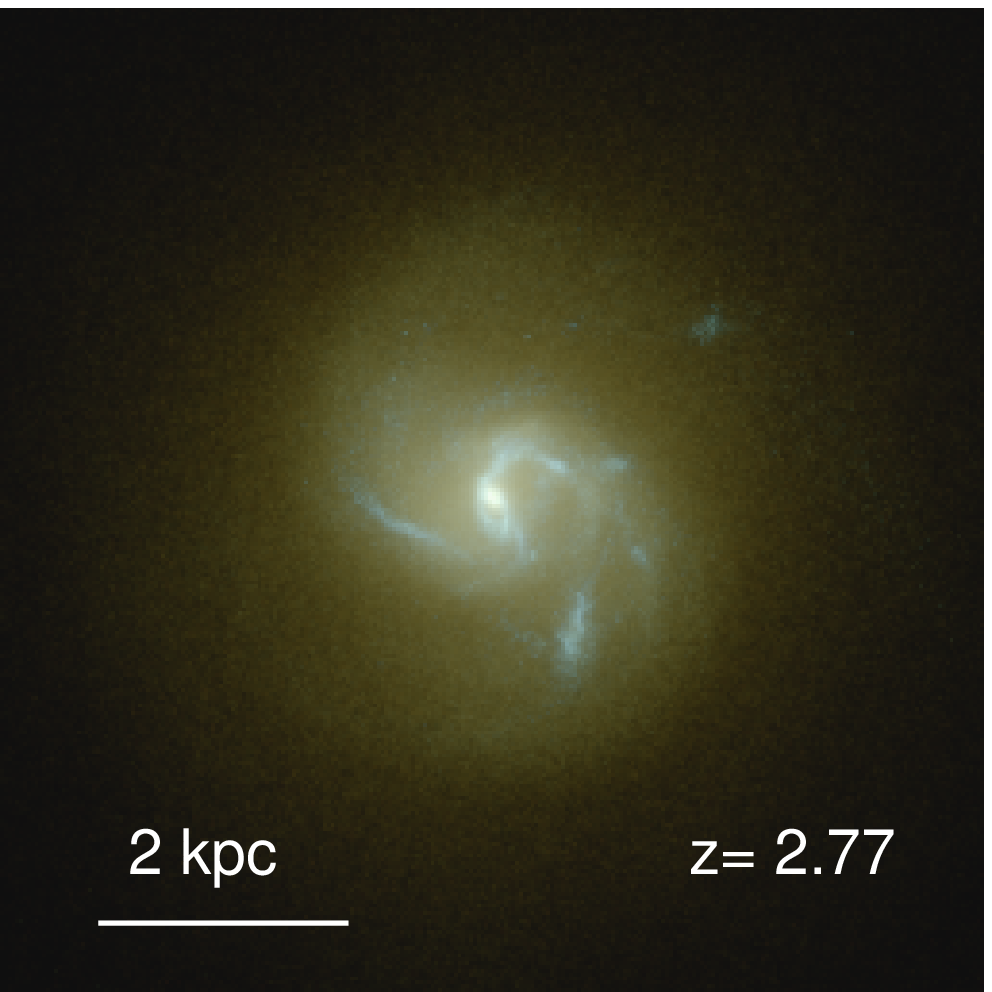}}}
  \centering{\resizebox*{!}{4.2cm}{\includegraphics{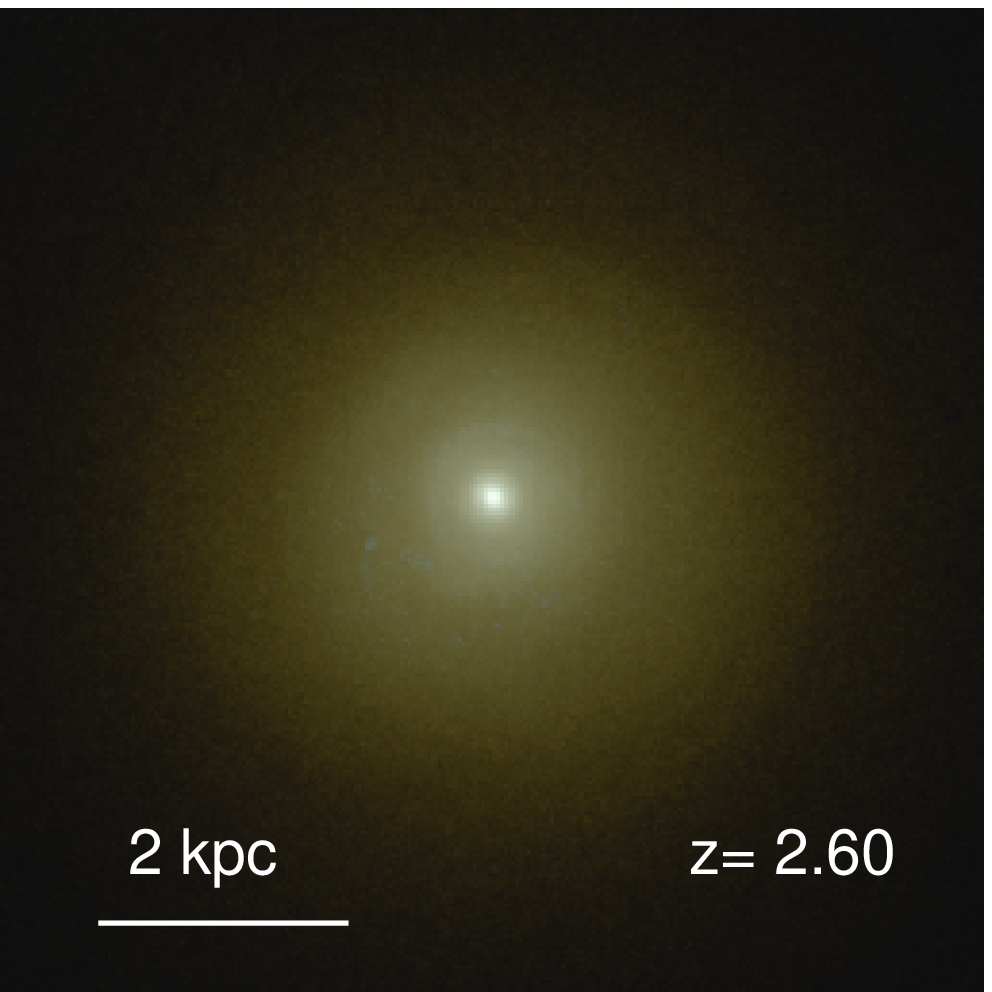}}}
  \centering{\resizebox*{!}{4.2cm}{\includegraphics{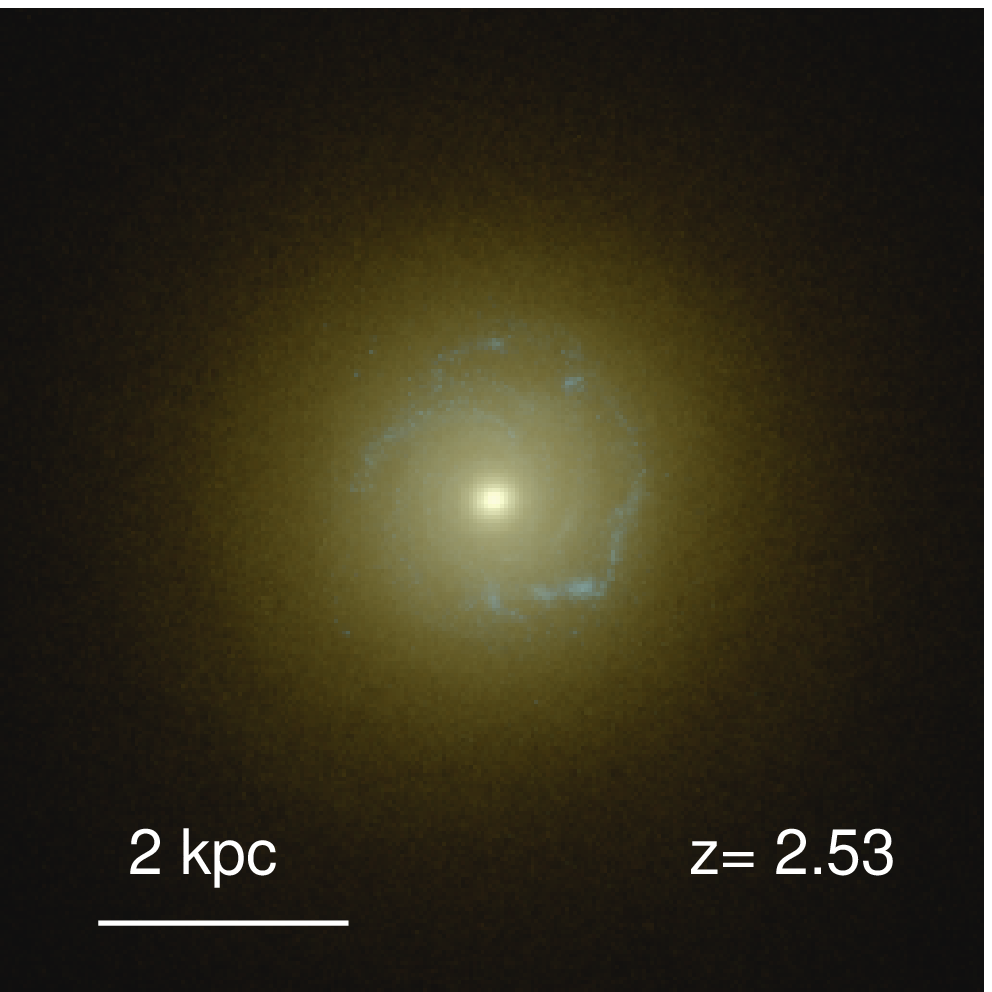}}}
  \centering{\resizebox*{!}{4.2cm}{\includegraphics{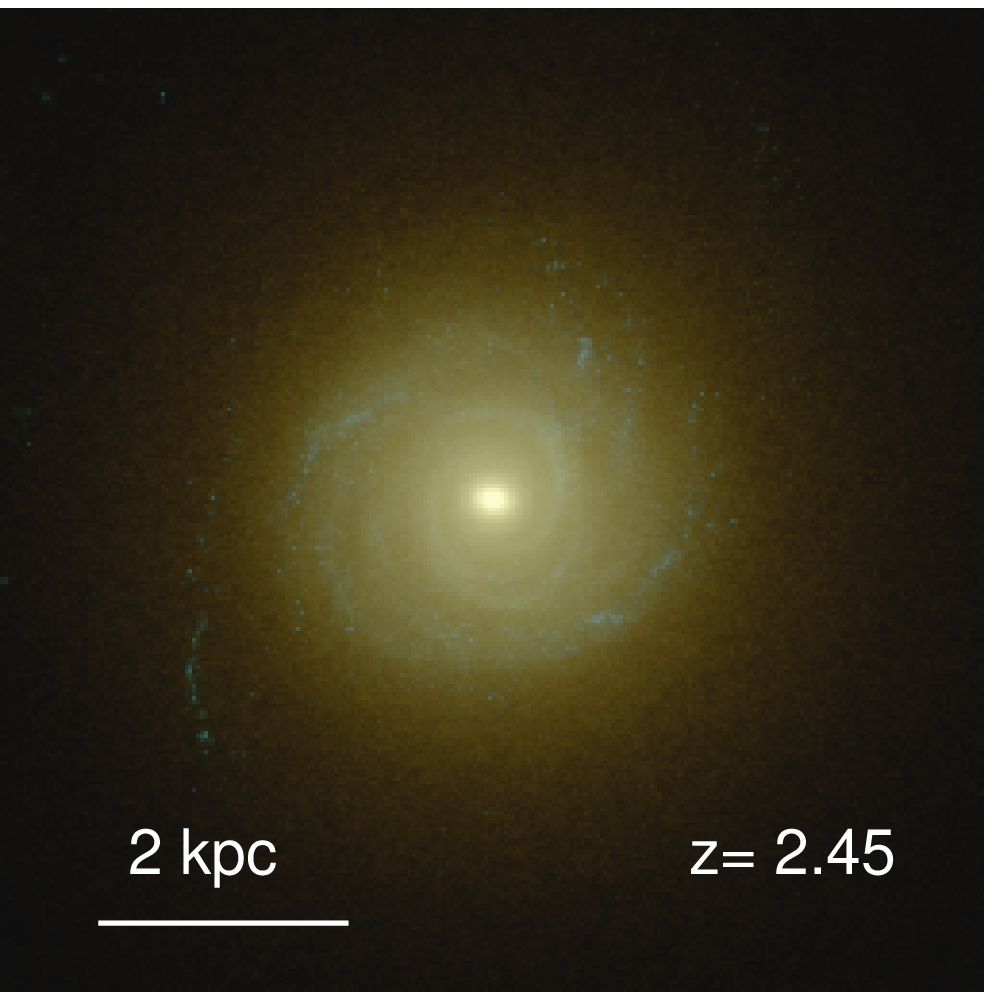}}}
  \centering{\resizebox*{!}{4.2cm}{\includegraphics{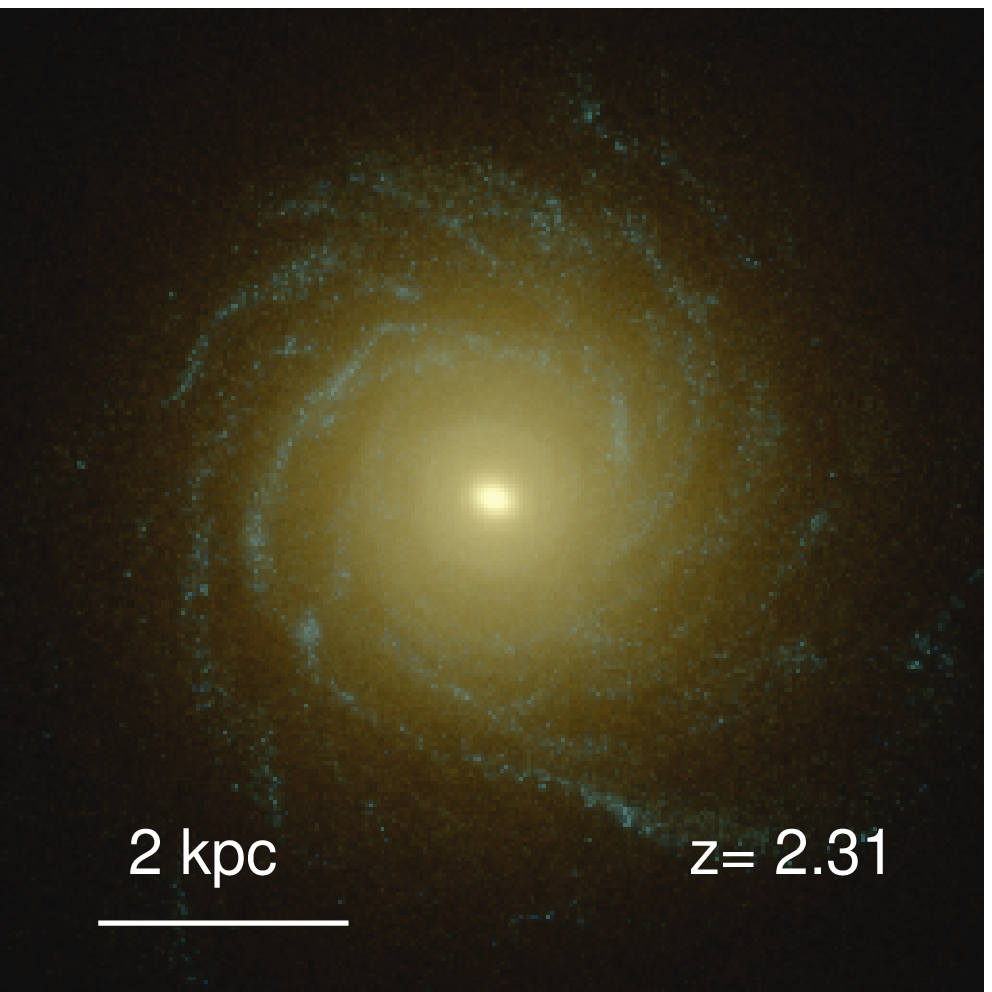}}}
  \centering{\resizebox*{!}{4.2cm}{\includegraphics{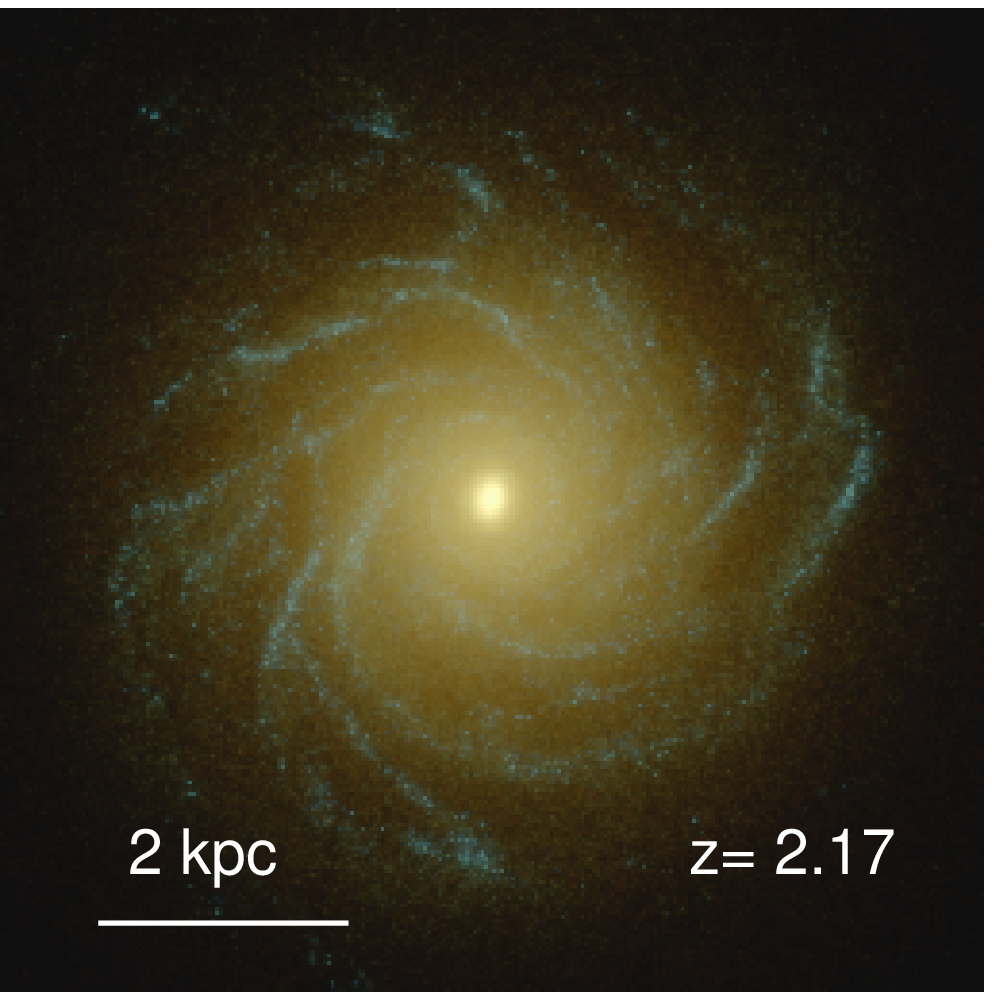}}}
  \centering{\resizebox*{!}{4.2cm}{\includegraphics{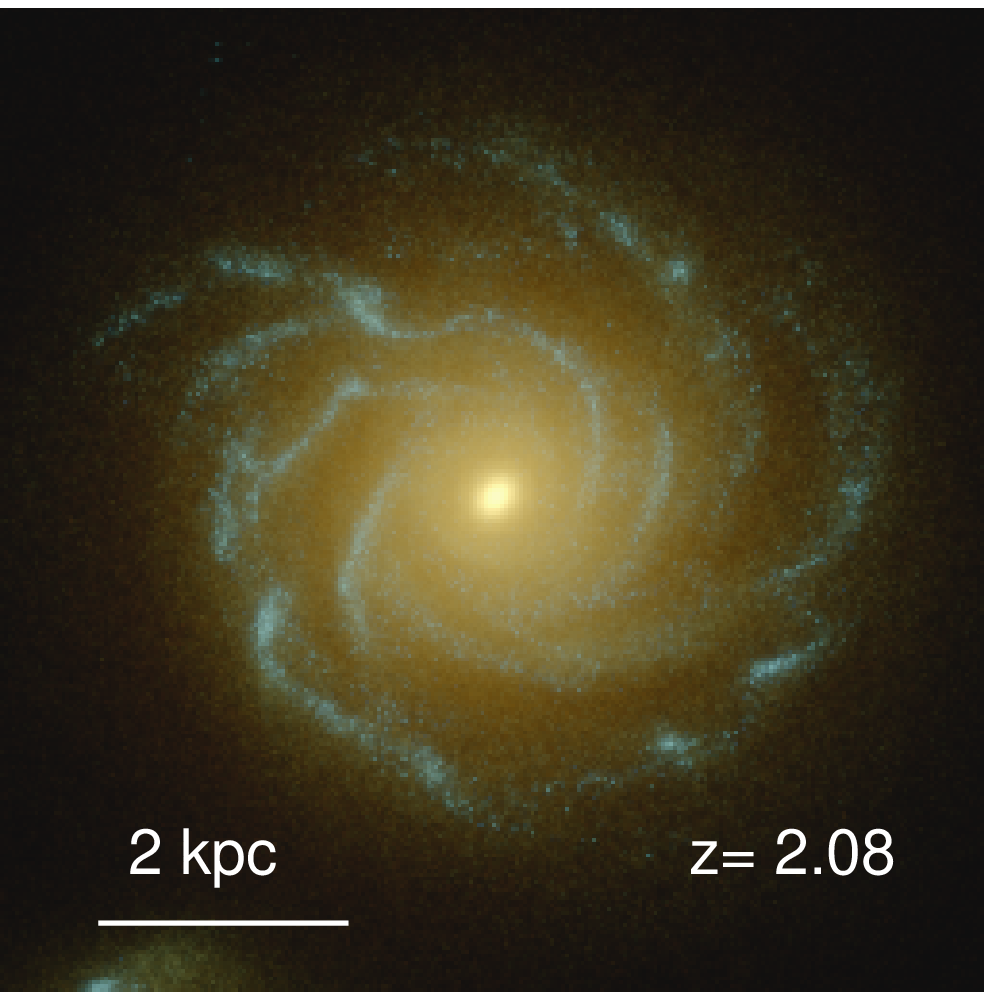}}}
  \centering{\resizebox*{!}{4.2cm}{\includegraphics{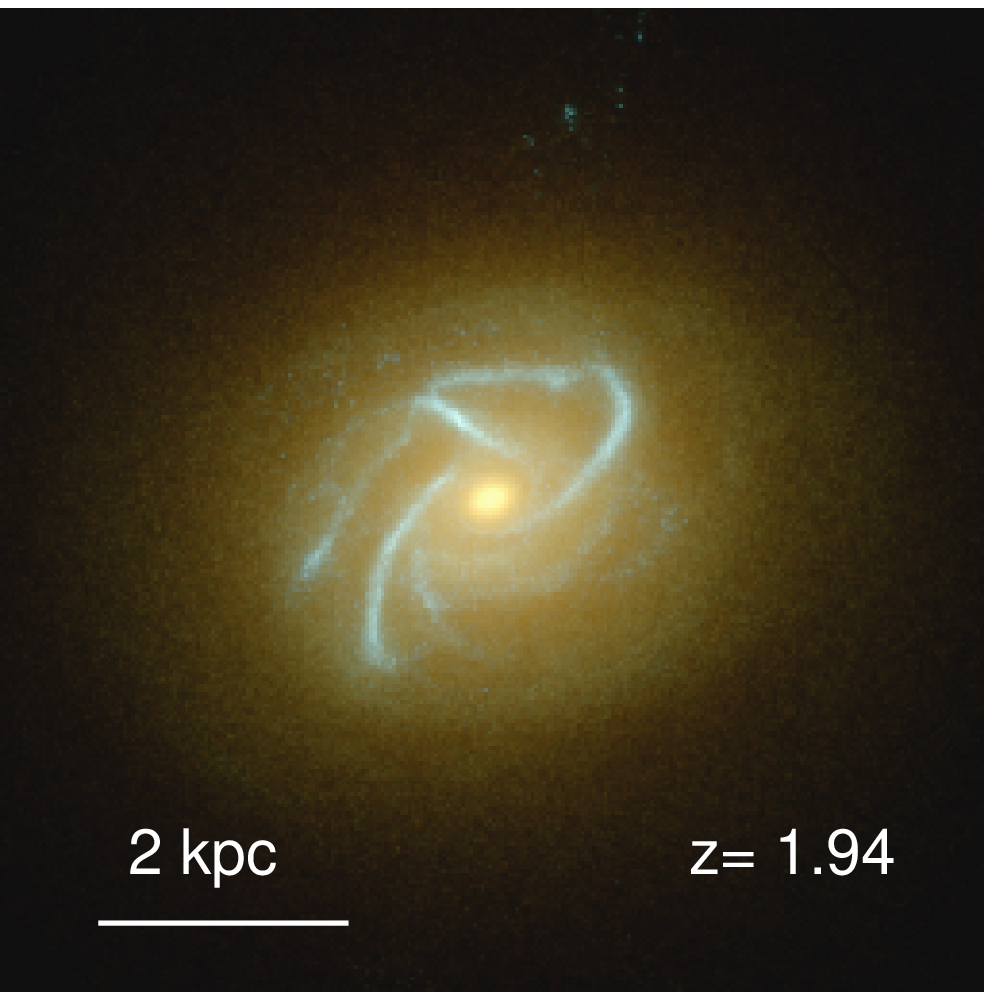}}}
  \caption{Same as Fig.~\ref{fig:shda} for the low-resolution run (SLDA). We have extended the stellar images to lower redshifts that we have been able to reach due to the lower resolution of the simulation. Note that though the resolution is now $\Delta x=34.8 \, \rm pc$, the galaxy looks alike the SHDA run (Fig.~\ref{fig:shda}). We also see the formation of a disc spiral galaxy  in the redshift range not probed by SHDA ($z\le2.31$).}
    \label{fig:slda}
\end{figure*}

In Fig.~\ref{fig:slda}, we show the stellar emission in the SLDA run with lower resolution ($\Delta x = 34.8 \, \rm pc$) than the SHDA run ($\Delta x = 8.7 \, \rm pc$) presented in Fig.~\ref{fig:shda}.
The morphology of the galaxy follows qualitatively  the same behavior than the high-resolution run with the morphologically disturbed shape at $z>3.6$, the disc-like shape with a bar component (see e.g. panels $z=3.44$ and $3.26$ in Fig.~\ref{fig:slda}) during $3.1<z<3.8$ (see Fig.~\ref{fig:barvsz}), the red nugget phase at $2.53\le z\le 2.77$, and the formation of a spiral disc galaxy at $z\le 2.45$ with a red bulge.
As for the SHDA run, the near-Eddington BH growth starts around $z=3.5$, once the galaxy quits the morphological disturbed phase and that the massive merger triggers the formation of a bar and nuclear inflows.

\end{document}